\pdfoutput=1
\documentclass[twocolumn,preprintnumbers,amsmath,amssymb]{revtex4-1}
\usepackage[utf8]{inputenc}
  \DeclareUnicodeCharacter{2265}{$\geq$}
  \DeclareUnicodeCharacter{2264}{$\leq$}
\usepackage{graphicx}
\usepackage{dcolumn}
\usepackage{bm}
\usepackage[caption=false]{subfig}
\usepackage{floatrow}
\usepackage{float}
\usepackage{array}
\usepackage{physics}
\usepackage{xcolor}
\usepackage{appendix}
\usepackage{bbm}
\usepackage[normalem]{ulem} 

\graphicspath{{./}{IMAGES/}}

\begin{document}

	\title[Physics Review B]{Momentum Signatures of Site Percolation in Disordered 2D Ferromagnets 
	}
	
	\author{Daniel Tay}
\affiliation{Laboratory for Solid State Physics, ETH Z\"urich, Otto-Stern-Weg 1, Zurich 8093, Switzerland}

	\author{Beno\^it Gr\'emaud}
\affiliation{Aix Marseille Univ, Université de Toulon, CNRS, CPT, IPhU, AMUTech, Marseille, France}

\affiliation{MajuLab, CNRS-UCA-SU-NUS-NTU International Joint Research Unit, Singapore}

\affiliation{Centre for Quantum Technologies, National University of Singapore, 117543 Singapore, Singapore}

	\author{Christian Miniatura}
\affiliation{MajuLab, CNRS-UCA-SU-NUS-NTU International Joint Research Unit, Singapore
}
\affiliation{
Centre for Quantum Technologies, National University of Singapore, 117543 Singapore, Singapore
}
\affiliation{
Department of Physics, National University of Singapore, 2 Science
Drive 3, Singapore 117542, Singapore
}
\affiliation{
School of Physical and Mathematical Sciences, Nanyang Technological University, 637371 Singapore, Singapore
}
\affiliation{
 Universit\'e C\^ote d'Azur, CNRS, INPHYNI, France
}

	\date{\today}
	
	\begin{abstract}
Since real devices necessarily contain defects, understanding wave propagation in disordered systems has proven a deep and important issue that led to several important developments in the field of electronic transport and metal-insulator transitions, in particular Anderson localization. In this work, we consider a two-dimensional square lattice of pinned magnetic spins with nearest-neighbour interactions and we randomly replace a fixed proportion of spins with nonmagnetic defects carrying no spin. We focus on the linear spin-wave regime and address the propagation of a spin-wave excitation with initial momentum ${\bm k}_0$. We compute the disorder-averaged momentum distribution obtained at time $t$ and show that the system exhibits two regimes. At low defect density, typical disorder configurations only involve a single percolating magnetic cluster interspersed with single defects essentially and the physics is driven by Anderson localization. In this case, the momentum distribution features the emergence of two known emblematic signatures of coherent transport, namely the coherent backscattering (CBS) peak located at $-{\bm k}_0$ and the coherent forward scattering (CFS) peak located at ${\bm k}_0$. At long times, the momentum distribution becomes stationary. However, when increasing the defect density, site percolation starts to set in and typical disorder configurations display more and more disconnected clusters of different sizes and shapes. At the same time, the CFS peak starts to oscillate in time with well defined frequencies. These oscillation frequencies represent eigenenergy differences in the regular, disorder-immune, part of the Hamiltonian spectrum. This regular spectrum originates from the small-size magnetic clusters and its weight grows as the system undergoes site percolation and small clusters proliferate. Our system offers a unique spectroscopic signature of cluster formation in site percolation problems.

\end{abstract}
	
\maketitle

\section{Introduction}
It is well known that two-dimensional (2D) ferromagnets can exhibit a collective behavior known as spin waves (magnons) which propagate throughout the  entire lattice \cite{serga2010yig}. As a rule of thumb, real solid-state systems always depart from a clean idealized situation. As is now well known, wave transport in such disordered media host a bunch of phenomena called weak localization effects \cite{Berg84,AkkMon2007}. Though, the most dramatic (and iconic) effect is Anderson localization, the complete suppression of transport through destructive interference \cite{Anderson1958absence,Gang4,Kramer93,MirlinEvers,50Years}, and its many-body version in the presence of interactions \cite{Abanin19}. In this context, it is particularly important to understand how disorder affect these spin systems \cite{Laflo2013}, and their wave propagation properties in particular \cite{Bruinsma1986,Monthus2010}. 

In 2D ferromagnets, disorder appears essentially under the form of point defects (vacancies, interstitial atoms and impurities), dislocations or grain boundaries \cite{MerminDefects,Kleinert,ChaLub95}. We will consider here the case of point defects: Starting from a clean 2D spin square lattice, a certain fraction $\rho$ of magnetic atoms are replaced by non-magnetic impurities (site percolation model). A similar situation has been considered in \cite{arakawa2018inplane,arakawa2018weak} for a disordered
Heisenberg antiferromagnet where defects were introduced on a square lattice using a ``partially substituted"  model. In marked contrast with our work however, the breaking of the lattice into independent clusters does not occur in that model since the ``partially substituted"   defects are still coupled to the rest of the lattice. The problem of spin wave propagation in disordered 2D square ferromagnets has been studied in the limit of relatively low defect densities in \cite{evers2015spin}. In this case, the impact of cluster formation is almost negligible and can be ignored: The usual predictions of Anderson localization theory apply. In particular, when analyzed in momentum space, coherent transport, localization and critical effects are revealed by the now emblematic coherent backscattering (CBS) \cite{CherCBS2012,JosseCBS2012,ghosh2015cbs} and coherent forward scattering (CFS) interference peaks \cite{CherCFS2012,lee2014cfs,ghosh2014coherent, ghosh2017cfs,HainautERO2017,Lemarie2017,HainautSym2018,martinez2020}. In this paper, we expand on the discussion in \cite{evers2015spin} in two important, and different, ways. First, we consider the limit of small fluctuations around the ground state, setting any magnetic anisotropy to zero. This allows us to study localization properties of linear magnon waves instead of having to deal with the more involved nonlinear Landau–Lifshitz–Gilbert (LLG) equation. We are thus able to derive some analytical results for microscopic transport parameters like the scattering mean free path, etc, in the low-density regime. Second, we also consider the regime of higher defect densities where cluster formation has a significant impact on transport properties. In particular, we show that cluster formation gives rise to periodic time oscillations of the CFS peak height.

The paper is organized as follows. First, we describe the effective disordered linear spin-wave Hamiltonian under study, discuss its main features and give the expression of the disorder-averaged momentum distribution. For the rest of the paper, we consider the case of uniform hopping amplitudes and highlight some important properties of the Hamiltonian inferred by its Laplacian matrix form. We next present numerical studies of the momentum distribution at  low defect densities for a spin wave with some initial momentum ${\bf k}_0$. In this case, cluster formation is negligible and we recover the expected known properties of wave propagation in momentum space: A CBS peak develops at the scattering mean free time scale on top of an isotropic diffusive background reached at the Boltzmann time scale and a CFS peak develops later at the Heisenberg time scale, signalling Anderson localization in the bulk. In particular, we show that the time behavior of the CFS contrast is given, as expected, by the spectral form factor. We also recover the $(ka)^4$ dependence of the scattering mean free rate at low momenta ($k$ is the wave number and $a$ the lattice constant). We then proceed to the higher defect density regime. In this case, cluster formation is no longer negligible, which dramatically impacts the time behavior of the CFS peak. We numerically show that the CFS peak height exhibit time oscillations. These oscillations originate from disorder-immune frequency differences associated to the small-cluster eigenspectra, thus allowing for a spectroscopic study of clusters. We conclude by giving some perspectives on the interplay between percolation and localization. Details of the calculations can be found in the Appendices.

\section{Effective disordered Hamiltonian for Spin Wave systems}

\subsection{Clean Hamiltonian}

We consider here a 2D ferromagnetic spin $S$ lattice system $\mathcal{L}$ with nearest-neighbor interactions described by the quantum Heisenberg Hamiltonian
\begin{equation}
    \mathcal{H}_S = - \sum_{(ij) \in \mathcal{L}} J_{ij} \, \bm{S}_i \cdot \bm{S}_j
\end{equation}
where $(ij)\equiv (ji)$ denotes the link that connects the {\it unordered pair} of nearest-neighbor sites $i$ and $j$ and where the coupling constants are all symmetric and positive $J_{ij} = J_{ji} >0$. At zero temperature, such a system exhibits a spontaneous magnetization where all spins are aligned along the same direction. We conveniently choose this (spontaneous symmetry-breaking) direction as the quantization axis $Oz$ of the system that we assume, for convenience, perpendicular to the lattice plane (this can be always achieved by adding an infinitesimal magnetic field along $Oz$ to help fix the direction of the spontaneous magnetization). 

We are interested in the linear dynamics of the long-wavelength excitations of the system (magnons) when disorder is present. Starting from the spin Heisenberg equations of motion, we derive in Appendix \ref{CleanH} the effective tight-binding clean Hamiltonian $H_0$ describing the linear spin-wave regime of the spin system. Introducing the positions states $\ket{\bm{r}_i}$ ($i \in \mathcal{L})$, satisfying $\langle \bm{r}_i | \bm{r}_j \rangle = \delta_{ij}$ and the closure relation $\sum_{i \in \mathcal{L}} \ket{\bm{r}_i} \bra{\bm{r}_i} = \mathbbm{1}_\mathcal{L}$, $H_0$ reads
\begin{equation}
    \begin{aligned}
        H_0 &= - \sum_{(ij) \in \mathcal{L}}  SJ_{ij} \, (\ket{\bm{r}_i} \bra{\bm{r}_j} + \ket{\bm{r}_j} \bra{\bm{r}_i}) + \sum_{i \in \mathcal{L}} U_i \, \ket{\bm{r}_i} \bra{\bm{r}_i} \\
        U_i & = \sum_{j \in \mathcal{N}(i)} SJ_{ij}
    \end{aligned}
\label{eq:cleanH}
\end{equation}
where $\mathcal{N}(i)$ denotes the set of all nearest-neighbor sites to site $i$. $H_0$ describes the dynamics of a spinless particle with nearest-neighbor hopping rates $t_{ij} = SJ_{ij}$ and onsite energies $U_i$. A crucial aspect of the effective model is that the properties of the onsite energy at a given site $i$ cannot be simply described by an independent (random) local variable. Indeed, $U_i$ being given by the sum of the hopping rates along all the links connected to site $i$, its (random) properties depend on the neighboring sites, attaining thereby a \textit{nonlocal} character.

\subsection{Model of Disorder and Disordered Hamiltonian}

We now introduce disorder in the system through a "site percolation" model: Starting from the clean system described by Eq.\eqref{eq:cleanH}, we replace at random a certain number $N_D$ of the $N$ magnetic sites by defects (non-magnetic sites), leaving $N_m = N-N_D = (1-\rho) \, N$ magnetic sites alive where $\rho = N_D/N$ is the defect density  \footnote{An alternative method to introduce disorder would be to turn each site of the lattice $\mathcal{L}$ into a defect with a fixed probability $p$. This flip-method gives similar results, see Appendix \ref{Perco}.}.

This random arrangement of defects within the lattice of magnetic sites drastically modifies the effective Hamiltonian of the whole system. Indeed, The physical effect of these defects on the system is threefold: First, {\it all} coupling terms connecting a pair of nearest-neighbor sites where at least one of the two sites is defective are set to 0: $J_{ij}=0$ when $i$ or $j$ or both are defective. This means that nonmagnetic defects {\it decouple} from magnetic sites and that the effective disorder Hamiltonian $H$ {\it only involves sums over magnetic sites}. Second, the onsite energy of a magnetic site $i$ depends now on the number of its nearest-neighbor defects. Thirdly, the onsite energy of nonmagnetic sites is set to zero. Defining the subset $\mathcal{M} \subseteq \mathcal{L}$ of magnetic sites, the disordered Hamiltonian $\mathcal{H}$ is readily obtained from $H_0$ by the replacement $\mathcal{L} \to \mathcal{M}$ and $U_i \to V_i$:
\begin{equation}
\label{eq:DisoH}
\begin{aligned}
 \mathcal{H} &= - \!\!\! \sum_{(ij) \in \mathcal{M}}  SJ_{ij} \, (\ket{\bm{r}_i} \bra{\bm{r}_j} + \ket{\bm{r}_j} \bra{\bm{r}_i}) + \sum_{i \in \mathcal{M}} V_i \, \ket{\bm{r}_i} \bra{\bm{r}_i} \\
 V_i &= \sum_{j \in \mathcal{N}(i) \cap \mathcal{M}} SJ_{ij}. 
\end{aligned}
\end{equation}

\subsection{Scattering Approach and Defect Hamiltonian}
\label{ScatApp}

Since $\mathcal{H}$ is acting on the $N_m$ sites of the random subspace $\mathcal{M}$ alone, diagrammatic expansions and related analyses of the disordered spin wave system based on $\mathcal{H}$ are not straightforward as we would have to deal with the random boundaries of $\mathcal{M}$. For this, a better suited approach is to extend $\mathcal{H}$ to a disorder Hamiltonian $H$ acting on the full regular lattice $\mathcal{L}$. 

The Hamiltonian $H$ is readily obtained from the clean Hamiltonian $H_0$ through the replacement $J_{ij} \to J_{ij} m_im_j$, where the random variable $m_i$ takes value $m_i = 0$ if site $i$ is a defect (probability $p=\rho$) and takes value $m_i = 1$ otherwise (probability $q=1-\rho$) \cite{evers2015spin}. A disorder configuration is then fully characterized by the set of values $\{m_i, i\in \mathcal{L}\}$ and there are $\genfrac(){0pt}{2}{N}{N_{\!\footnotesize{D}}} = \frac{N!}{N_{\!D}!(N-N_{\!D})!}$ possible such disorder configurations. 

The next step is to break $H = H_0 + H_d$ into the sum of the clean Hamiltonian $H_0$ and a defect Hamiltonian $H_d$ acting on the full regular lattice $\mathcal{L}$. We introduce the link random variable $m_{ij}=1-m_im_j$ with property $m_{ij} = 0$ if both endpoints of the link $(ij)$ are magnetic and $m_{ij}=1$ otherwise. Then, we have:
\begin{equation}
H_d = \sum_{(ij)} m_{ij} SJ_{ij}  \, (\ket{\bm{r}_i}\bra{\bm{r}_j} + \ket{\bm{r}_j}\bra{\bm{r}_i}) + \sum_{i \in \mathcal{L}} W_i \ket{\bm{r}_i}\bra{\bm{r}_i}, 
\end{equation}
where $W_i= - \sum_{j \in \mathcal{N}(i)} m_{ij} SJ_{ij}$. As easily checked, we do have $H_d=0$ and $H=H_0$ if all sites are magnetic and $H_d=-H_0$ and $H=0$ if all sites are defective. It can also be seen that the presence of defects reduces the onsite energy of their nearest-neighbor sites since $W_i \leq 0$. 

At this stage, it is advantageous to further break the defect Hamiltonian $H_d = \overline{H_d} + \delta H_d$ into a disorder-averaged part $\overline{H_d}$ and a fluctuating part $\delta H_d$ with zero mean $\overline{\delta H_d} = 0$. We next write $H = \widetilde{H}_0 + \delta H_d$ and introduce the disorder-renormalized clean Hamiltonian $\widetilde{H}_0 \equiv \overline{H} = H_0 + \overline{H_d}$. 

The scattering approach, based on $H_d$, and the diagrammatic expansion of the Green's function associated to $H = \widetilde{H}_0 + \delta H_d$, will be detailed in Section \ref{Diag} and Appendices \ref{Self}, \ref{ScatDef} and \ref{ScatTime}.

\subsection{$\mathcal{M}$ or $\mathcal{L}$ as Hilbert Spaces}

We begin by partitioning the full lattice $\mathcal{L} = \mathcal{M} \bigoplus \mathcal{D}$ into its (disjoint) magnetic $\mathcal{M}$ and defective $\mathcal{D}$ subspaces and we define the corresponding projectors on these subspaces:
\begin{equation}
    \mathcal{P} = \sum_{i \in \mathcal{M}} \ket{\bm{r}_i} \bra{\bm{r}_i} \hspace{1cm} \mathcal{Q} = \sum_{i \in \mathcal{D}} \ket{\bm{r}_i} \bra{\bm{r}_i}
\end{equation}
with $\mathcal{P}+\mathcal{Q} = \mathbbm{1}_\mathcal{L}$. 

From a physical point of view, the relevant Hilbert space and Hamiltonian are the magnetic subspace $\mathcal{M}$ and $\mathcal{H}$, Eq.\eqref{eq:DisoH}, since the defects do not carry any spin. However, as seen in the previous paragraph, it is also useful to embed $\mathcal{M}$ in $\mathcal{L}$ and work with Hamiltonian $H=H_0+H_d$ and $\mathcal{L}$ as the Hilbert space.

From the identity $H = (\mathcal{P} + \mathcal{Q})) H (\mathcal{P} + \mathcal{Q}))$, we infer $H=\mathcal{P}H\mathcal{P} + \mathcal{Q}H\mathcal{Q}$ since $H$ does not couple the subspaces $\mathcal{M}$ and $\mathcal{D}$. We readily have $\mathcal{H} = \mathcal{P} H \mathcal{P}$ and thus $H = \mathcal{H} + H_{\mathcal{D}}$. By construction, $H_{\mathcal{D}} = \mathcal{Q} H \mathcal{Q} =0$. It is important however to keep track of $H_{\mathcal{D}}$ in Green's function calculations, see paragraph~\ref{Diag}.

\section{Momentum Distribution}

\subsection{Plane Wave States}

We first define the plane wave states through
\begin{equation}
    \ket{\bm{k}} = \frac{1}{\sqrt{N}} \, \sum_{i \in \mathcal{L}} e^{\mathrm{i} \bm{k} \cdot \bm{r}_i} \, \ket{\bm{r}_i}.
\label{eq:PWstates}
\end{equation}
They are normalized to $\langle \bm{k}' | \bm{k} \rangle = \delta_{\bm{k}\bm{k}'}$ and resolve the identity on $\mathcal{L}$, namely $\sum_{\bm{k} \in \Omega} \ket{\bm{k}} \bra{\bm{k}} = \mathbbm{1}_\mathcal{L}$, where $\Omega$ is the first Brillouin zone of $\mathcal{L}$. The two following identities prove particularly useful:
\begin{equation}
\sum_{i \in \mathcal{L}} e ^{\mathrm{i} (\bm{k}-\bm{k}')\cdot\bm{r}_i} = N \delta_{\bm{k}\bm{k}'} \hspace{0.5cm} \sum_{\bm{k} \in \Omega} e ^{\mathrm{i} \bm{k}\cdot (\bm{r}_i-\bm{r}_j)} = N \delta_{ij}.
\label{eq:Useful}
\end{equation}

\subsection{Truncated Plane Wave States and On-Shell Energy}

We define a normalised truncated plane wave state by
\begin{equation}
    \ket{\Phi_{\bm{k}}} = \frac{\mathcal{P}\ket{\bm{k}}}{\sqrt{1-\rho}}.
    \label{eq:Trunc}
\end{equation}
It represents an initial plane wave state $\ket{\bm{k}}$ projected onto $\mathcal{M}$. The denominator $\sqrt{1-\rho}$ is introduced to normalise the state to $\langle \Phi_{\bm{k}} \ket{\Phi_{\bm{k}}} = 1$ for each disorder configuration. Indeed, it is easy to check that $\langle \bm{k} | \mathcal{P} | \bm{k} \rangle = 1-\rho$ for each disorder configuration. We further have $\overline{\langle \bm{k}' | \mathcal{P} | \bm{k} \rangle} = (1-\rho) \delta_{\bm{k}\bm{k}'}$, where $\overline{(\cdot\cdot\cdot)}$ denotes the disorder average.

We next define the on-shell energy $E_{os}$ associated to $\ket{\Phi_{\bm{k}}}$ by
\begin{equation}
    E_{os} = \overline{\bra{\Phi_{\bm{k}}} \mathcal{H} \ket{\Phi_{\bm{k}}} } = \frac{\bra{\bm{k}} \overline{\mathcal{H}} \ket{\bm{k}}}{1-\rho}.
\label{eq:OnShell}
\end{equation}
It represents the average energy of a plane wave state projected onto $\mathcal{M}$. The dispersion of energies $\delta E$ around $E_{os}$ is further defined by $\delta E^2 = \overline{\bra{\Phi_{\bm{k}}} (\mathcal{H}-E_{os})^2 \ket{\Phi_{\bm{k}}} }$.

\subsection{Momentum Distribution}

In the rest of this paper, we are interested in the disorder-averaged momentum distribution $n(\bm{k},t)$ obtained from the time evolution of an initial plane wave state $\ket{\bm{k}_0}$ under $\mathcal{H}$, namely $\ket{\Phi(t)} = e^{- \mathrm{i} \mathcal{H} t/\hbar} \ket{\Phi_{\bm{k}_0}}$.

The disorder-averaged momentum distribution then reads:
\begin{equation}
    \begin{aligned}
        n(\bm{k}, t) &= \overline{| \langle \bm{k} | \Phi (t)\rangle|^2} =\overline{|\bra{\bm{k}} e^{- \mathrm{i} \mathcal{H} t/\hbar} \ket{\Phi_{\bm{k}_0}} |^2} \\
        & \equiv \frac{\overline{|\langle \bm{k} | \mathcal{P} e^{- \mathrm{i} \mathcal{H} t/\hbar} \mathcal{P}| \bm{k}_0 \rangle |^2}}{1-\rho}.
    \end{aligned}
\label{eq:MomDist}
\end{equation}
It is easy to check that $\sum_{\bm{k}} n(\bm{k},t) = 1$ and that this equality in fact holds at the level of each single disorder configuration as it should. In the following, we will numerically compute and analyze $n(\bm{k}, t)$ for different defect densities $\rho$.  

At this point, we introduce the $(N-N_D)$ eigenstates $\ket{\varphi_n}$ and eigenenergies $\epsilon_n$ of the Hamiltonian $\mathcal{H}$ seen as a $(N-N_D)\times (N-N_D)$ square matrix acting on $\mathcal{M}$, namely $\mathcal{H}|\varphi_n\rangle = \epsilon_n |\varphi_n\rangle$ with $\langle \varphi_n | \varphi_m \rangle = \delta_{nm}$. This means that we perform a change of basis in the $\mathcal{M}$ subspace and write $\mathcal{P}=\sum_n \ket{\varphi_n}\bra{\varphi_n}$ ($n = 1, ..., N-N_D$) so that
\begin{equation}
    \mathbbm{1}_\mathcal{L} = \sum_n \ket{\varphi_n}\bra{\varphi_n} + \sum_{i\in \mathcal{D}} \ket{\bm{r}_i}\bra{\bm{r}_i}
\end{equation}
with the normalization: 
\begin{equation}
    \sum_{i \in \mathcal{L}} |\varphi_n(\bm{r}_i)|^2 \equiv \sum_{i \in \mathcal{M}} |\varphi_n(\bm{r}_i)|^2 = 1
\end{equation}
since $\varphi_n(\bm{r}_i) = 0$ if $i \in \mathcal{D}$. The momentum distribution, Eq.\eqref{eq:MomDist}, then reads:
\begin{equation}
    n(\bm{k}, t) = \frac{\overline{\sum_{n,m} e^{- \mathrm{i} \omega_{nm} t} \varphi_n(\bm{k}) \varphi^*_m(\bm{k}) \varphi_m(\bm{k}_0) \varphi^*_n(\bm{k}_0)}}{1-\rho},
    \label{eq:nkt}
\end{equation}
where $\omega_{nm} = \omega_n-\omega_m$, $\omega_n=\epsilon_n/\hbar$ and where $\varphi_n(\bm{k}) \equiv \langle \bm{k}|\varphi_n\rangle$ is given by
\begin{equation}
    \varphi_n(\bm{k}) = \frac{1}{\sqrt{N}} \, \sum_{i \in \mathcal{M}} e^{-\mathrm{i} \bm{k} \cdot \bm{r}_i} \varphi_n(\bm{r}_i).
\end{equation}
It is easy to see that:
\begin{equation}
    \sum_{\bm{k} \in \Omega} \varphi^*_m(\bm{k}) \varphi_n(\bm{k}) = \delta_{nm} \hspace{0.5cm}  \sum_{n} |\varphi_n(\bm{k})|^2 = 1-\rho
\end{equation}

At this stage, it is important to note that, because the onsite energies depend on the neighboring sites properties, $\mathcal{H}$ {\it is not} the restriction of $\mathcal{P} H_0 \mathcal{P}$ to $\mathcal{M}$. If it were the case, the eigenenergies of $H$ would not be random and its eigenfunctions would be simply related to the plane waves states.  Indeed, since $\mathcal{P} H_0 \mathcal{P} = \sum_{\bm{k}} \epsilon^0_{\bm{k}} \, \mathcal{P}\ket{\bm{k}}\bra{\bm{k}}\mathcal{P}$, the eigenenergies would be the non random clean ones $\epsilon^0_{\bm{k}}$ while the eigenstates $\mathcal{P}\ket{\bm{k}}$ would be random but with rather simple statistical properties.

\section{Hamiltonian with uniform hopping rates}

In the rest of our Paper, we consider the case of uniform hopping rates $J_{ij} = J$. Then, the clean onsite energies are uniform $U_i \equiv U = ZJS$ ($Z$ is the lattice coordination number) whereas the onsite disorder energy $V_{i} = Z_{i}JS$ depend on the {\it local} environment of defects ($Z_i \leq Z$ is the total number of {\it magnetic} nearest neighbors of site $i$). In this case, we have:
\begin{equation}
 \mathcal{H} = JS \Big[- \!\!\! \sum_{(ij) \in \mathcal{M}} (\ket{\bm{r}_i} \bra{\bm{r}_j} + \ket{\bm{r}_j} \bra{\bm{r}_i}) + \sum_{i \in \mathcal{M}} Z_i \, \ket{\bm{r}_i} \bra{\bm{r}_i}\Big]
\label{eq:DisorH}
\end{equation}
and 
\begin{equation}
H_d = JS \Big[\sum_{(ij)} m_{ij} (\ket{\bm{r}_i}\bra{\bm{r}_j} + \ket{\bm{r}_j}\bra{\bm{r}_i}) + \sum_{i \in \mathcal{L}} w_i \ket{\bm{r}_i}\bra{\bm{r}_i}\Big], 
\label{eq:effectivehwithdefects}
\end{equation}
with $w_{i} = (m_i Z_{i}-Z)$ and $Z_i = \sum_{j \in \mathcal{N}(i)} m_j \leq Z$.

For concreteness, we further consider the case of a two-dimensional square lattice ($Z=4$) made of $S=1/2$ spins (lattice constant $a=1$ set to unity) containing $N=50\times50 = 2500$ sites and we use periodic boundary conditions. In all our numerical simulations, we have used $J$ and $t_J = \hbar /J$ as the energy and time units of the system.

\subsection{Free Dispersion Relation}

Under the previous assumptions, the clean Hamiltonian reads
\begin{equation}
    H_0 = JS \Big[- \! \sum_{(ij)}  (|\bm{r}_i\rangle\langle \bm{r}_j| +  |\bm{r}_j\rangle\langle \bm{r}_i|) + Z\sum_{i \in \mathcal{L}} |\bm{r}_i\rangle\langle \bm{r}_i|\Big]
\label{eq:CleanHUni}
\end{equation}
and is readily diagonalized in the plane wave states basis
\begin{equation}
        H_0 = \sum_{\bm{k}} \epsilon^0_{\bm{k}} \, \ket{\bm{k}}\bra{\bm{k}},
\end{equation}
featuring the well-known free dispersion relation given by
\begin{equation}
    \begin{aligned}
         \varepsilon^0_{\bm{k}} & = 2JS \, \big[2 - \cos (k_x a) - \cos (k_y a) \big] \\
         & \approx JS \, (ka)^2 \hspace{1cm} (ka \ll 1)
    \end{aligned}
    \label{eq:DispRel}
\end{equation}
in dimension two. 

\subsection{Renormalized Clean Dispersion Relation}
Since the disorder average restores the original translation invariance properties of the system, it is obvious that $\overline{H_d}$, and in turn $\widetilde{H}_0$, are diagonal in $\bm{k}$.

As shown in Appendix \ref{RenormH}, it is easy to compute $\overline{H_d}$ from the statistical properties of the link random variable $m_{ij}$ and we find $\overline{H_d} = -\rho(2-\rho) H_0$. As a consequence, $\widetilde{H}_0$ is diagonal in the plane wave basis with a renormalized clean dispersion relation $\varepsilon_{\bm{k}}$:
\begin{equation}
    \begin{aligned}
        \widetilde{H}_0 &= \sum_{\bm{k}} \varepsilon_{\bm{k}} |\bm{k}\rangle\langle \bm{k}| \\
        \varepsilon_{\bm{k}} &= (1-\rho)^2 \, \varepsilon^{0}_{\bm{k}}.
    \end{aligned}
\label{eq:RenormH0}
\end{equation}

Note that, in this case, the on-shell energy, Eq.~\eqref{eq:OnShell}, simply reads
\begin{equation}
    E_{os} = \frac{\overline{\sum_n \epsilon_n \, |\varphi_n(\bm{k})|^2}}{1-\rho} \\
    = \frac{\varepsilon_{\bm{k}}}{1-\rho} = (1-\rho) \, \varepsilon^0_{\bm{k}}
\label{eq:OnShell2}
\end{equation}

\subsection{Laplacian Matrix}
\label{LapMat}

From Eq.~\eqref{eq:DisorH}, we can write $\mathcal{H} = JS \, \mathcal{R}$. In the position basis, the operator $\mathcal{R}$ takes the form of a Laplacian matrix~\cite{Merris1994}:
\begin{equation}
\label{eq:Laplacian}
	\mathcal{R}_{ij} = \left\lbrace
		\begin{aligned}
			&\textrm{deg}(i) \quad i=j\\
			& -1 \quad i\neq j, \, \textrm{{\it i} and {\it j} adjacent}\\
			& 0 \quad \textrm{otherwise}
		\end{aligned}
	\right.
\end{equation}
and where $\textrm{deg}(i)$ represents the degree of site $i$, the number of edges emanating from it. In our case, the simple graph, characterized by its vertices and edges, associated to $\mathcal{R}$ identifies with the magnetic lattice $\mathcal{M}$ and $\textrm{deg}(i)$ is simply the number of magnetic sites coupled to site $i$. 

The properties of Laplacian matrices on graphs are well-studied \cite{Godsil2001,Brouwer2012}. Of particular relevance to us are the following ones: 
\begin{itemize}
    \item $\mathcal{R}$ is positive semi-definite (all its eigenvalues are positive) 
    \item The sum of entries in every column and row being zero, the lowest eigenvalue of $\mathcal{R}$ is thus zero
    \item Its multiplicity is the number of connected components of $\mathcal{M}$, i.e. the number of isolated magnetic clusters in our context.
\end{itemize}

\section{Momentum Distribution at Small Defect Densities $\rho \ll 1$}

\begin{figure*}[!htbp]
\sidesubfloat[]{\includegraphics[width = 0.45\linewidth]{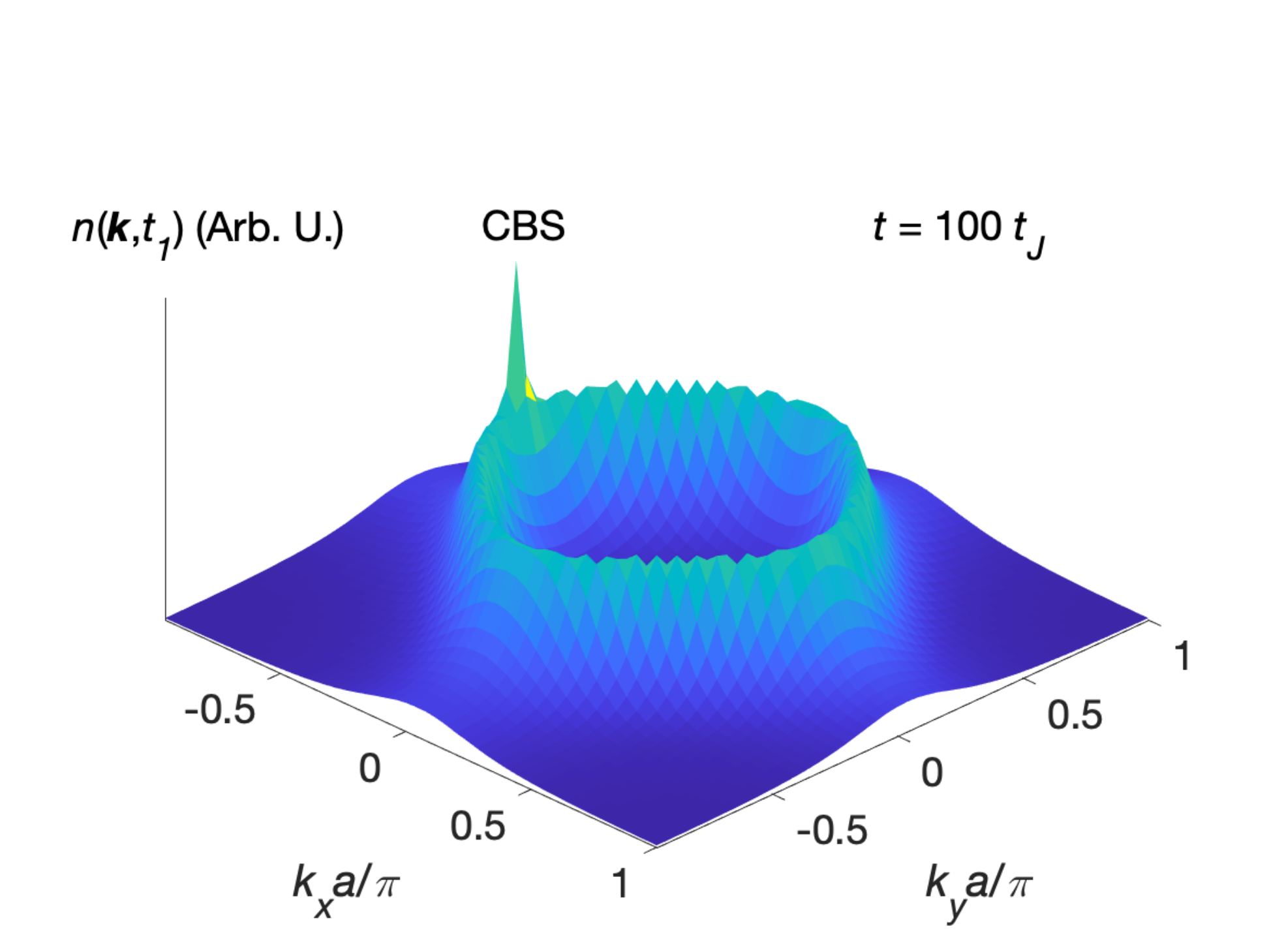}\label{fig:kdistlow}}
\sidesubfloat[]{\includegraphics[width = 0.45\linewidth]{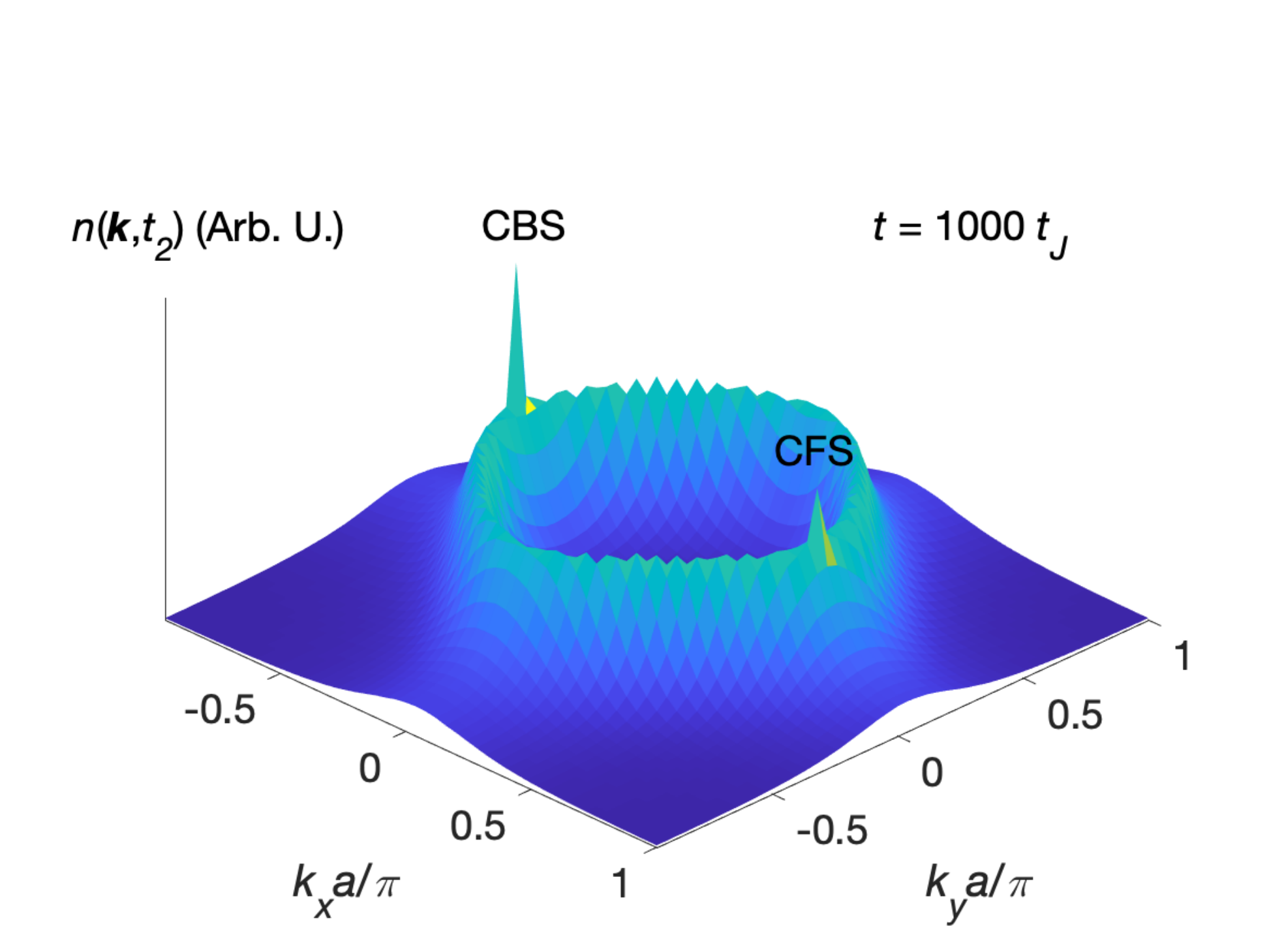}\label{fig:kdistlong}}
\medskip

\sidesubfloat[]{\includegraphics[width = 0.3\linewidth]{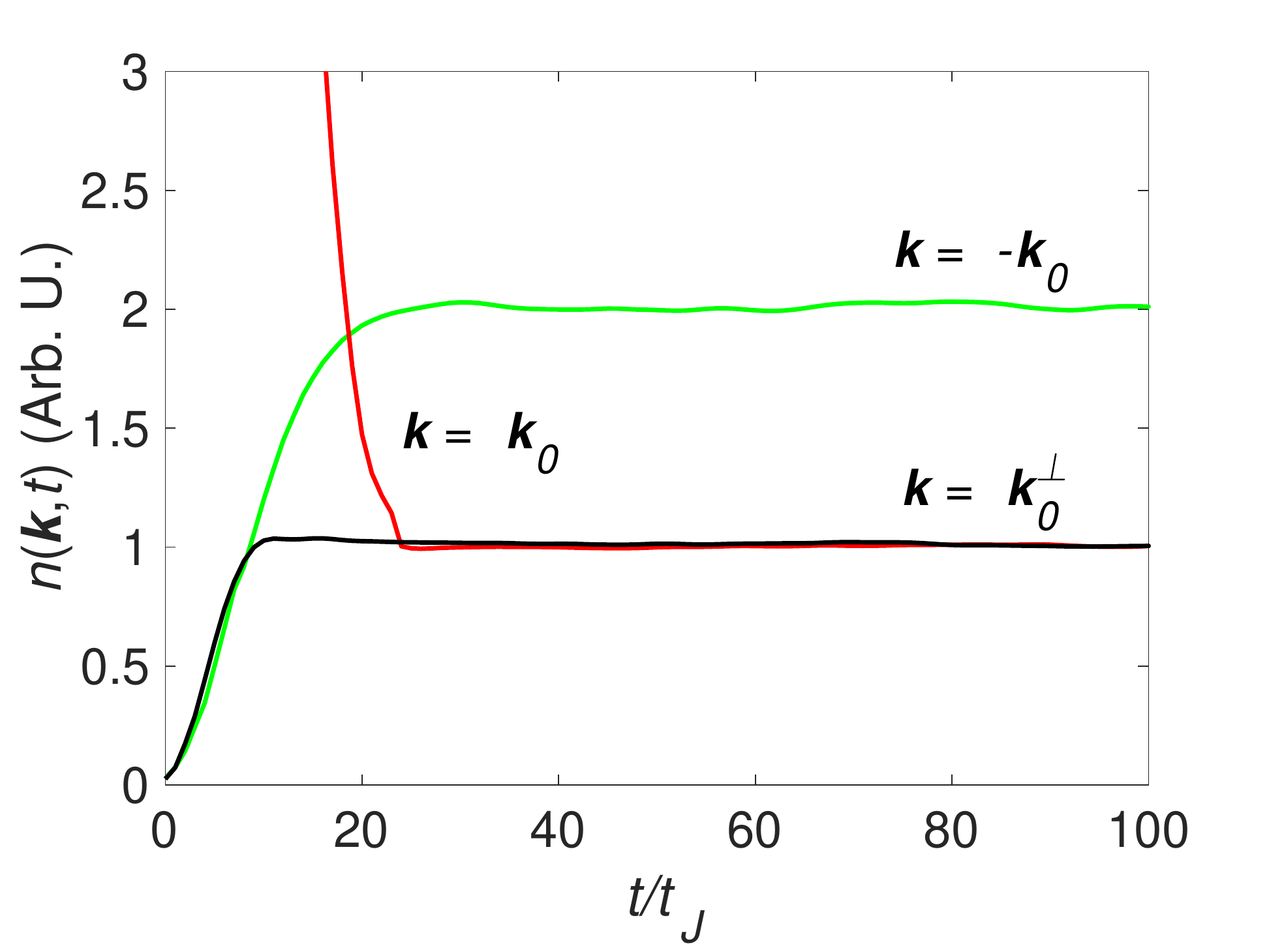}\label{fig:CFSat10low}}
\sidesubfloat[]{\includegraphics[width = 0.3\linewidth]{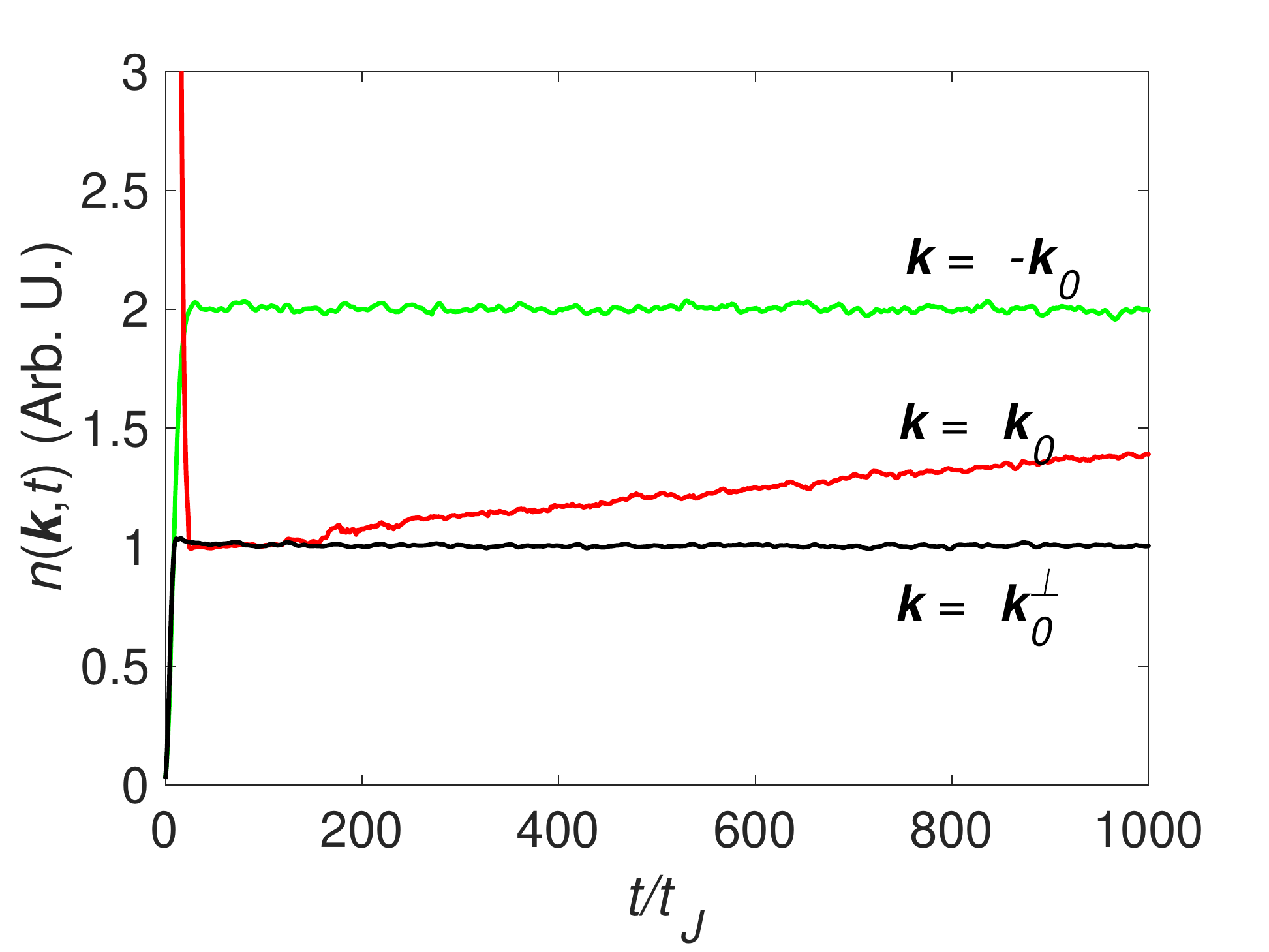}\label{fig:CFSat10med}}
\sidesubfloat[]{\includegraphics[width = 0.3\linewidth]{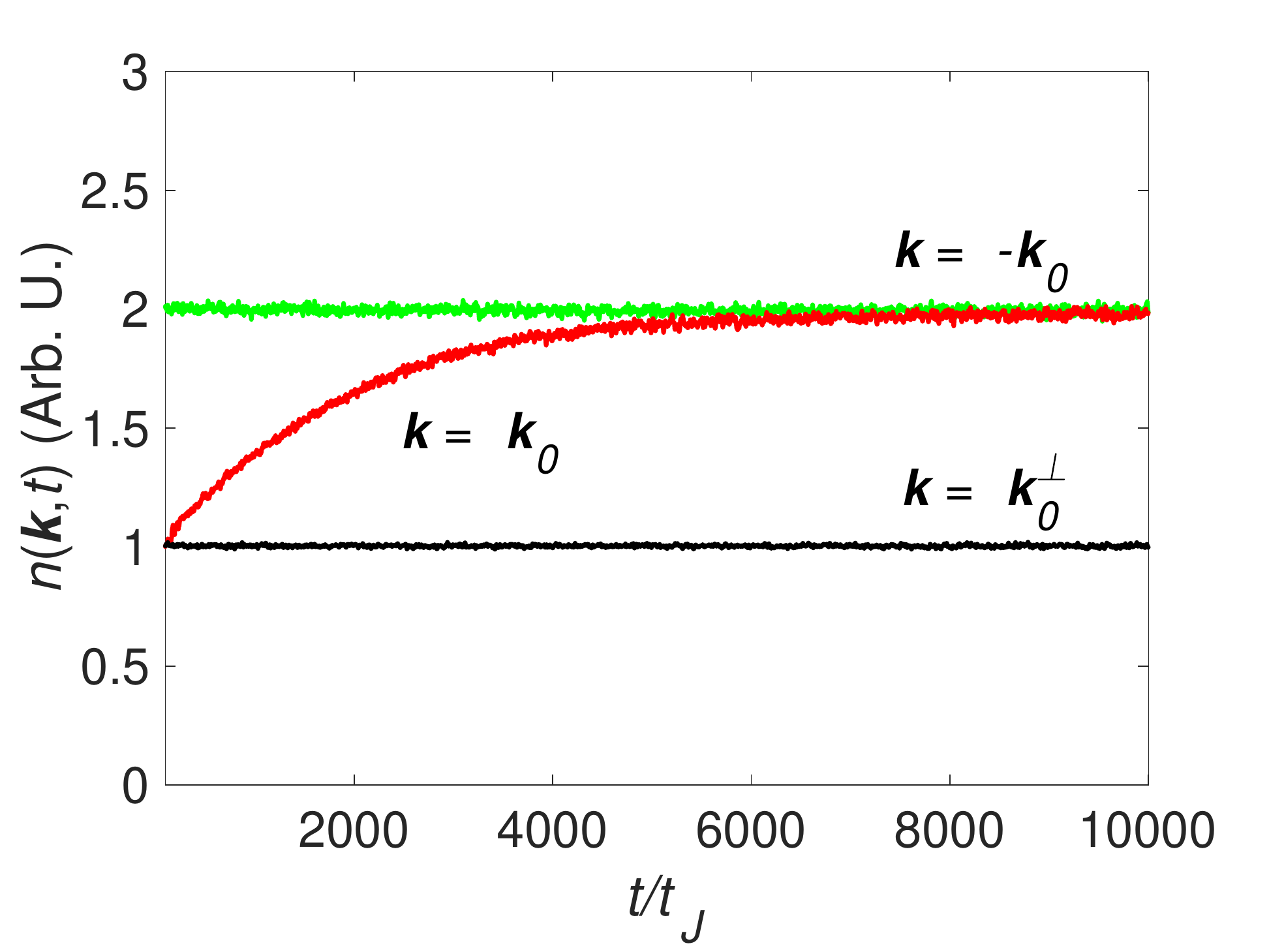}\label{fig:CFSat10long}}
\medskip

\caption{Disorder-averaged momentum distribution $n(\bm{k},t)$ at $\rho = 0.1$. Panel (a):  
Momentum distribution at time $t_1 = 100 t_J$. An isotropic (diffusive) ring-shaped structure has fully developed with a fully contrasted CBS peak on top of it and located at $-\bm{k}_0$. The CBS peak signals that disorder-immune constructive interference effects at play in the system. We have $n(-\bm{k}_0,t_1) = 2 n(\bm{k}^\perp_0,t_1)$ where $\bm{k}^\perp_0 = k_0 \hat{\bm{e}}_y$.
Panel (b): Momentum distribution at time $t_2 = 1000 t_J$. The CFS peak starts to emerge, signalling the onset of Anderson localization.  At longer time scales, the CFS and CBS peaks achieve equal peak values twice that of the diffusive background. 
Panel (c): Small time evolution of $n(\bm{k},t)$ at the CBS point $\bm{k}= -\bm{k}_0$, at the CFS point $\bm{k}=\bm{k}_0$ and at $\bm{k}= \bm{k}^\perp_0$ (background).  The initial momentum distribution $n(\bm{k}_0, t=0)$ decays over a time scale given by the scattering mean free time $\tau_s$. Meanwhile the diffusive background grows and the two curves achieve the same stationary  value $n_B(\bm{k}_0)$, at a time scale given by the Boltzmann time $\tau_B$. Concomitantly, the CBS peak grows to reach its stationary value. We note that the CFS peak not emerged yet and that both $\tau_s$ and $\tau_B$ are  of the same order of magnitude  (several $t_J$). Panel (d): Same as (c) but at intermediate times. The CFS peak at $\bm{k}_0$ emerges around $t \sim 150 t_J$, signalling the onset of Anderson localization.  
Panel (e): Same as (c) but over much larger times. The CBS peak and background no longer evolve at $t \gg \tau_B$. Conversely,the CFS peak evolves at a time scale given by the Heisenberg time $\tau_H$ (in the range of a few thousands $t_J$ here). At  $t \gg \tau_H$, the CFS peak achieves the same height as the CBS peak, twice that of the background. }
    \label{fig:CBSandCFS}
\end{figure*}

\subsection{Time Scales and Expected General Behavior of the Momentum Distribution}

Several physical time scales characterize the propagation of waves in random media \cite{AkkMon2007, Sheng1995, Kuhn2005, Kuhn2007}. The first one is the scattering mean free time $\tau_s$ which gives the average time interval separating two successive scattering events suffered by an initial plane wave $\ket*{\bm{k}_0}$. Over time, the wave momenta are being randomized by scattering events and the system reaches isotropization after the transport (or Boltzmann) mean free time $\tau_{B}$. At low momenta $k_0a \ll 1$ where geometrical lattice effects can be discarded, the disorder-averaged momentum distribution achieves a ring-shaped structure of radius $k_0$, width given by $\tau_s^{-1}$ and constant ridge height $n_B(\bm{k}_0)$. During the isotropization process, and in the absence of any dephasing phenomena that could break phase-coherent effects, a narrow coherent backscattering (CBS) peak emerges around the direction $-\bm{k}_0$, signalling that disorder-immune constructive interference effects are at play. After $\tau_B$, the CBS peak has fully developed on top of the diffusive background with a stationary peak value $2n_B(\bm{k}_0)$. Wave transport, apart from the CBS peak, has entered the ergodic regime and the system explores all of its accessible energy shell through a diffusion process. If the conditions are right, then the system enters a localization regime after some localization time $\tau_{loc}$: The diffusion process slows down and stops. This is the celebrated Anderson localization phenomenon. Finally, for times much longer than the Heisenberg time $\tau_H$, the quantum limit where energy levels are resolved is reached and the system no longer evolves \cite{ghosh2014coherent}. During this process, a narrow coherent forward-scattering (CFS) peak develops at $\bm{k}_0$, twining the CBS peak in the long run. As it turns out, CFS is a smoking gun of bounded motion and, thus, of Anderson localization in the bulk. Both the CBS and CFS angular sizes $\Delta\theta \sim \xi^{-1}$ are given, in this regime, by the inverse of the localization length $\xi$ of the system \cite{ghosh2014coherent,ghosh2015cbs,ghosh2017cfs}. 

All in all, for time reversal symmetric systems like the one we consider here, the following picture emerges for the disorder-averaged momentum distribution $n(\bm{k},t)$: At small times ($t \lesssim \tau_B$), the initial momentum distribution $n(\bm{k},t=0)$, peaked at $\bm{k}_0$, is depleted by scattering events and a diffusive shell forms with mean radius $|\bm{k}_0|$ while the CBS peak emerges at $-\bm{k}_0$. After isotropization is reached, this distribution does not evolve significantly until the localization threshold is crossed ($\tau_B < t < \tau_{loc}$). In turn, a CFS peak starts to develop at $\bm{k}_0$ ($t > t_{loc}$) and twins the CBS peak over a time scale given by $\tau_H$. In the long-time limit ($t \gg \tau_H$), the momentum distribution does not evolve any more (quantum limit) and is given by a perfectly contrasted twin peak interference structure on top of an otherwise isotropic diffusive-like background.

\subsection{Actual Behavior of the Linear Spin-Wave System at $\rho \ll 1$}

\begin{figure}[htbp]
    \centering
    \includegraphics[width=1\linewidth]{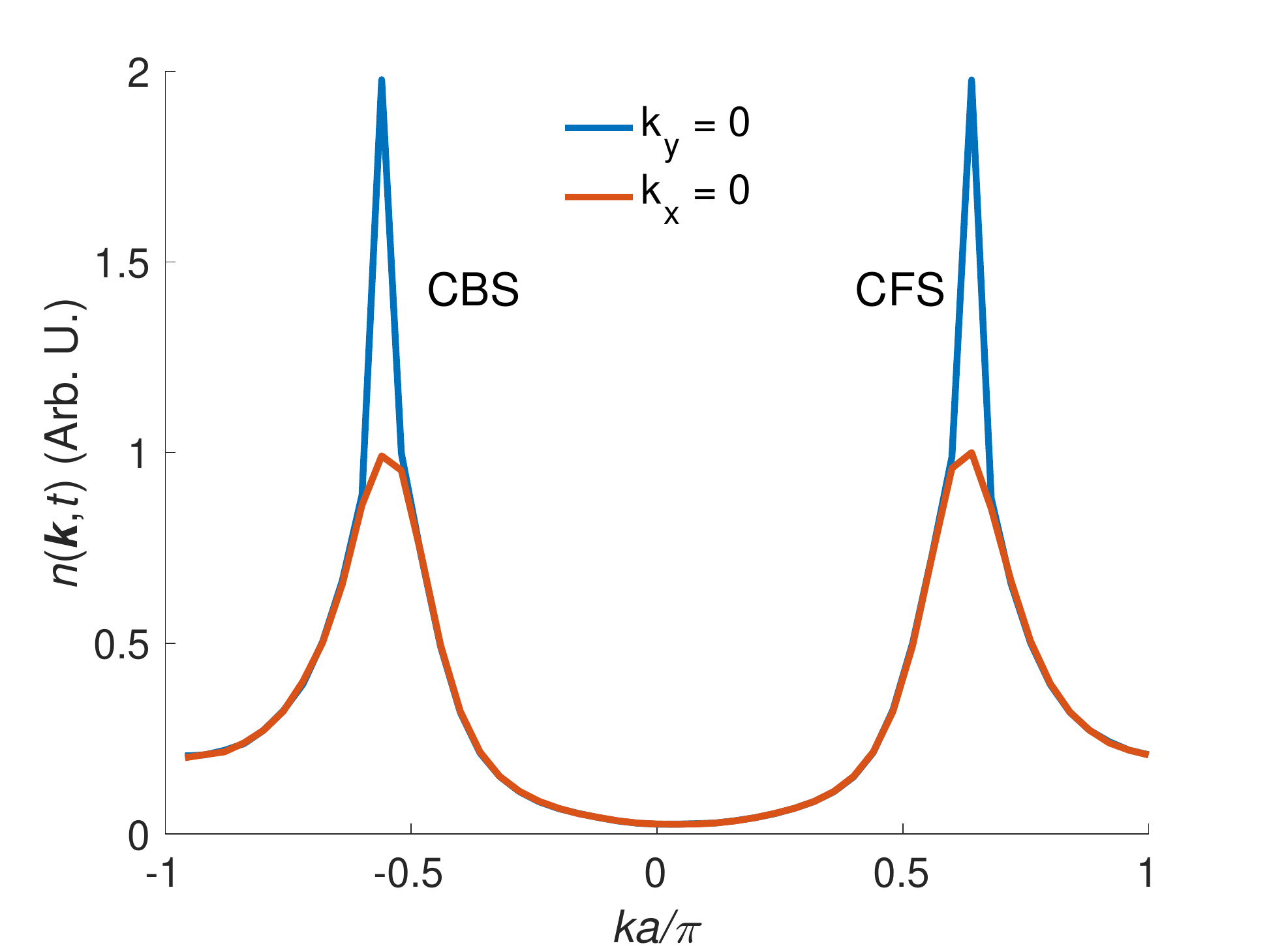}
    \caption{Cuts of the momentum distribution $n(\bm{k},t)$ at time $t = 10000 \, t_J \gg \tau_H$. The parameters are $\rho =0.1$ and $\bm{k}_0 = k_0 \, \hat{\bm{e}}_x$ with $k_0a = 0.6\pi$. Red color: Cut along $k_x=0$. This profile, obtained along a direction orthogonal to $\bm{k}_0$, represents a good approximation of what a cut of the actual isotropic background would be. We arbitrarily set the rim value of this background at unity. Blue color: Cut along $k_y =0$ showing the CBS and CFS peaks. In this long-time limit, both the CBS and CFS peaks have become twin peaks and thus have same heights and widths. Their equal width $\Delta k = 2\pi/\xi$ defines the localization length. We find $\Delta k \approx 0.145/a$ and thus $\xi \approx 43.33 \, a$, to be compared to the linear size of the system $L= \sqrt{N} \, a = 50 \, a$. Since $L/\xi \sim 1$, the emergence of the CFS peak here is in fact dominated by finite-size effects and not by genuine Anderson localization (see text). 
    }
    \label{fig:PeakWidth}
\end{figure}

In Fig.\ref{fig:CBSandCFS}, we plot the disorder-averaged momentum distribution $n(\bm{k},t)$ and its temporal behavior for the linear spin-wave system described by Eqs.\eqref{eq:DisorH} on a 2D square lattice at low defect density $\rho= 0.1$ and an initial plane wave momentum $\bm{k}_0 = k_0 \hat{\bm{e}}_x$ with $k_0a = 0.6\pi $ ($a$ is the lattice constant). The total number of lattice sites is $N=50 \times 50$ and the total number of magnetic sites is $N_m = (1-\rho) N = 2250$. The numerical results are averaged over $1000$ disorder configurations in panels (a) and (b) and over $10000$ disorder configurations in panels (c), (d) and (e). In panels (c), (d) and (e), we have arbitrarily fixed the background rim value to unity. As seen from the data obtained in Fig.\ref{fig:CBSandCFS} at defect density $\rho = 0.1$, the linear spin-wave system does exhibit the expected signatures of localization theory in momentum space at low defect densities $\rho \ll 1$. As predicted, at intermediate times $\tau_B \lesssim t \lesssim \tau_{loc}$, only the CBS peak is seen on top of an isotropic ring-shaped  background (Fig.\ref{fig:kdistlow}) whereas the CFS peak starts to grow at $t \gtrsim t_{loc}$, after the localization onset has been reached (Fig.\ref{fig:kdistlong}). Note that the bell-shaped features visible at the edges of the contour plot of the momentum distribution were also observed in \cite{evers2015spin}. They can be attributed to lattice effects (Brillouin zone boundaries).

From Fig.\ref{fig:CFSat10low}, we see that $\tau_{s} \sim \tau_{B}$, both being in the range of a few $t_J$ whereas, from Fig.\ref{fig:CFSat10long}, we see that the CFS peak grows with a much larger time scale in the range of a few thousands of $t_J$. Do note that the CBS peak value is also reached after a time scale of the order of $\tau_B$. Do also note that the background value (measured at a momentum $\bm{k}_{\perp} \perp \bm{k}_0$) and the CBS peak value do not change over time after isotropization has been fully reached: The only visible dynamics happen for the CFS peak height. As the two peaks become mirror images of each other in the long-time limit, the wings of the 2 peaks also change over time (not shown here) \cite{ghosh2015cbs}.
Last but not least, we observe that, in the long-time limit, the CFS peak height reaches the same height as the CBS peak, twice the height of the diffusive background like predicted by theory \cite{ghosh2014coherent}. We will see later that this expected temporal picture changes dramatically when the defect density $\rho$ is increased. 

Fig.\ref{fig:PeakWidth} shows the CBS and CFS peaks on top of the diffusive background at times $t \gg \tau_H$. In this long-time limit, the CBS and CFS structures become twin peaks and their equal widths $\Delta k$ relate to the size of the localization length through $\Delta k = 2\pi / \xi$. For the parameters of the numerical computation, we find $\Delta k \, a \simeq 0.145$ and thus $\xi \simeq 43.33 \, a$ which is comparable to the linear size of the system $L= \sqrt{N} \, a = 50 \, a$. This does not come as a complete surprize here. As is well known from the scaling theory of localization \cite{Gang4, Kramer93, MirlinEvers}, two-dimensional systems are always localized in the infinite-size limit in the absence of spin-orbit coupling, which is the case here. Furthermore, as shown in \cite{ghosh2014coherent}, a proper analysis of the localization dynamics in momentum space requires energy filtering. Indeed, the localization length actually depends on both the energy $E$ at which the dynamics is analyzed and on the disorder strength. As a consequence, the size $L$ of the system provides a natural cut-off: At energies and disorder strengths such that $\xi < L$, the system is genuinely Anderson localized and develops a CFS peak at long times. However, at energies and disorder strengths such that $\xi > L$, the system appears extended but still develops a CFS peak at long times, the boundaries of the system playing the role of a classical localization box. As is also the rule, the weaker the disorder strength, the larger $\xi$. However, the catch is that, for two-dimensional systems, $\xi$ increases exponentially when the disorder strength decreases. With the parameters chosen in our numerical simulations ($S=1/2$, $\rho=0.1$, $k_0a=0.6\pi$, $\varepsilon_{\bm{k}_0} = 1.31 J$), we have $E_{os} = 1.178 J$ and $\delta E = 0.003 J$, where $\delta E$ is the dispersion of eigenvalues). 

It seems that, in the energy range ($E_{os} \pm \delta E/2$) accessible to the system, we have not reached the regime where genuine Anderson localization dominates the dynamics for the system size explored. As a consequence, the emergence of the CFS peak here is mainly due to finite-size effects. On the other hand, we would like to stress that this does not affect the momentum signature of the percolation, which, as explained below, appears in the temporal behavior of the CFS peak after it has emerged.

\begin{figure*} [!htbp] 
        \sidesubfloat[]{%
        \includegraphics[width=.45\linewidth]{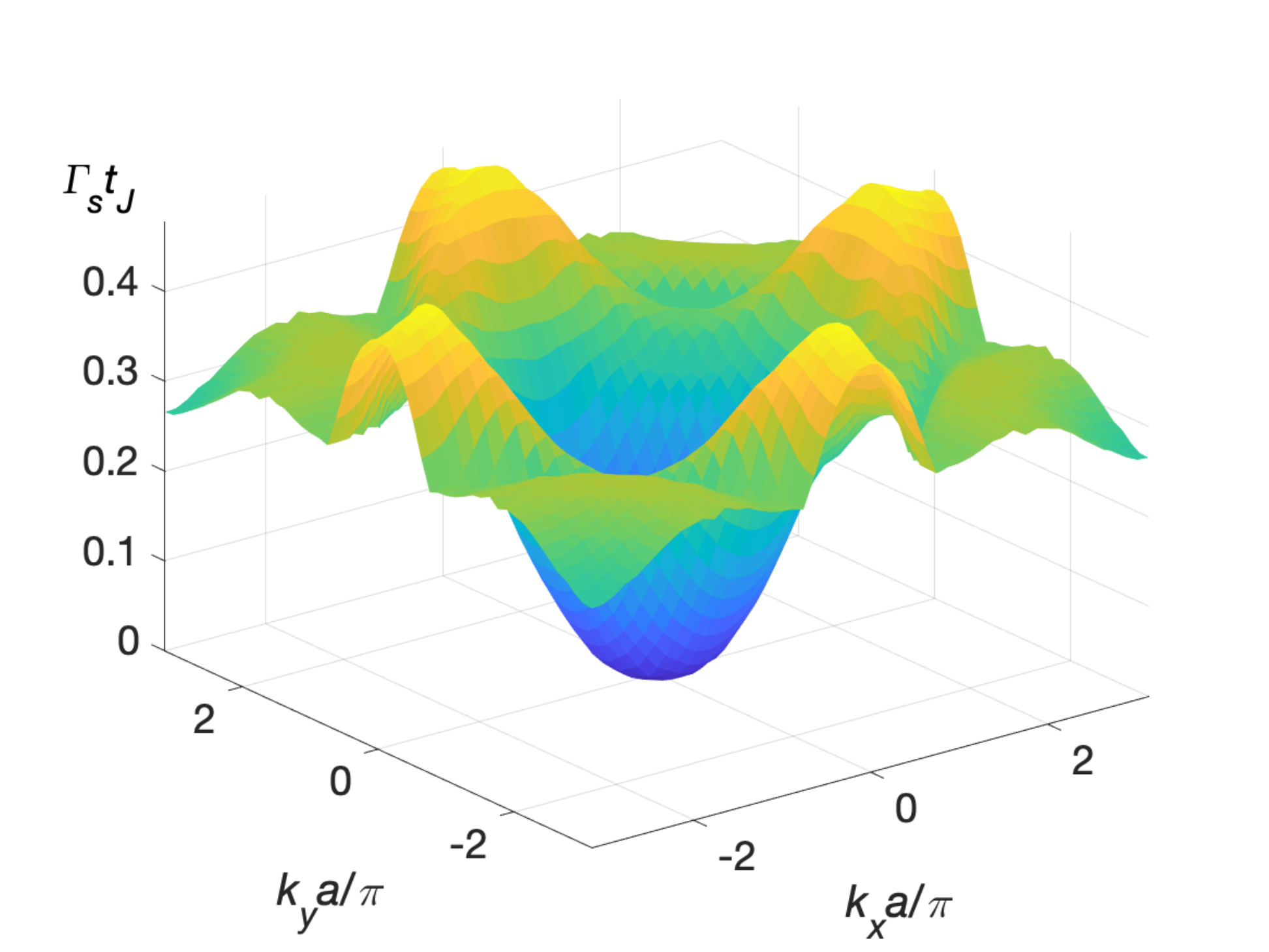}\label{fig:sumimsigma}}\quad
        \sidesubfloat[]{%
        \includegraphics[width=.45\linewidth]{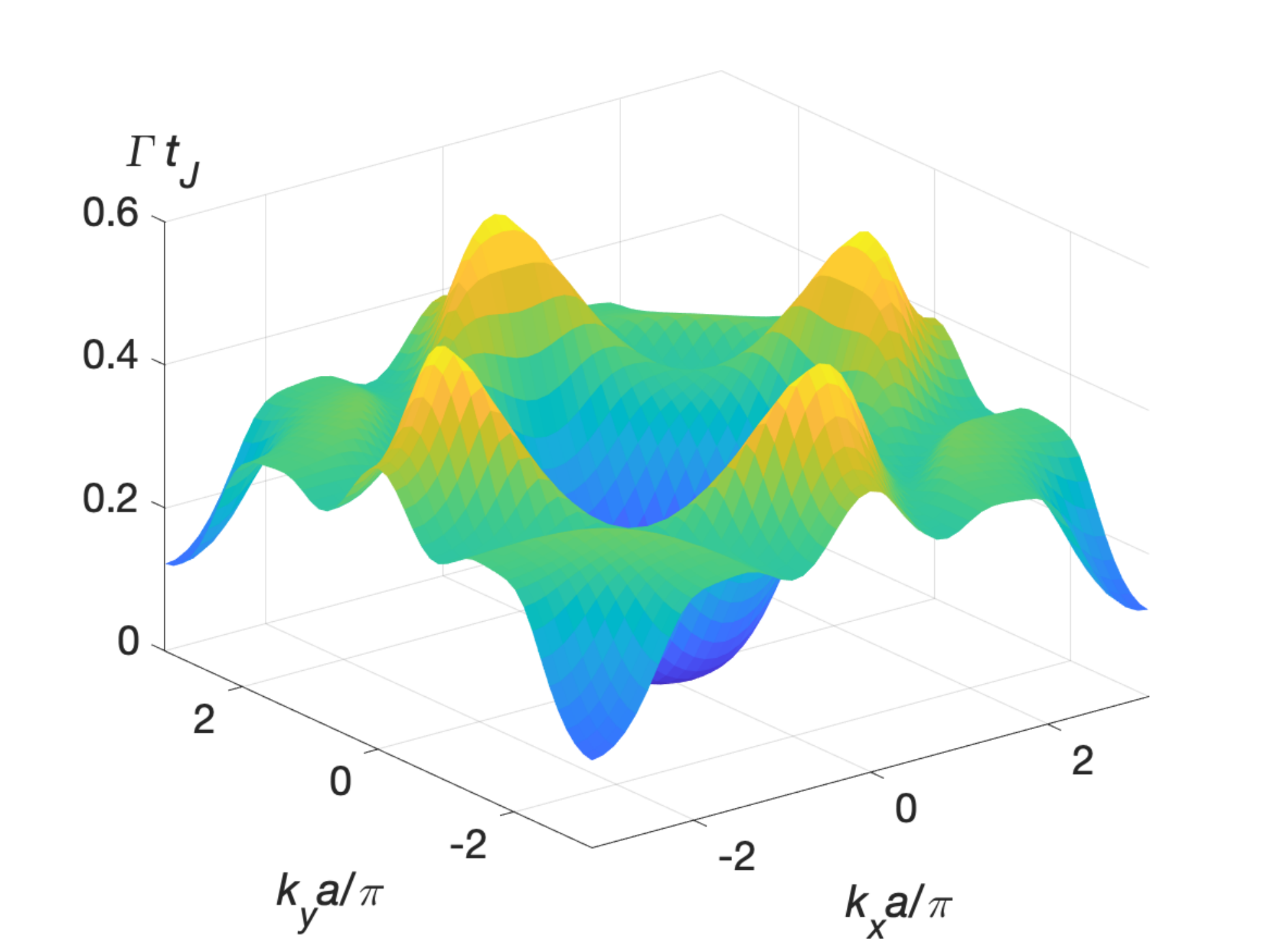}\label{fig:inversescatteringtime}}
		\caption{\label{fig:assortedplots}(a) On-shell scattering mean free rate $\Gamma_s t_J = -2\textrm{Im}\Sigma(E_{os},\bm{k})/J$ obtained numerically from Eqs.~(\ref{eq:computeG}-\ref{eq:ImSigma}) and plotted as a function of $\bm{k}$ in the Brillouin zone. The on-shell energy $E_{os}$ is given by Eq.~\eqref{eq:OnShell2}. 
		(b) Scattering mean free rate $\Gamma t_J$ numerically extracted from the exponential fit of Eq.\eqref{eq:ExpFit} and plotted as a function of $\bm{k}$ in the Brillouin zone.  
		For both plots, the defect density has been set to $\rho = 0.1$ and the data have been averaged over $2000$ disorder configurations.}
\end{figure*}

\subsection{Self-Energy and Scattering Mean Free Time}
\label{Diag}

The retarded Green's function associated to our Hamiltonian $H$ (see Appendix \ref{Self} for general definitions) can be expanded over the $\mathcal{P}$ and $\mathcal{Q}$ subspaces and we find $G(E) = \mathcal{P} G(E) \mathcal{P} + \mathcal{Q} G(E) \mathcal{Q}$ since $H$ does not couple these two subspaces. Since $\mathcal{Q}H\mathcal{Q} = 0$, we have $\mathcal{Q} G(E) \mathcal{Q}= \mathcal{Q}/(E+\mathrm{i}0^+)$. As a consequence, $\overline{\mathcal{Q} G(E) \mathcal{Q}}= \rho/(E+\mathrm{i}0^+) \, \mathbbm{1}_\mathcal{L}$ and $\textrm{Im}[\overline{\mathcal{Q} G(E) \mathcal{Q}}]= - \pi \rho \delta (E) \, \mathbbm{1}_\mathcal{L}$. This shows that, {\it for $E \neq 0$}, $\textrm{Im}\overline{G(E)} = \textrm{Im}[\overline{\mathcal{P}G(E)\mathcal{P}}]$: Both $H$ and $\mathcal{H}$ give rise to the same scattering mean free time as long as $E\neq 0$.

To numerically compute the scattering mean free time, we use two methods. In the first one, we expand $G_{\mathcal{P}}(E) \equiv \mathcal{P}G(E)\mathcal{P} = [E- \mathcal{H} + \mathrm{i} 0^+]^{-1}$ over the eigenstates and eigenenergies of $\mathcal{H}$:
\begin{equation}
    G_{\mathcal{P}}(E,\bm{k}) = \sum_{n} |\varphi_n(\bm{k})|^2 \frac{(E-\epsilon_{n}- i\eta)}{(E-\epsilon_{n})^{2} + \eta^{2}}
\label{eq:computeG}
\end{equation}
with $\eta = 10^{-3}$ and we get $\overline{G_{\mathcal{P}}}(E,\bm{k})$ by  averaging over 2000 disorder configurations. 
We next obtain 
\begin{equation}
\label{eq:ImSigma}
    \textrm{Im}\Sigma(E,\bm{k}) = \frac{\textrm{Im}\overline{G_{\mathcal{P}}}(E,\bm{k})}{ |\overline{G_{\mathcal{P}}}(E,\bm{k})|^2},
\end{equation}
compute it for the on-shell energy $E=E_{os}$, Eq.~\eqref{eq:OnShell} and \eqref{eq:OnShell2}, and finally get the on-shell scattering mean free rate $\Gamma_s$. 

In the second method, we compute
\begin{equation}
    \big|\langle\bm{k}\overline{\ket{\Psi(t)}}\big|^2 \propto \Big|\overline{\sum_n \, |\varphi_n(\bm{k})|^2 \exp (-\mathrm{i} \omega_n t)}\Big|^2 \sim e^{-\Gamma t}
\label{eq:ExpFit}
\end{equation}
for $t \leq 50 t_J$ (averaged over $20$ disorder configurations) and perform an exponential fit to extract the decay rate $\Gamma$. We then compare $\Gamma$ to the numerically-computed on-shell weak-disorder prediction $\Gamma_s$.

\begin{figure} [!htbp] 
\sidesubfloat[]{%
        \includegraphics[width=1\linewidth]{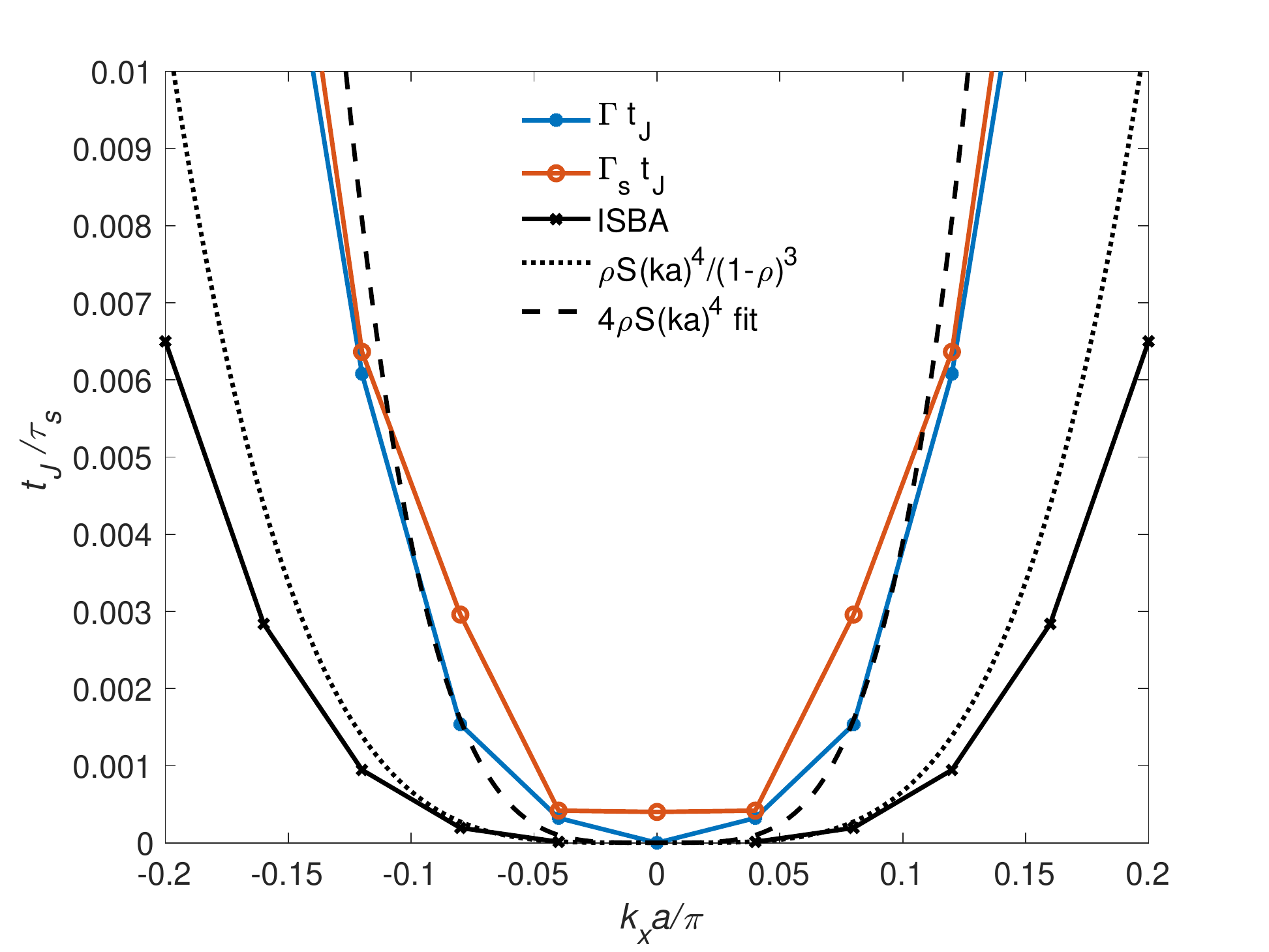}}
        \caption{Inverse scattering time and comparison to theoretical predictions. Red solid curve: Cut of $\Gamma_s t_J$ in Fig.\ref{fig:sumimsigma} along the line $k_y=0$. Blue solid curve: Cut of $\Gamma t_J$ in Fig.\ref{fig:inversescatteringtime} along the line $k_y=0$. Black solid curve: Independent Scattering Born Approximation (ISBA) prediction, Eq.\eqref{eq:ScatPred}. Black dotted curve: $ka \to 0$ limit of the ISBA prediction, Eq.\eqref{eq:ScatPredLowk}. The ISBA prediction reproduces the expected $(ka)^4$ dependence at very low $ka$ but fails to reproduce the scattering rates $\Gamma_s t_J$ and $\Gamma t_J$ at larger $ka$. However, a fit using the function $\alpha(\rho) \rho S (ka)^4$ with $\alpha(\rho) = 4$ at $\rho = 0.1$ shows that both $\Gamma_s t_J$ and $\Gamma t_J$ are well reproduced by the $(ka)^4$ dependence for a larger range of $ka$ values.
        }
\label{fig:k4dependence}
\end{figure}

Our results at small defect density $\rho = 0.1$ are given in Fig.\ref{fig:assortedplots} and Fig.\ref{fig:k4dependence}. It is observed that both dimensionless quantities $\Gamma_s t_J$ and $\Gamma t_J$ display approximately the same functional shape, except near the edges of the Brillouin zone, and differ by less than 10$\%$. This discrepancy is not surprising since $\Gamma_s$ is calculated on-shell whereas $\Gamma$ represents the resulting "average" exponential decay rate obtained after integration over all possible energies. As such, Fig.\ref{fig:inversescatteringtime} is somehow a "smoothed" version of Fig.\ref{fig:sumimsigma}.

As seen in Fig.\ref{fig:k4dependence}, when $ka \to 0$, both quantities agree well with the theoretically predicted $(ka)^4$ dependence. This sharp drop when $ka \to 0$ means that the scattering mean free time $\tau_s$ diverges like $\varepsilon^{-2}_{\bm{k}}$ and that the disorder is less and less effective in the long wavelength limit.

\subsection{CFS Contrast, Spectral Form Factor and Heisenberg Time}

As seen in Fig.\ref{fig:CBSandCFS}, the CFS peak appears at times much larger than the isotropization time scale $\tau_B$ and thus grows on top of a stationary ring-shaped diffusive background of rim height $n_B(\bm{k}_0)$. To quantify the time dynamics of the CFS peak height, it is convenient to introduce the CFS contrast $\Lambda (\bm{k}_0,t)$. It is defined as the ratio between the CFS peak height above the stationary diffusive background and this same background for $t \gtrsim \tau_B$:
\begin{equation}
    \Lambda (\bm{k}_0,t) = \frac{n(\bm{k}_0,t) - n_B(\bm{k}_0)}{n_B(\bm{k}_0)}. 
    \label{eq:cfscontrast}
\end{equation}
Saliently, the CFS contrast embeds the critical properties of the Anderson transition in momentum space \cite{ghosh2017cfs}. 

At this point, we introduce the spectral form factor \cite{haake91} associated to Hamiltonian $\mathcal{H} = \mathcal{P} H \mathcal{P}$ and its $N_m$ eigenenergies $\epsilon_n = \hbar\omega_n$:
\begin{equation}
K_N(t) = \frac{1}{N_m} \, \overline{|\sum_n e^{-\mathrm{i} \omega_n t}|^2}. 
\label{eq:formfactor}
\end{equation}  

It satisfies $K_N(t=0) = N_m$ and $\lim_{t\to\infty}K_N(t)=1$. In the continuum limit $N\to \infty$ at fixed $\rho$, we have $K_N(t) \to \delta(\tau) + K_{\textrm{reg}}(\tau)$ where $K_{\textrm{reg}}$ is the regular part of the form factor and $\tau = t/\tau_H $.

\begin{figure}
    \centering
    \includegraphics[width=\linewidth]{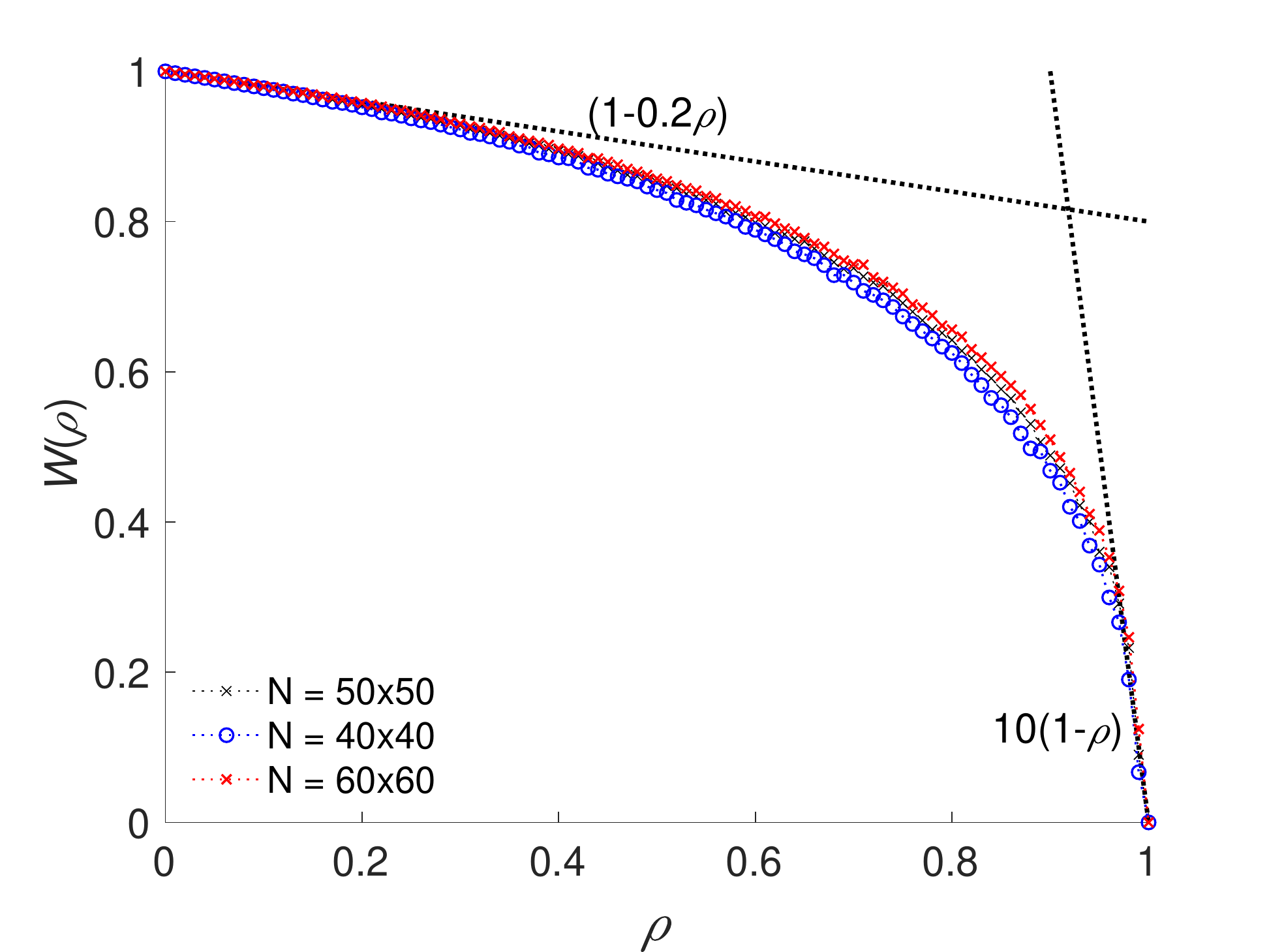}
    \caption{Typical dimensionless range $W$ as a function of the defect density $\rho$ for different system sizes. Here, $8W$ is the disorder-averaged value of the largest eigenvalue of the Laplacian matrix $\mathcal{R}$ associated to the disorder Hamiltonian $\mathcal{H}$, see Eq.~\eqref{eq:Laplacian}. The respective limiting behaviors are well fitted by $W(\rho) \approx (1-\rho/5)$ when $\rho \to 0$ and by $W(\rho) \approx 10(1-\rho)$ when $\rho \to 1$.
    \label{fig:W_vs_rho}}
\end{figure}

The Heisenberg time $\tau_H$ that sets the temporal variations of the form factor is defined by $\tau_H = 2\pi\hbar/ \Delta$ where $\Delta$ is the mean level spacing for a system of linear size $\sqrt{N_m}a$. At this stage, it is important to recall that eigenvalues of Laplacian matrices are always positive with the lowest one being always $0$. This means that one can operationally define $\Delta = 8JS W/N_m$ where $8W$ is the disorder-averaged value of the largest eigenvalue $R=8w$ of the Laplacian matrix $\mathcal{R}$, see Eq.~\eqref{eq:Laplacian} (we have introduced the factor $8$ for convenience). 
We thus have:

\begin{equation}
\label{eq:tauH}
    \frac{\tau_H}{t_J} = \frac{2\pi}{8SW} \, N_m = \frac{\pi (1-\rho)}{4SW} \, N
\end{equation}
From Appendix \ref{CleanH}, we see that $W=1$ when $\rho=0$ and Fig.~\ref{fig:W_vs_rho} shows $W(\rho)$ when $\rho$ is varied for different $N$. For $\rho \to 0$, we expect $W$ to decrease linearly with the defect density, $W \sim 1 - c_0 \rho$ ($c$ being some constant), a result consistent with a perturbation argument starting from the clean Hamiltonian and removing magnetic sites as the defect density increases. On the other hand, for $\rho \to 1$, we also expect a linear behavior, $W \sim c_1 (1-\rho)$, again consistent with a perturbation argument starting from the null Hamiltonian (all sites defective) and increasing the number of magnetic sites. We have not developed these perturbative arguments and have rather resorted to numerical calculations. The question of the limiting behaviors of the constants $c_0$ and $c_1$ in the thermodynamic limit is left open. 

For $N=2500$, $S=1/2$ and $\rho =0.1$, we find $\tau_H \approx 3606 t_J$, a value consistent with the CFS evolution time scale in Fig.~\ref{fig:CFSat10long}. Note that the $2\pi$ in the definition of $\tau_H$ is somewhat arbitrary, so this calculated numerical value carries over this arbitrariness. More important physically is in fact the scaling of $\tau_H$ with $N_m$ (or with $N$).

\begin{figure}[!htbp]
\includegraphics[width = 1 \linewidth]{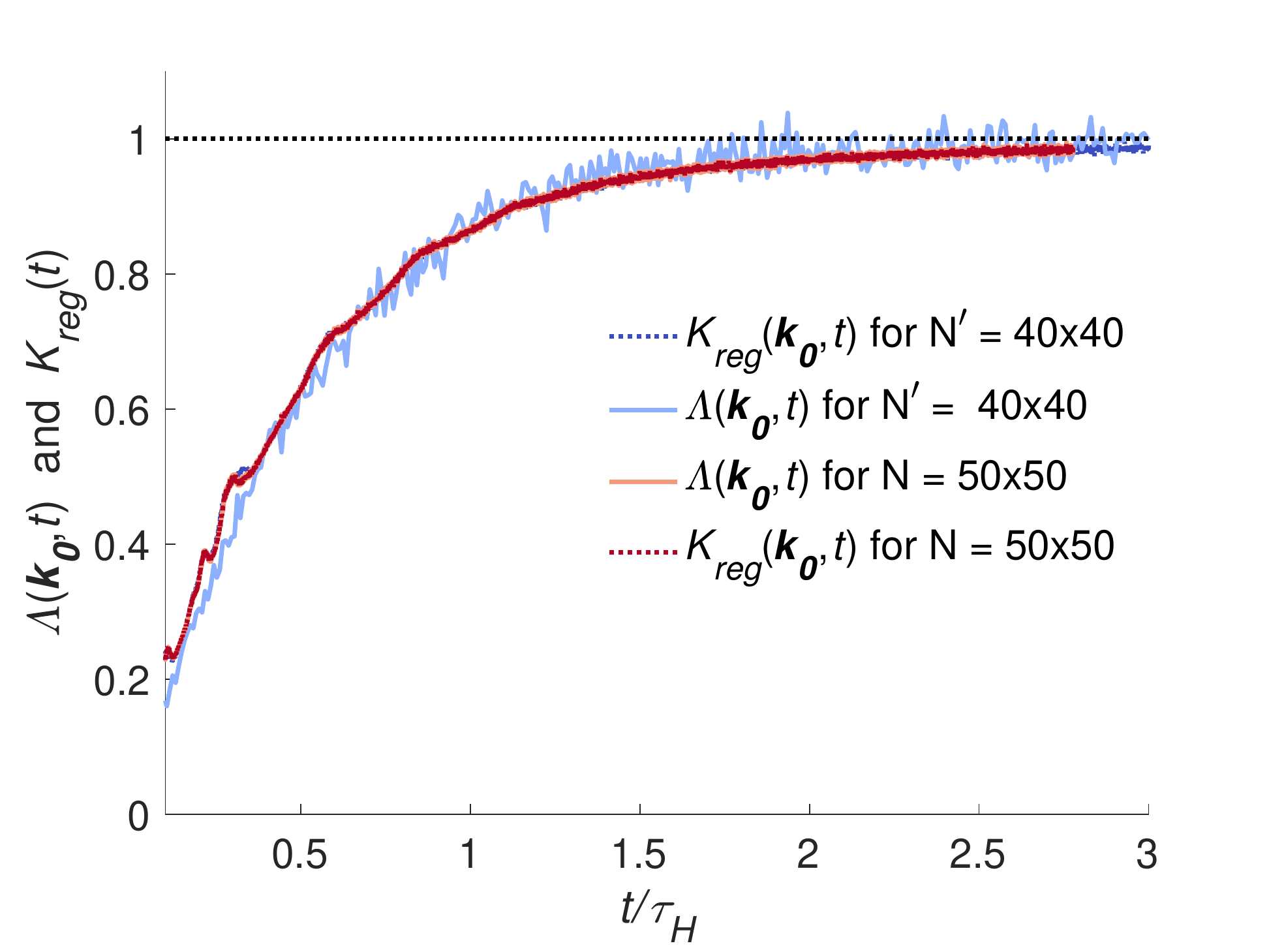}
\caption{Plots of the CFS contrast $\Lambda (\bm{k}_0,t)$, Eq.\eqref{eq:cfscontrast}, and of the regular form factor $K_{\textrm{reg}}(t)$ as a function of time $t/\tau_H$ for $N=50\times 50$ and $N'=40\times 40$ ($\rho=0.1$). The Heisenberg times $\tau_H$, computed with Eq.~\eqref{eq:tauH} and Fig.\ref{fig:W_vs_rho}, are $\tau_H \approx 3606 t_J$ for $N$ and $\tau_H \approx 2308 t_J$ for $N'$. The parameters are $\rho =0.1$ and $\bm{k}_0 = k_0 \, \hat{\bm{e}}_x$ with $k_0a = 0.6\pi$. $\Lambda (\bm{k}_0,t)$ is extracted from the time dynamics of the disorder-averaged momentum distribution, see Fig.\ref{fig:CFSat10long}, and averaged over $10^{4}$ disorder configurations. $K_{\textrm{reg}}(t)$ is obtained from $K_N(t)$, Eq.\eqref{eq:formfactor}, at $t>0$ by numerically computing the eigenvalues of the disordered Hamiltonian $\mathcal{H}$ and averaging the results over $10^{5}$ disorder configurations. As expected, all quantities collapse onto each other, at least when $t/\tau_H \gtrsim 1$. 
\label{fig:formfactor}}
\end{figure}

In Appendix \ref{Form}, we show the important (scaling) result:
\begin{equation}
    \Lambda(\bm{k}_0,t) \approx K_{\textrm{reg}}(t/\tau_H) \quad  (t \gtrsim \tau_H),
    \label{eq:HeisTime}
\end{equation}
see also \cite{ghosh2014coherent,lee2014cfs,martinez2020}. 
Fig.\ref{fig:formfactor} shows the numerically-computed regular form factor $K_{\textrm{reg}}$ and its comparison to the time evolution of the CFS contrast $\Lambda$ when plotted against $t/\tau_H$ for 2 different sizes $N=50\times 50$ and $N'=40\times 40$. As expected, the agreement is very good for $t\gtrsim \tau_H$, at least in the small-$\rho$ limit considered up to now. 

To summarize the results of the previous Sections and related Appendices, the linear spin-wave system subjected to site percolation disorder in the low defect density regime $\rho \ll 1$ satisfies perfectly well the usual predictions of quantum transport theory. 

\section{Momentum Distribution at Larger Defect Densities $\rho$}

\subsection{Formation of polyomino clusters}
\begin{figure}[ht]
    \centering
    \includegraphics[width=0.9\linewidth]{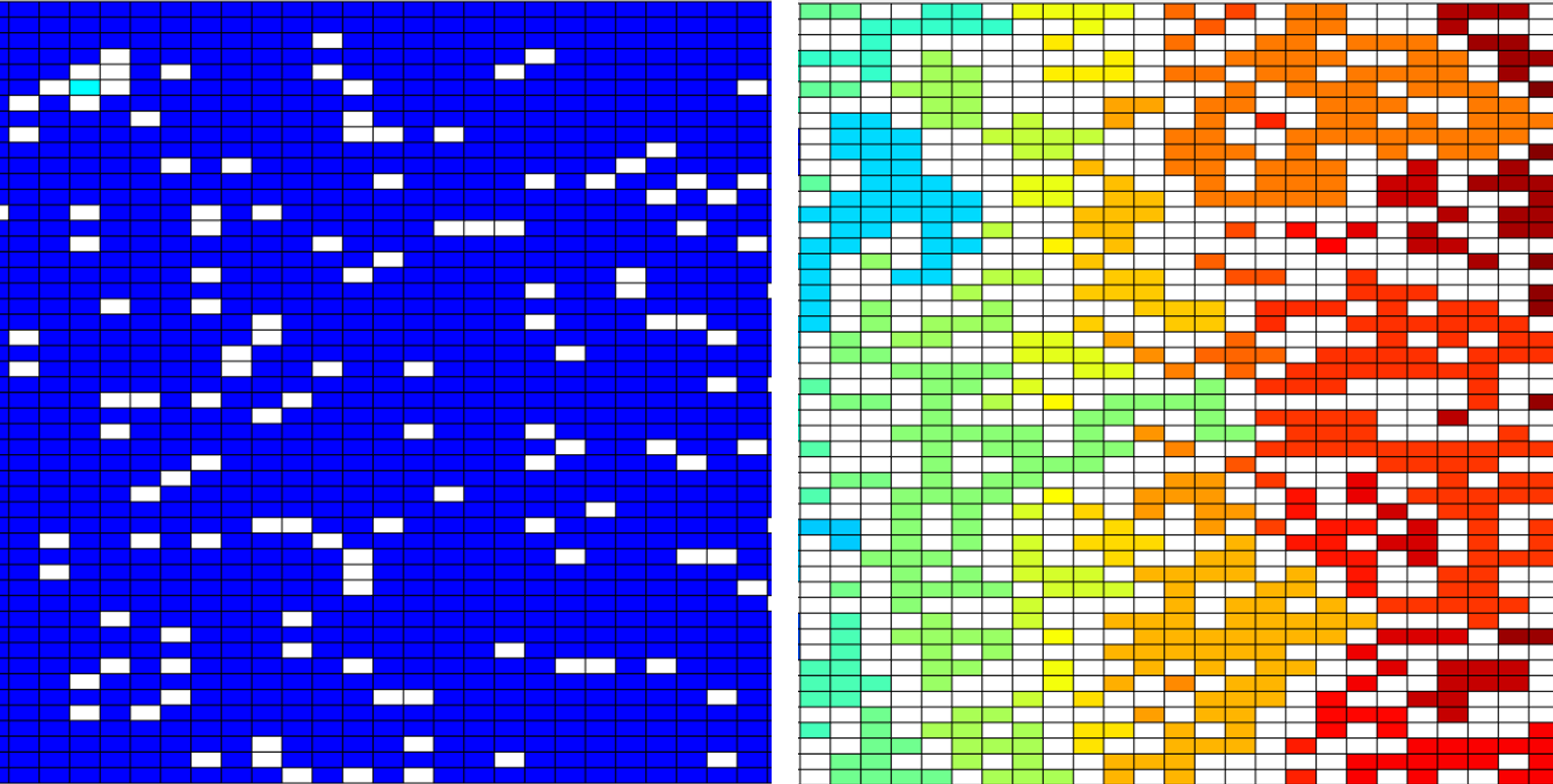}
    \caption{ \label{fig:percolationexample}Disordered magnetic configurations obtained at defect density $\rho$ by randomly removing sites (in white) from a 2D clean lattice. The percolation threshold is $\rho^*=0.41$. Sites within a given connected magnetic $n$-polyominoes have the same color (note that different clusters may have the same color when one can easily distinguish one from the other). Left panel: $\rho=0.1$. For the configuration obtained, we have a single percolating (macroscopic) magnetic $n$-polyomino with $n \sim (1-\rho)L^2$, $L$ being the linear lattice size, and one isolated cluster consisting of only one site (in green). Note that the defect configuration is dominated by single defective sites, corroborating our theoretical model at $\rho \ll 1$. Right panel: $\rho=0.5 >\rho^*$. For the configuration obtained, the system breaks into a collection of magnetic $n$-polyominoes of many different sizes. By the same token, the defect configuration is made of defective clusters of many different sizes.}
   
\end{figure}

A disorder configuration is obtained from the clean lattice by punching holes: One randomly removes sites and their attached Z links. The net effect of this procedure is to replace the initial uniform and connected 2D magnetic lattice grid with a collection of separated, independent, connected magnetic clusters, see Fig.\ref{fig:percolationexample}. In the literature connected clusters comprising $n$ sites are often referred to as $n$-polyominoes. The statistics of $n$-polyominoes is given in Appendix \ref{Perco}. 

In the dilute regime $\rho \ll 1$, the probability of aggregated defects (defective islands) is very small and drops very quickly with their size. We thus expect that the typical random configuration is essentially made of sparse and isolated defects, the magnetic sites forming a single macroscopic $n$-polyomino with $n = (1-\rho) L^2$. This was the basis of our theoretical analysis at $\rho \ll 1$. 

The situation changes dramatically as $\rho$ increases: Bigger and bigger defective islands become more and more probable and these aggregated defects can break the system into more and more isolated magnetic $n$-polyominoes with smaller and smaller $n$. In other words, we face a percolation problem. For our 2D system, the percolation threshold where the systems breaks into isolated magnetic $n$-polyominoes of any size (in the thermodynamic limit) is $\rho^* \approx 0.41$~\cite{newman2000efficient}. When $\rho$ increases further beyond $\rho^*$, defective sites take over and we get macroscopic defective islands interspersed with magnetic $n$-polyominoes where $n$ is small. 

\subsection{Temporal oscillations of the CFS peak}
\begin{figure*}[!htbp] 

\sidesubfloat[]{\includegraphics[width = 0.40\linewidth]{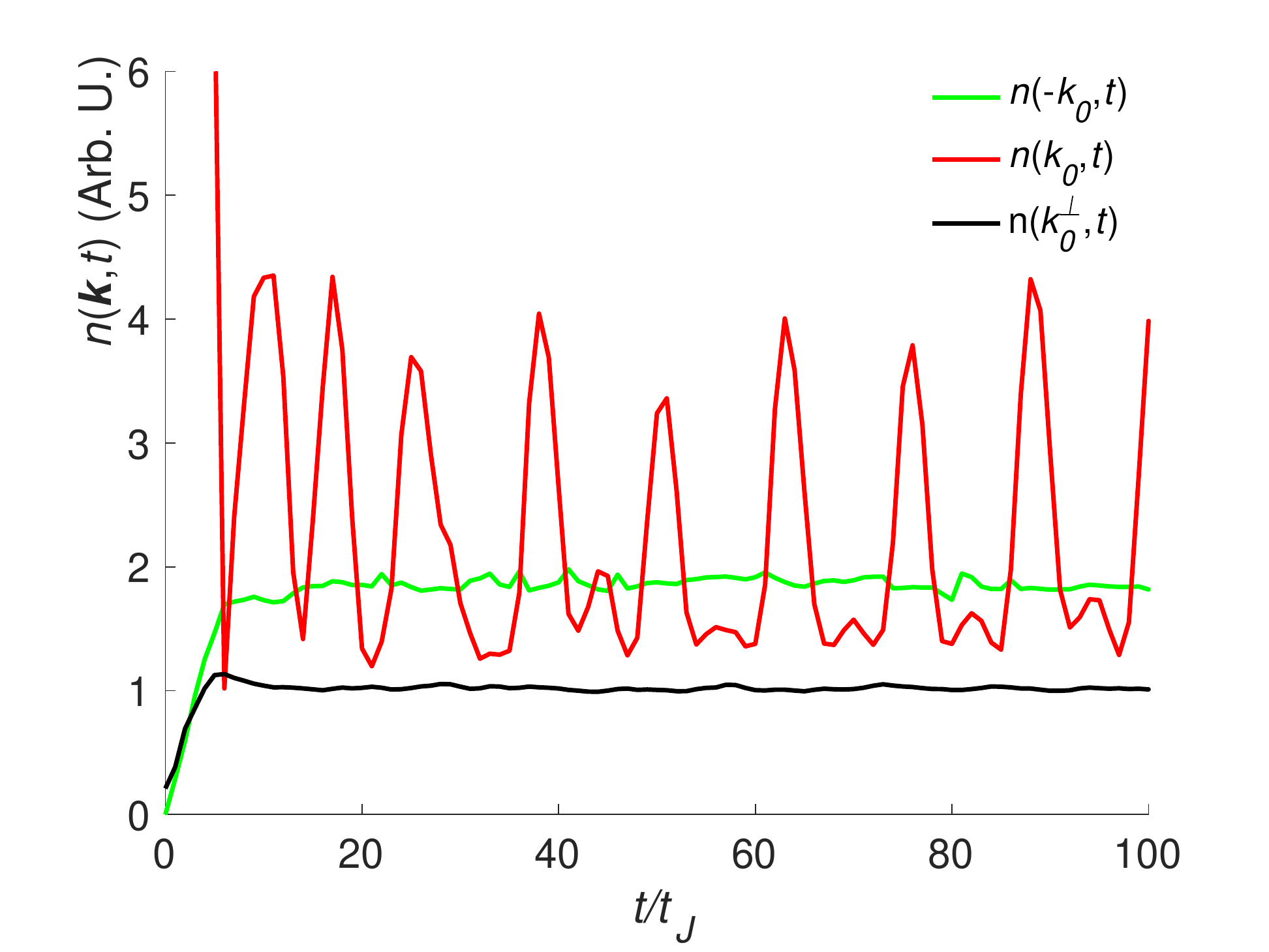}\label{fig:CFSat34low}}
\sidesubfloat[]{\includegraphics[width = 0.40\linewidth]{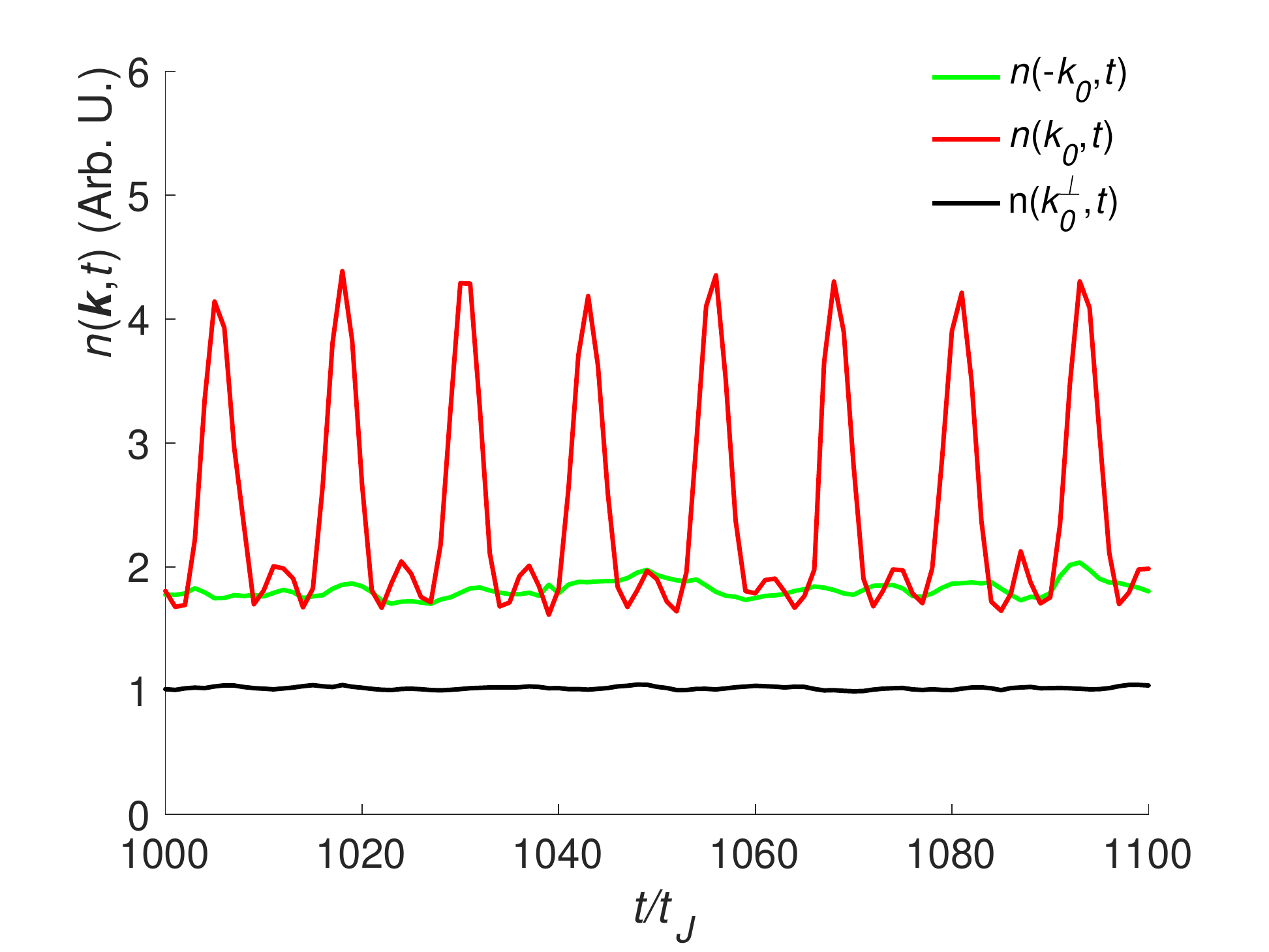}\label{fig:CFSat34long}}

\sidesubfloat[]{\includegraphics[width = 0.40\linewidth]{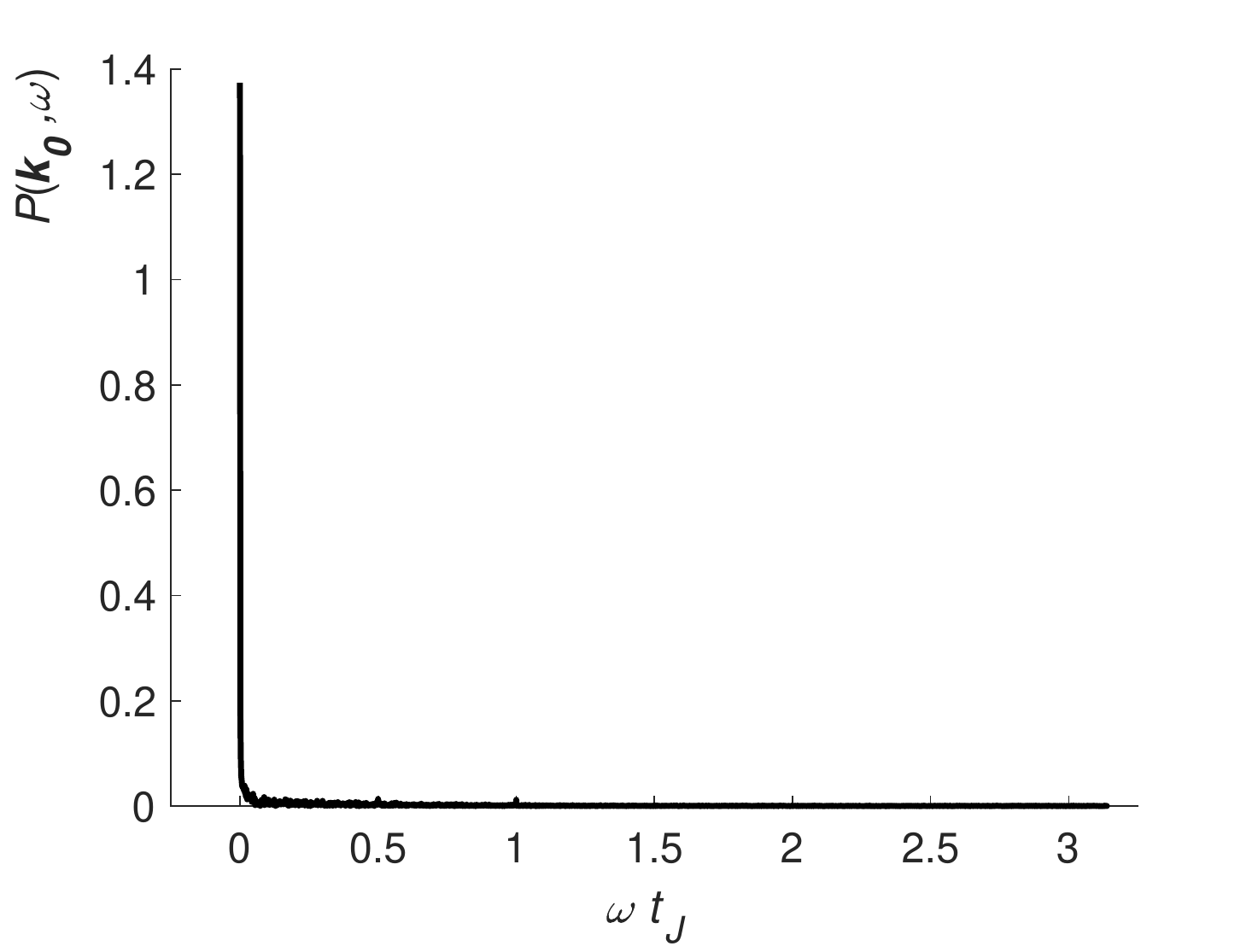}\label{fig:spectralsteady20}}
\sidesubfloat[]{\includegraphics[width = 0.40\linewidth]{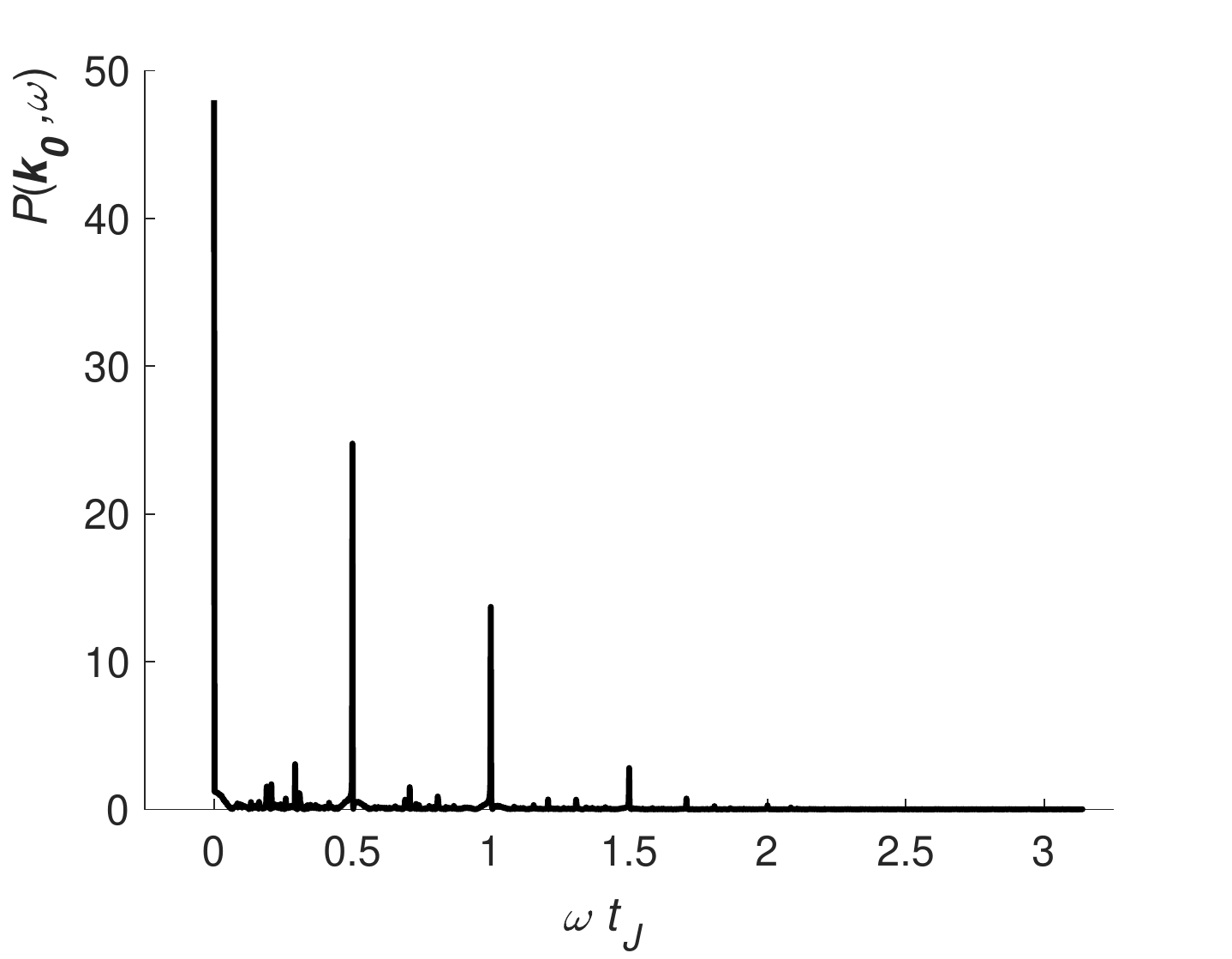}\label{fig:spectralsteady50}}

\sidesubfloat[]{\includegraphics[width = 0.40\linewidth]{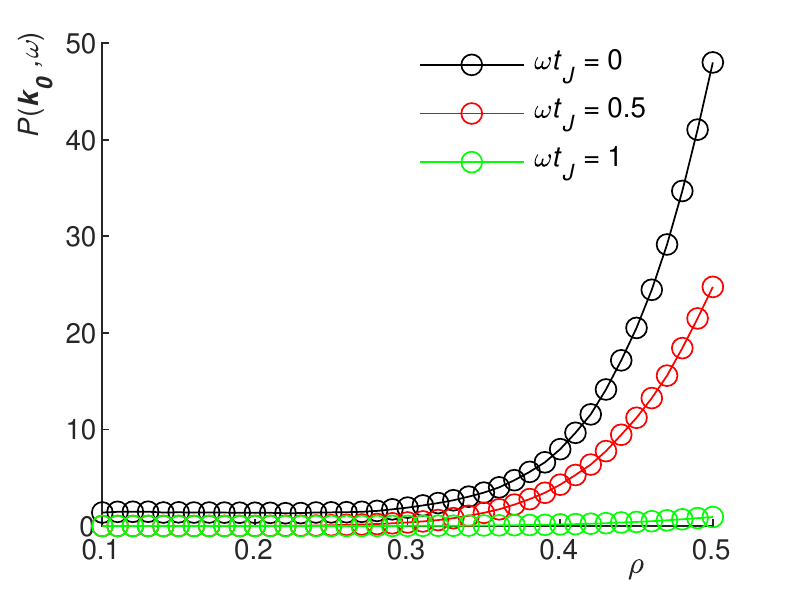}\label{fig:spectralsteadyall}}
\sidesubfloat[]{\includegraphics[width = 0.40\linewidth]{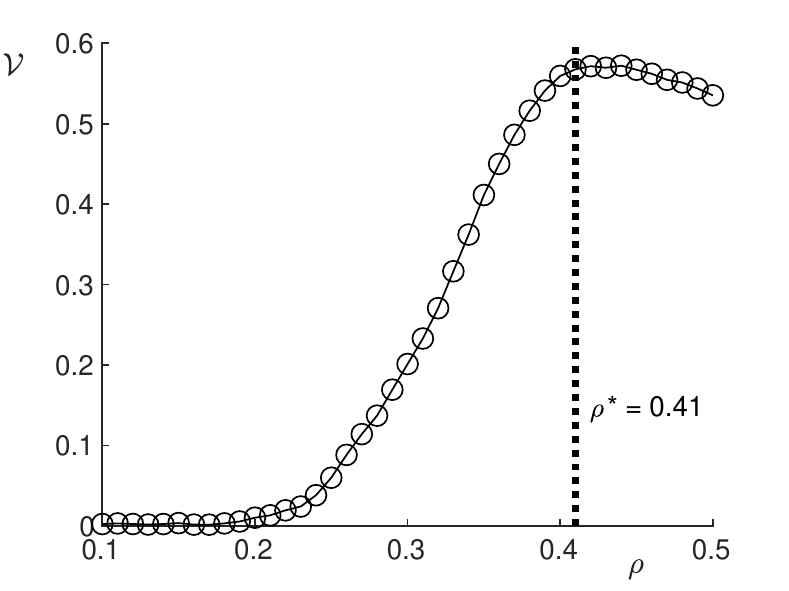}\label{fig:spectralcontrastall}}

\caption{\label{fig:momrho}Behavior of the momentum distribution at higher defect densities $\rho$ Panel (a): Background $n(\bm{k}_{0}^\perp,t)$ (black curve), CBS $n(-\bm{k}_{0},t)$)(green curve) and CFS $n(\bm{k}_{0},t)$ (red curve) signals at short time scales $t \leq 100 t_J$ and $\rho=0.34$. Panel (b):  Same as (a) but at long time scales $t \geq 1000 t_J$. Panel (c): CFS Fourier spectrum $P(\bm{k}_{0},\omega)$ at defect density $\rho=0.2$ obtained at $t \geq 1000 t_{J}$, see Eq.\eqref{eq:fourieramplitudes}. The spectrum is completely dominated by the static component at $\omega =0$: The CFS displays the same behavior as at small $\rho$ and CFS time oscillations are too small to be visible. Panel (d): Same as (c) but at $\rho=0.5$. This time, clear visible discrete peaks, present at $\omega t_J = 0.5$, $1$ and $1.5$ in the spectrum, flag a CFS signal oscillating with time. The main oscillation period comes from $\omega t_J = 0.5$ and is thus $T= 4\pi \, t_J$. Panel (e): Fourier Amplitude $P(\bm{k}_{0},\omega)$ as a function of $\rho$ at $\omega t_J=0$ (black solid curve and symbols), $\omega t_J=0.5$ (red solid curve and symbols) and $\omega t_J=1$ (green solid curve and symbols). Panel (f): CFS time oscillation visibility $\mathcal{V}$ as a function of $\rho$, see Eq.\eqref{eq:Visib}. It increases until it reaches the percolation threshold $\rho^* \approx 0.41$ before decreasing. 
}
    \label{fig:CBSandCFS2}
\end{figure*}

In Fig.\ref{fig:momrho}, we show how the disorder averaged momentum distribution changes when we increase $\rho$. First, we remark that the behavior of the isotropic background and CBS peak in the high and low defect density regimes are quite similar, see the black and green curves in Fig.\ref{fig:CFSat10low} and Fig.\ref{fig:CFSat34low}. However, the behavior of the CFS peak is markedly different, see red curves in Fig.\ref{fig:CFSat10low} and Fig.\ref{fig:CFSat34low}. There is no visible Heisenberg time $\tau_{H}$ at which a CFS peak forms. Instead, we observe an oscillatory behavior taking place already at short time scales. For the defect density $\rho = 0.34$ considered in Fig.\ref{fig:CFSat34low}, the period of oscillations is $T=4\pi \, t_J$. Furthermore, the CFS signal oscillates in time around a mean height of about $3$, i.e. almost $50\%$ more than the peak height of $2$ found at low $\rho$. 

To better quantify this behavior, We perform a spectral analysis by expanding of the CFS signal into Fourier amplitudes

\begin{equation}
n(\bm{k}_{0},t) = \int \frac{d\omega}{2\pi} \, P(\bm{k}_{0},\omega) \, e^{\mathrm{i}\omega t},
\label{eq:fourieramplitudes}
\end{equation}
and by computing the visibility $\mathcal{V}$ of the CFS signal

\begin{equation} \label{eq:Visib}
    \mathcal{V} = \frac{\textrm{Max}[n(\bm{k}_{0},t)] - \textrm{Min}[n(\bm{k}_{0},t)]}{\textrm{Max}[n(\bm{k}_{0},t))] + \textrm{Min}[n(\bm{k}_{0},t)]}.
\end{equation}
The $\omega$-dependence of $P(\bm{k}_{0},\omega)$ is shown in Fig.\ref{fig:CBSandCFS2}c for $\rho=0.2$ and in Fig.\ref{fig:CBSandCFS2}d for $\rho=0.5$. We see that the CFS time oscillations have not appeared at $\rho=0.2$ where the $P({\bm k}_0, \omega=0)$ component, giving the mean value of the CFS signal, completely dominates the spectrum. On the contrary, the oscillations show up clearly at $\rho=0.5 > \rho^*$ with well visible discrete peak components in the spectrum at angular frequencies $\omega t_J = 0.5$, $1$ and $1.5$.

The $\rho$-dependence of both $P({\bm k}_0, \omega=0)$ and $\mathcal{V}$ on $\rho$ can be seen in Fig.\ref{fig:spectralsteadyall} and Fig.\ref{fig:spectralcontrastall}
respectively: They both increase with the defect density. Since the CFS signal cannot become negative, the visibility is bounded by $\mathcal{V} \leq 1$. We see that $\mathcal{V}$ increases with $\rho$ and reaches its maximum value around the percolation threshold $\rho^* \approx 0.41$ where the system breaks into cluster components smaller than the full lattice size. For $\rho > \rho^*$, $\mathcal{V}$ decreases again. Note that we find that the maximum is obtained for $\rho \approx 0.44$ instead of $\rho^*$ because of finite lattice size effects. 

\subsection{Origin of the CFS temporal oscillations}

The preceding discussion shows that, very generally, each disorder configuration is a collection of different independent polyominoes, the probability of getting a polyomino of size $n$ increasing with the defect density $\rho$. As such, the Hamiltonian on the magnetic lattice $\mathcal{H} = \sum_C \mathcal{H}(C)$ breaks into a sum of independent Hamiltonians $H(C)$ on each independent polyomino $C$ and one has:

\begin{equation}
\label{eq:PhiC}
\begin{aligned}
    &\ket{\Phi (t)} = \sum_{C} \ket{\Phi_C(t)} \\
    &\ket{\Phi_C(t)} = \sum_a e^{-i\omega_a(C)t} \, \varphi^*_a(C, {\bm k}_0) \, \ket{\varphi_a(C)}
\end{aligned}
\end{equation}
where $\ket{\varphi_a(C)}$ and $\hbar\omega_a(C)$ are the eigenvectors and eigenvalues of Hamiltonian $\mathcal{H}(C)$. We note that the normalization conditions for these eigenvectors are given as

\begin{equation}
\begin{aligned}
    |\bra*{\Phi (t)}\ket*{\Phi (t)}|^{2} & = \sum_{C} |\bra*{\Phi_{C}(t)}\ket*{\Phi_{C}(t)}|^{2}
    = \frac{1}{1-\rho}\\ 
    |\bra*{\varphi_a(C)}\ket*{\varphi_a(C)}|^{2} & = 1\\
    \sum_{a}|\varphi_a(C, {\bm k}_0)|^{2} & = \frac{1}{1-\rho}
\end{aligned}
\end{equation}

Note that a disorder configuration can host several polyominoes, located at different places in the magnetic lattice, which can be superposed by an appropriate translation followed or not by a rotation or a reflection. The Hamiltonians associated to these polyominoes in the decomposition $\mathcal{H} = \sum_C \mathcal{H}(C)$ have obviously the same energy eigenvalues and eigenfunctions related by the relevant previous transformations. For example, for two polyominoes $C$ and $D$ having the same shape and simply related by translation vector ${\bf d}$, we would have $\ket{\varphi_a(D)} = e^{-i{\bm p}\cdot {\bm d}/\hbar} \, \ket{\varphi_a(C)}$. Actually, one can associate a graph to each polyomino by mapping the sites to vertices and by connecting vertices by an edge when the sites are connected by hopping. It is easy to see that two different polyominoes that are graph-equivalent have exactly the same eigenvalue spectrum.

More precisely, one can partition each cluster configuration into distinct equivalence classes $\{C_0 \}$ grouping all polyominoes $C = C_0 + \bm{d}$ which can be obtained from polyominoe $C_0$, positioned at some $\bm{R}_{C_0}$ in the lattice, by a translation vector $\bm{d}$ (note that $\bm{d}$ is a random vector that changes with the disorder configuration and that the possible choices are subject to $C_0$-dependent "excluded volume" constraints). Then, Eq.\eqref{eq:PhiC} can be rewritten as $\ket{\Phi (t)} = \sum_{C_0} \ket{\Phi_{\{C_0 \}}(t)}$ where $\ket{\Phi_{\{C_0\}}(t)} = \sum_{\bm{d}}\ket{\Phi_{C_0+\bm{d}}(t)}$. Since all polyominoes in $\{C_0\}$ have the same energy spectra and since the corresponding eigenfunctions are simply obtained by translation, we further have
\begin{equation}
    \Phi_{\{C_0\}}(\mathbf{k},t)=
    S_{C_0}(\bm{k}-\bm{k}_0) \ \Phi_{C_0}(\mathbf{k},t),
\end{equation}
where
\begin{equation}
\label{eq:PhiC0k}
    \Phi_{C_0}(\mathbf{k},t) = \sum_{a} e^{-i\omega_a(C_0) t} \, \varphi^*_a(C_0,\bm{k}_0)\varphi_a(C_0,\bm{k})
\end{equation}
and where
\begin{equation}
    S_{C_0}(\bm{q}) = \sum_{\bm{d}} e^{-i \bm{q} \cdot \bm{d}}
\end{equation}
plays the role of a $C_0$-dependent structure factor (remember that the origin of the translation vectors depend on $C_0$ as well as its possible values).

Finally, the disorder-averaged momentum density reads:
\begin{equation}
    n({\bm k},t) = \overline{|\sum_{C_0} S_{C_0}(\bm{k}-\bm{k}_0)
    \, \Phi_{C_0}(\mathbf{k},t)|^2} 
    \label{eqn:averagedmomentumdensity}
\end{equation}

One obtains quite different results whether 
$\mathbf{k}$ is equal to $\mathbf{k}_0$ (CFS peak height value), or far away from it.  Indeed, for
$\mathbf{k}=\mathbf{k}_0$, the phase factors in the structure factor cancel out and we have $S_{C_0}(\bm{q}=0) = N(C_0)$ where $N(C_0)$ is the number of polyominoes with the same shape (and orientation) as $C_0$ (cardinal of the set $\{C_0\}$). We have:
\begin{equation}
    n({\bm k_0},t) = \overline{|\sum_{C_0} N(C_0)
    \Phi_{C_0}(\mathbf{k}_0,t)|^2}
\end{equation}

Since the statistical properties of eigenenergies and eigenfunctions smoothly go from regular to fully random as the size of $C_0$ grows, one can ``artificially" partition the polyominoes $C_0$ into small ones ($\mathbb{S}$) and large ones ($\mathbb{L}$). For $C_0 \in \mathbb{S}$, we assume that the eigenenergies and eigenfunctions are fully regular, while for $C_0 \in \mathbb{L}$, we assume that they are fully random. In this case, we have:
\begin{equation}
\begin{aligned}
n(\bm{k_0},t) &= \overline{|\sum_{C_0 \in \mathbb{S}} (\cdot\cdot\cdot) + \sum_{C_0 \in \mathbb{L}} (\cdot\cdot\cdot)|^2}\\
& = \overline{|\sum_{C_0 \in \mathbb{S}} (\cdot\cdot\cdot)|^2} + \overline{|\sum_{C_0 \in \mathbb{L}} (\cdot\cdot\cdot)|^2}.
\end{aligned}
\end{equation}
Owing to the statistical independence of small and large polyominoes, the cross product terms $\overline{\sum_{C_0 \in \mathbb{S}, D_0 \in \mathbb{L}} (\cdot\cdot\cdot)}$ cancels and the momentum distribution splits into 2 independent components, $n(\bm{k_0},t) = n_S(\bm{k_0},t) + n_L(\bm{k_0},t)$, one related to $\mathbb{S}$ and the other one to $\mathbb{L}$. For $C_0 \in \mathbb{L}$, the average over disorder leads to the usual diagonal approximation in the long time limit and we get: 
\begin{equation}
    n_L({\bm k_0},t) \approx \overline{\sum_{C_0,a} N^2(C_0)
    |\varphi_a(C_0,\mathbf{k_0})|^4}.
\end{equation}
On the other hand, for $C_0 \in \mathbb{S}$, we expect:
\begin{equation}
    \begin{aligned}
        n_S(\bm{k}_0,t) &= \sum_{C_0 \in \mathbb{S}} \overline{N^2(C_0)} \, |\Phi_{C_0}(\bm{k}_0,t)|^2 \\
        &+ \sum_{C_0\neq D_0 \in \mathbb{S}} \overline{N(C_0)} \, \overline{N(D_0)} \Phi^*_{D_0}(\bm{k}_0,t) \Phi_{C_0}(\bm{k}_0,t)
    \end{aligned}
\end{equation}

The first term of the right-hand side involves $|\Phi_{C_0}(\bm{k}_0,t)|^2$ terms which correspond to the CFS signal associated with each polyomino $C_0$: It features intra-cluster terms oscillating in time with nonzero intra-spectrum frequency differences $\Delta_{ab}(C_0) = \omega_a(C_0)-\omega_b(C_0)$ ($a \neq b$), see Eq.\eqref{eq:PhiC0k}. The second term however involves inter-cluster interference terms and inter-spectra frequency differences $\omega_a(C_0)-\omega_b(D_0) (a \neq b)$.

Even if it is clear that the random nature of the energy spectrum depends on the size and shape of $C_0$, addressing the regular-to-random transition of the spectrum of $\mathcal{H}$ when the size and shape of $C_0$ changes is beyond the scope of this work and we leave it to future studies.
We can nevertheless very generally argue that the Fourier power spectrum breaks into a discrete component $P_d$ and a smooth continuous one $P_s$, $P({\bm k}_0,\omega) = P_d({\bm k}_0,\omega) + P_s({\bm k}_0,\omega)$, where \begin{equation} \label{eq:RegCFSPower}
    P_d({\bm k}_0,\omega) = \sum_s P_s({\bm k}_0) \, \delta (\omega - \Delta_s).
\end{equation}

Here $\Delta_s$ represents the frequency differences {\it that are immune to disorder average} and stem from small-size polyominoes having a regular spectrum which are thus responsible for the CFS temporal oscillations. These CFS temporal oscillations become sizeable only when the probability of small-size polyominoes becomes sizeable. They arise at sufficiently large defect densities when the sample splits into multiple connected magnetic clusters. Hence, the temporal CFS oscillations are a direct consequence of a percolation process at work. Finally, as mentioned previously, the temporal (oscillating) behavior of the CFS peak is actually independent of the physical mechanism triggering the appearance of the CFS peak (box confinement due to finite system size or genuine bulk Anderson localization). The point is that, because of the existence of a sizeable number of polyominoes, scaling like the system size, the discrete spectrum $P_d$ will remain essentially independent of the system size, leading to an almost size-independent oscillatory temporal behavior of the CFS.

In the case of the CBS peak height ($\mathbf{k}=-\mathbf{k}_0$), the situation is dramatically different. Invoking again the statistical independence of small and large polyominoes, the momentum distribution at $-\bm{k}_0$ also breaks into the sum of the small and large clusters contributions, $n(-\bm{k}_0,t) = n_S(-\bm{k}_0,t) + n_L(-\bm{k}_0,t)$. Since our system is time-reversal invariant, one has $n_L(-\bm{k}_0,t)= n_L(\bm{k}_0,t)$, i.e. the time-independent terms have the same value for both CBS and CFS peaks. We have:
\begin{equation}
    \begin{aligned}
    n_S(-\bm{k}_0,t) &= \overline{|\sum_{C_0\in \mathbb{S}} S_{C_0}(-\bm{k}_0) \, \Phi_{C_0}(-\bm{k}_0,t)|^2} \\
    &= \sum_{C_0 \in \mathbb{S}} (\overline{|S_{C_0}(-\bm{k}_0)|^2}-1) \, |\Phi_{C_0}(-\bm{k}_0,t)|^2 \\
    &+ |\sum_{C_0 \in \mathbb{S}} \Phi_{C_0}(-\bm{k}_0,t)|^2
    \end{aligned}
\end{equation}
where we have used, when $C_0 \neq D_0$, $\overline{S_{D_0}^*(-\bm{k}_0)S_{C_0}(-\bm{k}_0)} = \overline{S_{D_0}^*(-\bm{k}_0)} \ \overline{S_{C_0}(-\bm{k}_0)} = 1$
since $\overline{S_{C_0}(-\bm{k}_0)} = 1$. Furthermore:
\begin{equation}
    \overline{|S_{C_0}(-\bm{k}_0)|^2} = \overline{\sum_{\bm{d}\bm{d}'} e^{2i\bm{k}_0 \cdot (\bm{d}-\bm{d}')}} = \overline{\sum_{\bm{d}} 1} = \overline{N(C_0)}.
\end{equation}
As a consequence, the dominant contribution to $n_S(-\bm{k}_0,t)$ writes:
\begin{equation}
    n_S(-\bm{k}_0,t) \sim \sum_{C_0 \in \mathbb{S}} \overline{N(C_0)} \, |\Phi_{C_0}(-\bm{k}_0,t)|^2.
\end{equation}

As one can see, the CBS peak height also displays temporal oscillations at intra-cluster frequency differences $\Delta_{ab}(C_0)$ only but with a much reduced amplitude compared to the CFS oscillations. Indeed, the oscillation terms  are weighted by $\overline{N(C_0)}$ for the CBS sum while and $\overline{N^2(C_0)}$ for the CFS sum. In the limit of large system size, we expect the relative size of the CBS to CFS oscillations to go to zero. 

For example, from Fig.~\ref{fig:allclustersizedistributions}, one can see that, at size $L=50$ and defect density $\rho=0.34$, the number of small clusters is about $40$, such that the ratio $N_{C_0}/N^2_{C_0}\approx 1/N_{C0}$, is of order of $1/40$. The amplitude of the CFS oscillations being about $2$, this means that the amplitude of the CBS oscillations should be about $2/40=0.05$, in qualitative agreement with Fig.~\ref{fig:CBSandCFS2}b.

\subsection{CFS oscillation frequencies}

\begin{figure}[!htbp]
\includegraphics[width =1\linewidth]{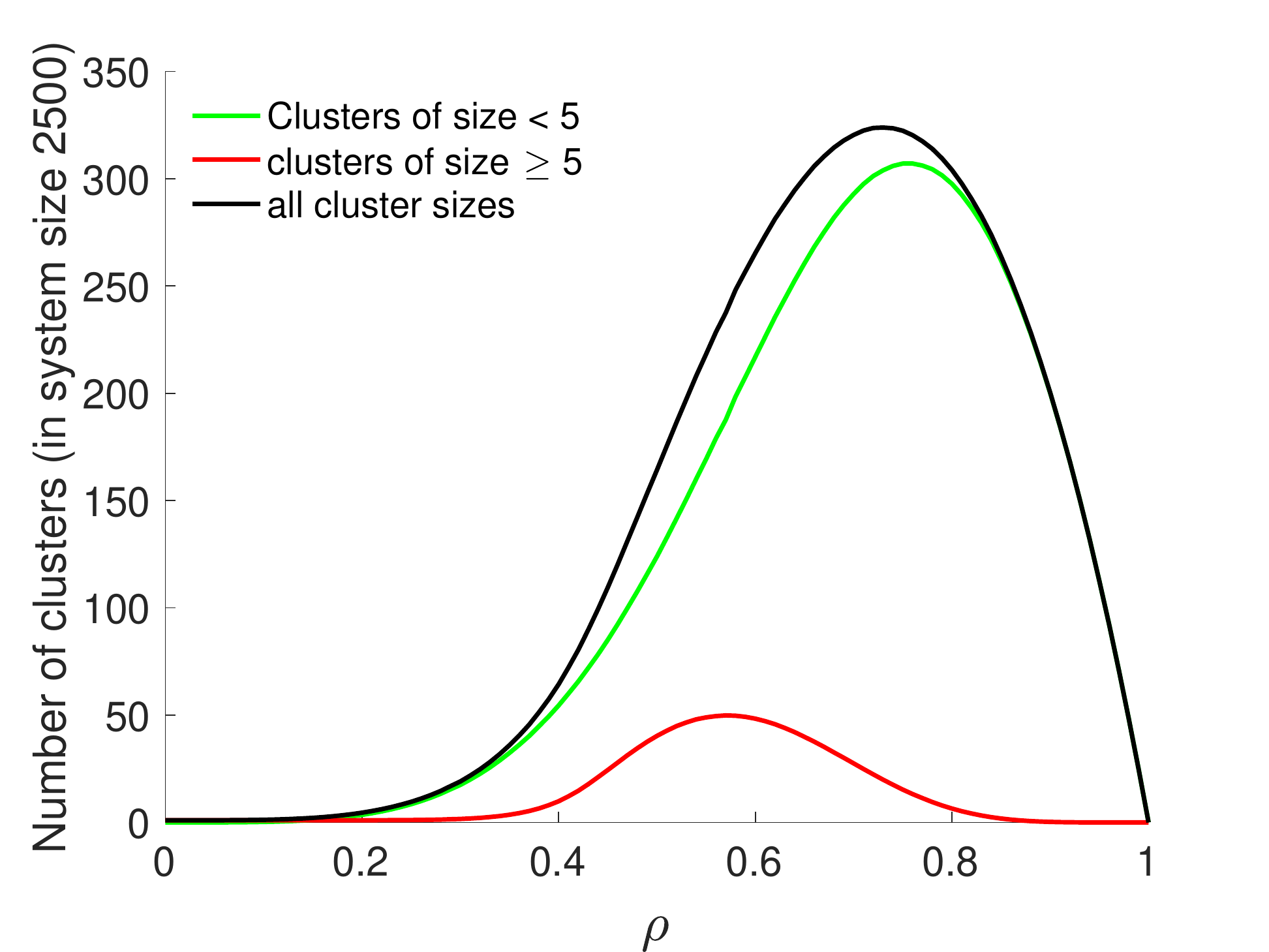}
\caption{\label{fig:allclustersizedistributions} Number of $n$-polyominoes as a function of defect density $\rho$ for our lattice system comprising $N=L^2=2500$ sites.
}\end{figure}

Fig.\ref{fig:allclustersizedistributions} shows that, in terms of number, the $n$-polyominoes with sizes $n \leq 4$ dominate the disorder configurations, see Appendix \ref{Perco} for more details. Obviously, such small-size polyominos have regular eigenspectra. From Fig.\ref{fig:spectralsteady50}, we see that the disorder-resisting frequency differences are $\Delta_s t_J=0.5$, $1$ and $1.5$. We now check how these $\Delta_s$ can be easily inferred from the spectra of small-size polyominoes. 

\begin{figure}[ht]
    \centering
    \includegraphics[width=0.8\linewidth]{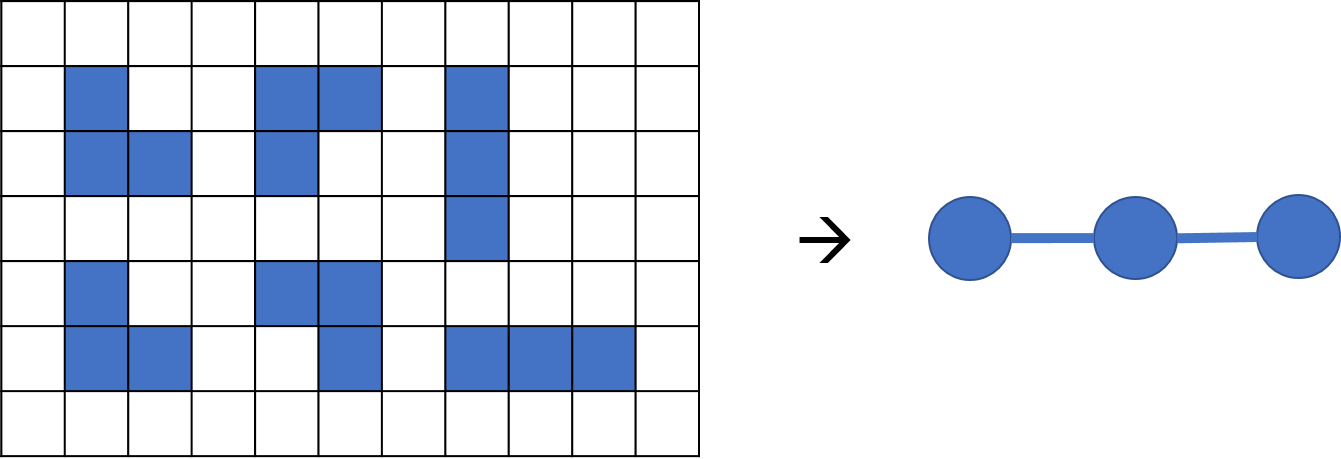}
    \caption{The 6 possible trominoes. Being all graph-equivalent, their associated Hamiltonians all have the same eigenvalues.}
    \label{fig:all3clusters}
\end{figure}

Using Eq. \eqref{eq:effectivehwithdefects}, it is easy to see that the spectrum of dominoes is $\{0,0.5\}$ and that of trominoes is $\{0, 0.5, 1.5\}$ (in units of $\hbar/t_J$). Fig.\ref{fig:all3clusters} gives the 6 possible trominoes. From this, we immediately see that the nonzero frequency differences are $\Delta_s t_J = 0.5$, $1$ and $1.5$, as witnessed in Fig.\ref{fig:spectralsteady50}. At this point, we remind the Reader that the $0$ eigenvalue always belongs to the spectrum of any Hamiltonian $\mathcal{H}_C$. This is because each of these Hamiltonians is represented by a Laplacian matrix, see Section \ref{LapMat}. 

\begin{figure}[ht]
    \centering
    \includegraphics[width=0.9\linewidth]{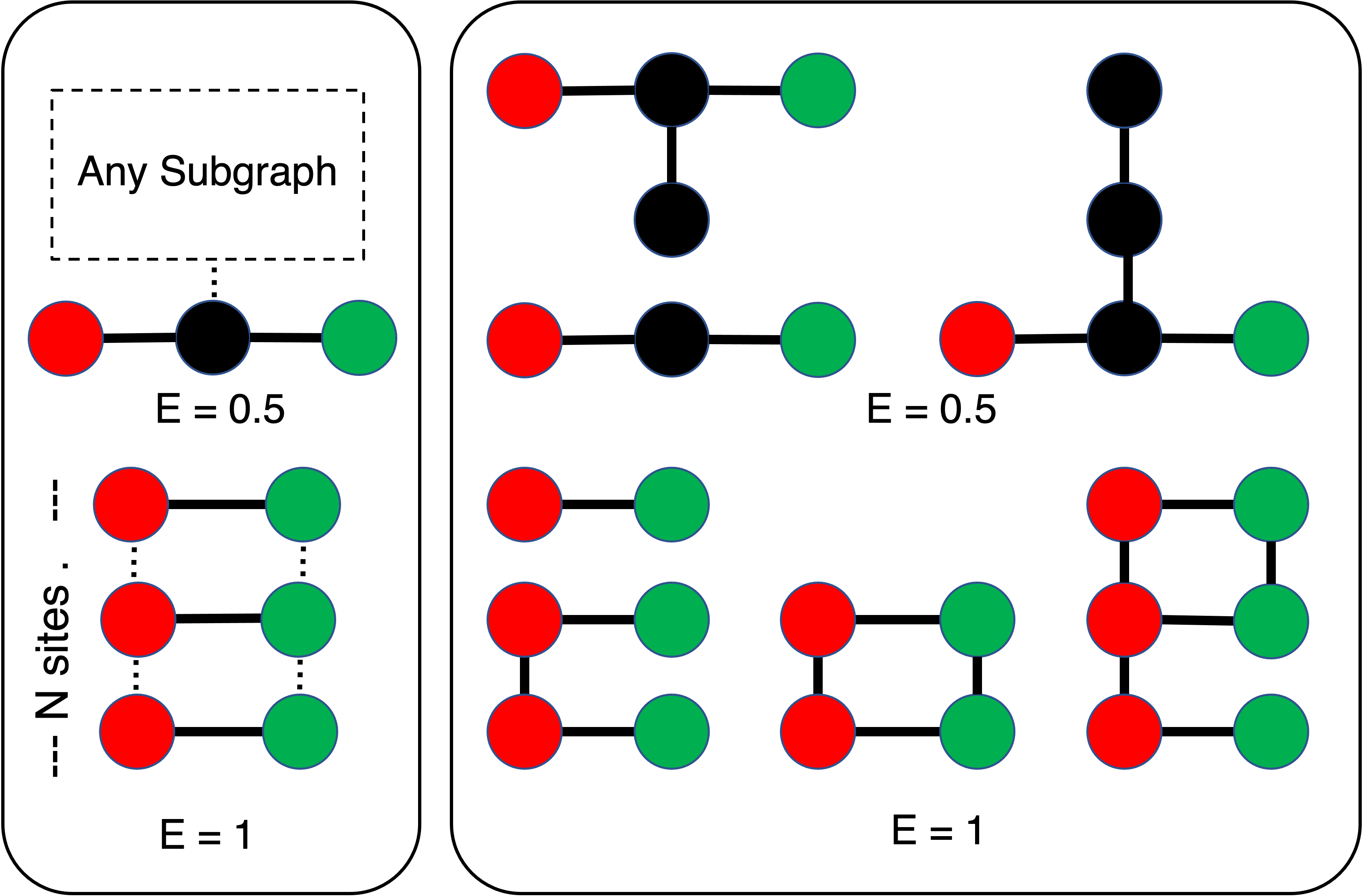}
    \caption{Graph-equivalent polyominoes that give rise to eigenvalues $E=0.5$ and $E=1$ (in units of $\hbar/t_J$). Left panel: Graph structure. The vertex is colored in red when the corresponding (unnormalized) eigenvector has site entry $1$, in black when it has site entry $0$ and in green when it has site entry $-1$. The dashed lines correspond to optional possible edges not changing the occurrence of the eigenvalues $0$ or $1$. Right panel: Some concrete examples of polyominoes giving rise to eigenvalue $E=0$ and $E=1$.
    }
    \label{fig:generalized_clusters}
\end{figure}

Actually, one can see that $E=0.5$ and $E=1$ (in units of $\hbar/t_J$) are two special graph-invariant eigenvalues, see Fig.\ref{fig:generalized_clusters}. Indeed, the eigenvalue $E=0.5$ always arises for polyominoes associated to graphs consisting of an arbitrary subgraph attached to the middle vertex of a $3$-vertex subgraph. The corresponding eigenvector for such a case has opposite components on the end vertices of the $3$-vertex subgraph and $0$ components elsewhere. The proof is simple: The hopping terms induce a destructive interference at the middle vertex which blocks spreading to the rest of the graph. By the same token, the eigenvalue $E=1$ always arises when the associated graph is build by connecting 2-vertex subgraphs, see Fig.\ref{fig:generalized_clusters}.

To conclude, the disorder-averaged CFS power spectrum $P({\bm k}_0, \omega)$ indeed exhibits discrete peaks growing with $\rho$ and mostly located at $\Delta_s t_J=0$ (static component), $\Delta_s t_J=0.5$ (temporal oscillation with period $T=4\pi \, t_J$) and $\Delta_s t_J=1$ (temporal oscillation with period $T=2\pi \, t_J$). 

In Appendix \ref{SpecFun}, we show the emergence of these discrete peaks signalling percolation in the disorder-averaged spectral function.

\section{Conclusion}

In this paper, we have considered a 2D ferromagnetic square lattice hosting randomly placed nonmagnetic defects and we have studied the time propagation of an initial plane wave ${\bm k}_0$ in the linear spin-wave limit. We have shown how the momentum distribution of the system changes when the defect density $\rho$ increases and site percolation sets in. We have documented the existence of two regimes. In the low defect density regime $\rho \ll 1$, typical disorder configurations are typically made of a macroscopic connected component essentially interspersed with single defects. In this case, the dynamics of the system falls into the usual category of wave propagation in random media and exhibits Anderson localization. Coherent transport effects in momentum space are revealed by the emergence of the emblematic CBS and CFS interference peaks, located at $-{\bm k}_0$ and ${\bm k}_0$ respectively, on top of an isotropic diffusive background. On the other hand, in the high defect density regime when $\rho$ is no longer much smaller than $1$, disorder configurations typically break up into many isolated clusters $C$ of different sizes and shapes called polyominoes. In this case, the CFS peak starts to oscillate in time. The total Hamiltonian of the system admits a cluster-component expansion $\mathcal{H} = \sum_C \mathcal{H}_C$ and a Fourier analysis reveals that the frequency spectrum of these CFS oscillations is given by energy differences between eigenenergies residing in the regular part of the spectrum of $\mathcal{H}$. These disorder-immune eigenenergies are associated to Hamiltonians $\mathcal{H}_C$ associated to small-size magnetic clusters $C$. Possible extensions of this work include (i) the regular-to-random transition of the eigenenergy spectrum of this system as $\rho$ increases, (ii) signatures of the percolation transition and of its critical properties in the CFS signal and (iii), the impact of interactions between magnons on the temporal evolution of the CFS peak and its nonlinear features.\\

{\it Acknowledgements --} C. M. would like to thank Sanjib Ghosh for useful discussions. The project leading to this publication has received
funding from Excellence Initiative of Aix-Marseille Uni-
versity - A*MIDEX, a French “Investissements d’Avenir”
program through the IPhU (AMX-19-IET-008) and
AMUtech (AMX-19-IET-01X) institutes.

\appendix
\section{Clean Linear Spin Wave Hamiltonian}
\label{CleanH}

We start from the ferromagnetic Heisenberg Hamiltonian on a lattice $\mathcal{L}$ with periodic boundary conditions:
\begin{equation}
        \mathcal{H}_S  = - \sum_{(ij)} J_{ij} \, \bm{S}_i\cdot \bm{S}_j
        \label{eq:Heis}
\end{equation}
where $(ij)\equiv (ji)$ denotes the link that connects the {\it unordered pair} of nearest-neighbor sites $i$ and $j$ and where $J_{ij} = J_{ji} >0$. For a 1D spin chain, we would have: 
\begin{equation}
        \mathcal{H}_S  = (\cdot\cdot\cdot) - J_{12} \, \bm{S}_1 \cdot \bm{S}_2 - J_{23} \, \bm{S}_2 \cdot \bm{S}_3 - J_{34} \, \bm{S}_3 \cdot \bm{S}_4 + (\cdot\cdot\cdot).
\end{equation}
Note that $\mathcal{H}_S$ can be rewritten as:
\begin{equation}
        \mathcal{H}_S = - \frac{1}{2}\sum_{i \in \mathcal{L}} \sum_{j \in \mathcal{N}(i)} J_{ij} \, \bm{S}_i \cdot \bm{S}_j
\end{equation}
where $\mathcal{N}(i)$ is the set of all nearest-neighbor sites to site $i$. The factor $1/2$ in front takes care of double counting the interaction terms.

Writing Eq.\eqref{eq:Heis} as $\mathcal{H}_S = - \bm{S}_i\cdot\bm{B}_i + \mathcal{H}'_S$, where $\bm{B}_i = \sum_{j \in \mathcal{N}(i)} J_{ij} \, \bm{S}_j $ and where $\mathcal{H}'_S$ does not involve spin $\bm{S}_i$, the Heisenberg equation of motion for spin $\bm{S}_i$ reads:
\begin{equation}
    \frac{d\bm{S}_i}{dt} = \mathrm{i} [\mathcal{H}_S, \bm{S}_i] = \bm{S}_i \times \bm{B}_i = \sum_{j \in \mathcal{N}(i)} \, J_{ij} \, \bm{S}_i \times \bm{S}_j
    \label{eq:HeiSpin}
\end{equation}
where we have used the commutation relations for spin components $[S_a,S_b] = \mathrm{i} \sum_c \epsilon_{abc} S_c$ ($\epsilon_{abc}$ is the fully anti-symmetric Levy-Civita tensor) \cite{quispel1982, Patterson2007}.

Note that these Heisenberg equations of motion are nonlinear in the spin operators. To extract the Hamiltonian describing the linear spin wave excitations of the system around its ferromagnetic ground state where all spins are aligned along $Oz$, we resort to the Holstein-Primakov transformation \cite{HolPrim}
\begin{equation}
    \begin{aligned}
        S^{z}_i & = S-a^{\dagger}_ia_i \\
        S^{+}_i & = S^x_i + \mathrm{i} S^y_i = \sqrt{2 S - a^{\dagger}_i a_i } \ a_i \\
        S^{-}_i & = S^x_i - \mathrm{i} S^y_i = a^{\dagger}_i\, \sqrt{ 2S-a^{\dagger}_i a_i}, 
        \label{eq:HolPrim}
\end{aligned}
\end{equation}
featuring the onsite {\it bosonic} creation and annihilation operators $a_i$ and $a^\dagger_i$ satisfaying $[a_i,a^\dagger_i]=1$. To lowest order in $a^{\dagger}_ia_i$, we have $S^{z}_i = S$, $S^{+}_i = \sqrt{2S} \, a_i$ and $S^{-}_i = \sqrt{2S} \, a^{\dagger}_i$ so that Eq.\eqref{eq:HeiSpin} reads:  
\begin{equation}
    \frac{da_i}{dt} = \mathrm{i} \sum_{j \in \mathcal{N}(i)} SJ_{ij} \, (a_j-a_i) = \mathrm{i} [H_0,a_i]
\end{equation}
with
\begin{equation}
    H_0 = - \sum_{(ij)}  SJ_{ij} \, (a^{\dagger}_i a_j + a^{\dagger}_j a_i) + \sum_{i \in \mathcal{L}} U_i \, a^{\dagger}_i a_i
\end{equation}
and $U_i = \sum_{j \in \mathcal{N}(i)} SJ_{ij}$. In first quantization language, we recover Eq.~\eqref{eq:cleanH} and Eq.\eqref{eq:CleanHUni} for the uniform case $J_{ij} = J$.

\section{Renormalized clean dispersion relation}
\label{RenormH}

\begin{figure}
    \centering
    \includegraphics[width=\linewidth]{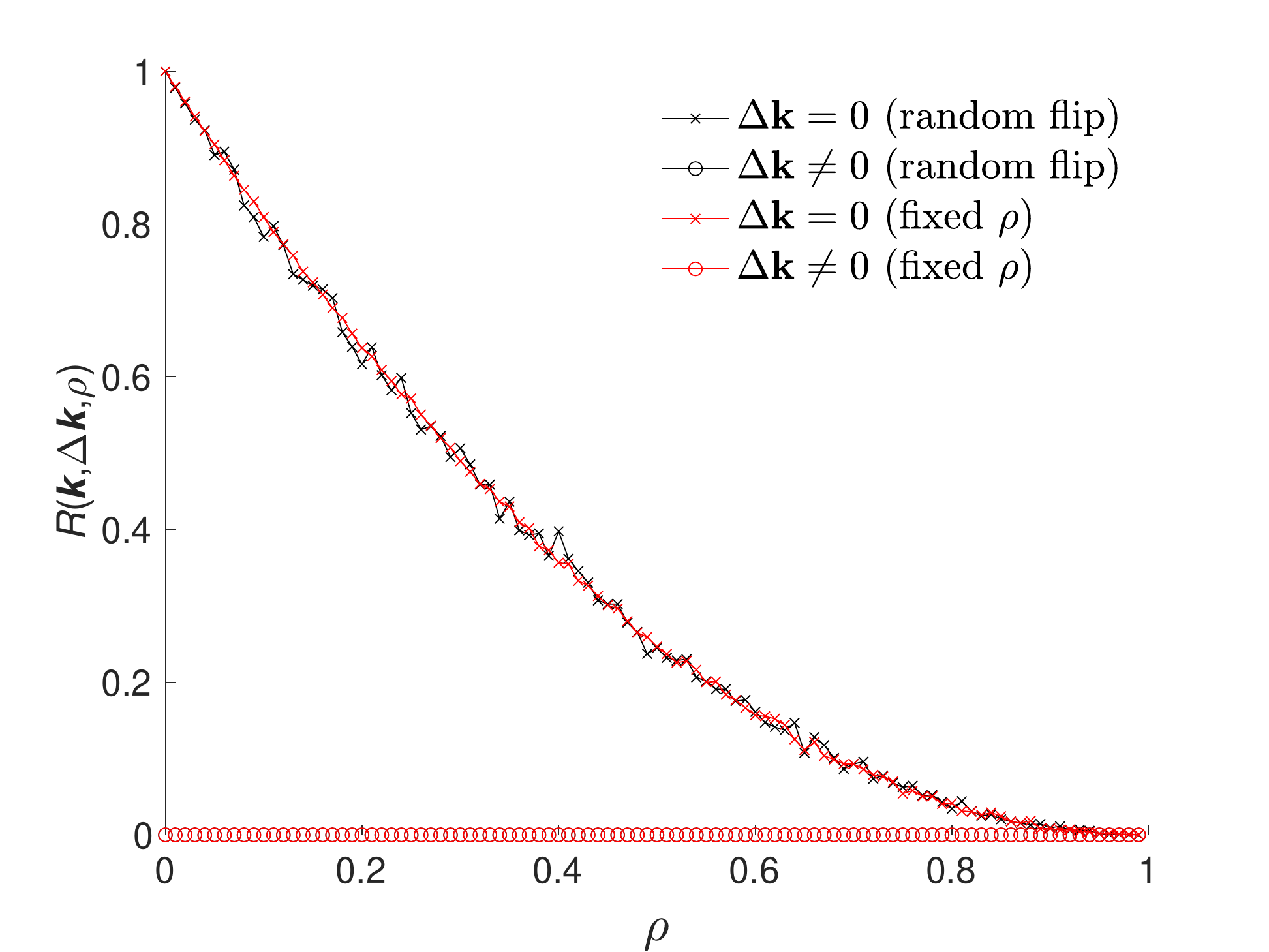}
    \caption{Plot of the ratio $R(\bm{k}, \Delta\bm{k}, \rho) = \langle \bm{k} + \Delta\bm{k}|\mathcal{H}|\bm{k}\rangle /\varepsilon^{0}_{\bm{k}}$ obtained for one disorder configuration and its disorder-averaged value $\overline{R}(\bm{k}, \Delta\bm{k},\rho)$ as a function of $\rho$ at $\bm{k}a = 0.4 \,  \pi \, \hat{e}_y$ and for $\Delta\bm{k} = \boldsymbol{0}$ and $\Delta\bm{k} = 0.1 \bm{k}$. As one can see, the diagonal element $R(\bm{k}, \Delta\bm{k} = \boldsymbol{0}, \rho)$ fluctuates around $(1-\rho)^2$ while the off-diagonal one $R(\bm{k}, \Delta\bm{k} \neq \boldsymbol{0}, \rho)$ fluctuates around $0$. The fluctuations themselves do average to zero after disorder average. This shows that $\widetilde{H}_0 = \overline{H} = \sum_{\bm{k}} \varepsilon_{\bm{k}} \ket{\bm{k}}\bra{\bm{k}}$ features the renormalized clean dispersion relation $\varepsilon_{\bm{k}} = (1-\rho)^2 \, \varepsilon^{0}_{\bm{k}}$. We have further checked that the two types of disorder (fixed defect density $\rho$ or randomly flipping each lattice site with probability $\rho$) give the same results.}
    \label{fig:E_vs_rho}
\end{figure}

To compute $\overline{H_d}$, we face the disorder average of the link random variable $\overline{m_{ij}} = 1 - \overline{m_im_j}$ for $j \neq i$. The random variable $m_im_j$ can only take two values, namely $1$ (with probability $p_1$) and $0$ (with probability $p_0=1-p_1$). Trivially, $\overline{m_im_j} = p_1$. Since $m_im_j =1$ ($j \neq i$) is obtained for $m_i =1$ and $m_j=1$, we have $p_1 =(N-N_D)(N-N_D-1)/[N(N-1)] \to (1-\rho)^2$ in the thermodynamic limit $(N,N_D) \to \infty$ at fixed $\rho = N_D/N$. We then conclude that $\overline{m_{ij}}= \rho(2-\rho)$ and thus $\overline{H_d} = -\rho(2-\rho) \, H_0$. As a consequence $\widetilde{H}_0 = \overline{H} =  H_0 + \overline{H_d} = (1-\rho)^2 \, H_0$, leading to the disorder-renormalized clean dispersion relation Eq.\eqref{eq:RenormH0}. We show in Fig.\ref{fig:E_vs_rho} that this predicted $(1-\rho)^2$ dependency is indeed satisfied.

\section{Green's Function, Self-Energy, and Transition Operator}
\label{Self}

\subsection{General Definitions}
We recapitulate here the general results about the retarded Green's function associated to some disorder Hamiltonian $H= H_0 +V$, where $H_0$ is the clean Hamiltonian and $V$ the disorder potential, assumed here to have a vanishing disorder average $\overline{V}=0$. It is defined by
\begin{equation}
    G(E) = [E-H +\mathrm{i}0^+]^{-1} = [E-H_0 -V +\mathrm{i}0^+]^{-1}
\label{eq:G}
\end{equation}
such that the time evolution operator reads
\begin{equation}
    U(t) = e^{-\mathrm{i} H t/\hbar} = \mathrm{i} \int_{-\infty}^{+\infty} \frac{dE}{2\pi} \, e^{-\mathrm{i} E t/\hbar} \, G(E)
\label{eq:U}
\end{equation}
for $t \geq 0$. 

The Green's function satisfies the recursive relation $G(E) = G_0(E) + G_0(E) V G(E)$, where $G_0(E)$ is the Green's function associated to 
the clean Hamiltonian $H_0$.

A first quantity of interest is the disorder-averaged Green's function $\overline{G}(E)$. It satisfies the Dyson equation $\overline{G}(E) = G_0(E) + G_0(E) \Sigma(E) \overline{G}(E)$ \cite{Rammer1998,Sheng1995} and reads:
\begin{equation}
    \overline{G}(E) = [E - H_0 - \Sigma(E)]^{-1}.
\label{eq:dyson}
\end{equation}

The Dyson equation in fact defines the self-energy operator $\Sigma(E)$. Since disorder average restores translation invariance of the system, $\overline{G}(E)$ and $\Sigma(E)$ are both diagonal in $\bm{k}$:
\begin{equation}
    \begin{aligned}
        \langle \bm{k}|\overline{G}(E)|\bm{k}'\rangle & = \overline{G}(E,\bm{k}) \ \delta_{\bm{k}\bm{k}'} \\
        \langle \bm{k}|\Sigma(E)|\bm{k}'\rangle & = \Sigma(E,\bm{k}) \ \delta_{\bm{k}\bm{k}'}, 
    \end{aligned}
\end{equation}
and we have
\begin{equation}
    \begin{aligned}
         \overline{G}(E,\bm{k}) & = [E - \varepsilon^0_{\bm{k}} - \Sigma(E,\bm{k}) ]^{-1} \\
         & = \Big[E - \varepsilon^0_{\bm{k}} - \textrm{Re}\Sigma(E,\bm{k}) + \mathrm{i} \frac{\hbar \Gamma_s(E,\bm{k})}{2}\Big]^{-1}.
    \end{aligned}
\end{equation}
where $\varepsilon^0_{\bm{k}}$ is the clean dispersion relation and $\Gamma_s(E,\bm{k})= -2 \textrm{Im}\Sigma(E,\bm{k})/\hbar >0$ is the scattering mean free rate at energy $E$ and wavenumber $\bm{k}$. The scattering mean free time is simply $\tau_s(E,\bm{k}) = \Gamma^{-1}_s(E,\bm{k})$.

The so-called coherent amplitude is given by the disorder-average state $\overline{\ket{\Psi(t)}}$. Starting from the initial plane wave $\ket{\Psi(t=0)} = \ket{\bm{k}}$, it is easy to see that $\langle \bm{k}'\overline{\ket{\Psi(t)}} = \langle \bm{k}\overline{\ket{\Psi(t)}} \, \delta_{\bm{k}\bm{k}'}$. Introducing the dispersion relation $E_{\bm{k}}$ of the disordered system, obtained by solving $E-\varepsilon^0_{\bm{k}} - \textrm{Re}\Sigma(E,\bm{k}) = 0$, we see that 
\begin{equation}
    \langle \bm{k}\overline{\ket{\Psi(t)}} \approx \mathrm{i} \int_{-\infty}^{+\infty} \frac{dE}{2\pi}  \, \frac{e^{-\mathrm{i} E t/\hbar}}{E - E_{\bm{k}}+ \mathrm{i} \frac{\hbar}{2 \tau_s}} = e^{-\frac{t}{2\tau_s}}
\end{equation}
provided $|\textrm{Re}\Sigma(E,\bm{k})-\textrm{Re}\Sigma_{\bm{k}}| \ll |E-E_{\bm{k}}|$ and $|\textrm{Im}\Sigma(E,\bm{k}) - \textrm{Im}\Sigma_{\bm{k}}| \ll |\textrm{Im}\Sigma_{\bm{k}}|$, with $\Sigma_{\bm{k}} = \Sigma(E_{\bm{k}}, \bm{k})$, hold over the whole energy range. When this is the case, we find that the initial coherent population peak decreases exponentially over the time scale $\tau_s \equiv \tau_s(E_{\bm{k}}, \bm{k})$. At weak enough disorder, we expect $E_{\bm{k}} \approx \varepsilon^0_{\bm{k}}$ (on-shell scattering).

The transition operator $T(E)$ is defined by $ G(E) = G_0(E) + G_0(E) T(E) G_0(E)$ and we have: 
\begin{equation}
    \begin{aligned}
        \overline{G}(E) & = G_0(E) + G_0(E) \Sigma(E) \overline{G}(E) \\
        & = G_0(E) + G_0(E) \overline{T}(E) G_0(E).
    \end{aligned}
    \label{eq:SigmaT}
\end{equation}
The disorder-averaged transition operator $\overline{T}(E)$ satisfies the iterative equation $\overline{T}(E) = \overline{V G(E)V}$ and is linked to the self-energy operator by
\begin{equation}
    \Sigma(E) = \overline{T}(E) \, [\mathbbm{1}+ G_0(E)\overline{T}(E) ]^{-1}.
\end{equation}
The self-energy is given by the sum of 1-particle irreducible diagrams \cite{Rammer1998,Sheng1995}. At lowest order in a perturbative expansion, one has:
\begin{equation}
   \begin{aligned}
         \Sigma(E) & = \overline{V G_0(E) V} + (...)\\
         \overline{T}(E) & = \overline{V G_0(E) V} + (...)
   \end{aligned}
\end{equation}
and thus $\Sigma(E) \approx \overline{T}(E) \approx \overline{V G_0(E) V}$. 

\subsection{Application to our System}

To match with the previous definitions, we need to write our system Hamiltonian $H= H_0 + H_d$ as $H=\widetilde{H}_0 + \delta H_d$ with $\widetilde{H}_0 = H_0 + \overline{H}_d$, see Section~\ref{ScatApp}, and use the previous definitions through the change $H_0 \to \widetilde{H}_0$, $V \to \delta H_d$, $G_0(E) \to \widetilde{G}_0(E)$ and $\varepsilon^0_{\bm{k}} \to \varepsilon_{\bm{k}}$, the renormalized clean dispersion relation.

To compute the self-energy and the scattering mean free time, we break $H_d$  into its {\it defect clusters} components
\begin{equation}
    H_d = \sum_{m=1}^{N_D} H^{(m)}_d
\end{equation}
and define $\delta H^{(m)}_d = H^{(m)}_d - \overline{H^{(m)}_d}$ is the disorder Hamiltonian associated to $m$-defects, that is clusters made of $m$ connected defects ($1$-defects are just single isolated defects). Note that, for a given configuration of $N_D$ defects, some of the $H^{(m)}_d$ may simply be zero.

At this point, it is difficult to proceed without approximations. In the dilute regime $\rho \ll 1$, the probability to get $m$-defects with sizes $m\geq 2$ should be extremely low so that one can discard them. Within this approximation, one has $\delta H_d \approx \delta H^{(1)}_d = \sum_{i_0} \delta H^{(1)}_d(i_0)$, with the sum running over isolated defective sites only, and $\Sigma (E) \approx \overline{T^{(1)}}(E)$. Since the average separation between defects is $\rho^{-1/2} a \gg a$, another approximation can be further made in this dilute regime by neglecting recurrent scattering events. This means one only keeps scattering paths where a given defective site is only visited once. Within this independent scattering approximation, we have $\Sigma(E) \approx \overline{T^{(1)}}(E) \approx N_D \overline{T^{(1)}}(E, i_0)$ where $T^{(1)}(E, i_0)$ is the transition operator associated to a {\it single} defect $i_0$ \cite{AkkMon2007}.

\section{Scattering by a Single Defect}
\label{ScatDef}
The disorder Hamiltonian $H^{(1)}_d(i_0)$ associated to a single isolated defect located at some lattice site $i_0$ labelled by $\bm{r}_0$ is obtained from Eq.\eqref{eq:effectivehwithdefects} by setting $m_i = 1- \delta_{ii_0}$ and thus $m_{ij} = \delta_{ii_0} + \delta_{ji_0} - \delta_{ii_0}\delta_{ji_0}$. Writing $H^{(1)}_d(i_0) = JS \sum_{j \in \mathcal{N}(i_0)} H^{(1)}_d(i_0,j)$, we get:
\begin{align}
   H^{(1)}_d(i_0)(i_0, j) & = \ket{\bm{r}_0} \bra{\bm{r}_j} + \ket{\bm{r}_j} \bra{\bm{r}_0} \nonumber \\
   & - \ket{\bm{r}_j} \bra{\bm{r}_j} - \ket{\bm{r}_0} \bra{\bm{r}_0}.
\end{align}
After simple algebra, we find:
\begin{equation}
   \bra{\bm{k}'} H^{(1)}_d(i_0) \ket{\bm{k}} = - \frac{2JS}{N}\, e^{\mathrm{i} (\bm{k}-\bm{k}')\cdot \bm{r}_0} \, F(\bm{k},\bm{k}'),
\end{equation}
where $F(\bm{k},\bm{k}') = \sum_{\alpha = x,y} f(k_\alpha,k'_\alpha)$ with \begin{equation}
    f(u,v) = 1+\cos{[a(u-v)]} -\cos{(a u)} -\cos{(a v)}.
\end{equation} 
From Eq.\eqref{eq:DispRel}, one may want to note that $\varepsilon^0_{\bm{k}} = JS \, F(\bm{k},\bm{k})$. It is easy to see that in the limits $(ka, k'a) \ll 1$, we have:
\begin{equation}
   F(\bm{k},\bm{k}') = a^2 \, \bm{k}\cdot\bm{k}'. 
\label{eq:H_d_kk'lowk}
\end{equation}
Since $\bm{r}_0$ can be anywhere in the lattice with equal probability, $\overline{e^{\mathrm{i} (\bm{k}-\bm{k}')\cdot \bm{r}_0}} = \delta_{\bm{k}\bm{k}'}$ and we have
\begin{equation}
   \bra{\bm{k}'} \overline{H^{(1)}_d(i_0)} \ket{\bm{k}} = - \frac{2JS}{N} \, F(\bm{k},\bm{k}) \, \delta_{\bm{k}\bm{k}'}.
\label{eq:avH_d_kk'}
\end{equation}
Defining $\delta H^{(1)}_d(i_0) = H^{(1)}_d(i_0) - \overline{H^{(1)}_d(i_0)}$, we thus find:
\begin{equation}
   \bra{\bm{k}'} \delta H^{(1)}_d(i_0) \ket{\bm{k}} = - \frac{2JS}{N}\, e^{\mathrm{i} (\bm{k}-\bm{k}')\cdot \bm{r}_0} \, F(\bm{k},\bm{k}') (1- \delta_{\bm{k}\bm{k}'}).
\end{equation}

\section{Scattering Mean Free Time}
\label{ScatTime}

Within the independent scattering approximation at the level of single isolated defects only, we have
\begin{equation}
    \Sigma (E) \approx N_D \, \overline{\delta H^{(1)}_d(i_0) \, \widetilde{G}_0(E) \, \delta H^{(1)}_d(i_0)}.
\end{equation}
Simple algebra then shows that $\langle \bm{k}'| \Sigma(E) | \bm{k}\rangle = \Sigma(E,\bm{k}) \, \delta_{\bm{k}\bm{k}'}$ where
\begin{equation}
    \Sigma(E,\bm{k}) = \frac{4\rho J^2S^2}{N} \sum_{\bm{q}} F^2(\bm{k},\bm{q}) \, \widetilde{G}_0(E,\bm{q}) (1- \delta_{\bm{k}\bm{q}}).
\end{equation}

With $t_J=\hbar /J$ and $\textrm{Im}\widetilde{G}_0(E,\bm{q}) = - \pi \, \delta(E-\varepsilon_{\bm{q}})$, we have 
\begin{align}
    \frac{t_J}{\tau_s(E,\bm{k})} & = - \frac{2\textrm{Im}\Sigma (E, \bm{k})}{J} \nonumber \\
    & = \frac{8\pi \rho J S^2}{N} \sum_{\bm{q}} F^2(\bm{k},\bm{q}) \, \delta(E-\varepsilon_{\bm{q}})(1- \delta_{\bm{k}\bm{q}}) \nonumber \\
    & = \frac{2\rho J S^2 a^2}{\pi} \, \int d\bm{q} \, F^2(\bm{k},\bm{q}) \, \delta (E-\varepsilon_{\bm{q}}) \label{eq:BornApprox},
\end{align}
where the last line is obtained in the continuum limit $N\to\infty$ with $\sum_{\bm{q}} \, (...) \to Na^2 \, \int d\bm{q}/(2\pi)^2 \, (...)$. Do note that the contribution of the $\delta_{\bm{k}\bm{q}}$ term reduces to $[-8\pi \rho J S^2 F^2(\bm{k},\bm{k})\delta(E-\varepsilon_{\bm{k}})]/N$ which vanishes in the limit $N\to\infty$. 

For on-shell scattering $E=E_{os}=\varepsilon_{\bm{k}}/(1-\rho)$, see Eq.\eqref{eq:OnShell}, we find: 
\begin{equation}
   \frac{t_J}{\tau_s(E_{os},\bm{k})}  = \frac{2\rho Sa^2}{\pi(1-\rho)^2} \, \int d\bm{q} \, F^2(\bm{k},\bm{q}) \, \delta \big[\frac{F(\bm{k},\bm{k})}{1-\rho}-
        F(\bm{q},\bm{q})\big].
\label{eq:ScatPred}
\end{equation}
In the limit $ka \ll \sqrt{1-\rho}$, we get 
\begin{equation}
    \frac{t_J}{\tau_s(E_{os},\bm{k})} \approx \frac{\rho S}{(1-\rho)^3} \, (ka)^4 \sim \rho S (ka)^4 \hspace{0.5cm} (\rho \ll 1)
\label{eq:ScatPredLowk}
\end{equation}
Since $\epsilon_{\bm{k}} \propto (ka)^2$ for $ka \ll 1$, we recover the well-known fact that $\tau_s \propto \varepsilon^{-2}_{\bm{k}}$ when $ka \to 0$ \cite{evers2015spin}. A plot of this independent scattering Born approximation (ISBA) prediction, Eq.\eqref{eq:ScatPred}, is shown in Fig.\ref{fig:k4dependence} as a function of $k_x a$ for $k_ya=0$ and compared to numerical data obtained for the scattering mean free rate. This ISBA prediction could be further improved by resorting to the Self-Consistent Born Approximation \cite{Vollhardt1980,Wolfle2010,Lee2013}. 

\section{Form Factor at $\rho \ll 1$}
\label{Form}

At very small defect densities, we can assume that a typical disorder configuration consists of a macroscopic connected magnetic cluster of size $N_m= (1-\rho)N$ randomly filled with $N_D=\rho N$ {\it isolated single defects}. Then, from Eq.\eqref{eq:nkt}, we see that:
\begin{equation}
   n(\bm{k}_0,t) = \overline{\sum_{nm} e^{-\mathrm{i}\omega_{nm}t} \, |\varphi_n(\bm{k}_0)|^2 |\varphi_m(\bm{k}_0)|^2}. 
\end{equation}
Writing $n(\bm{k}_0,t) = n_\infty(\bm{k}_0) + \delta n(\bm{k}_0,t)$ with $n_\infty(\bm{k}_0) \equiv  n(\bm{k}_0, t= \infty) = \overline{\sum_n |\varphi_n(\bm{k}_0)|^4}$, we have: 
\begin{equation}
        \delta n(\bm{k}_0,t) =\overline{\sum_{n \not= m} e^{-\mathrm{i}\omega_{nm}t} \, |\varphi_n(\bm{k}_0)|^2 |\varphi_m(\bm{k}_0)|^2}
\end{equation}
We now use the usual random matrix type assumption that eigenvalues fluctuations and eigenfunctions fluctuations are independent. This implies that for large enough times, the complex phase factors reach complete randomization and we get the decoupling:
\begin{equation}
        \delta n(\bm{k}_0,t) \approx \sum_{n \not= m} \overline{e^{-\mathrm{i}\omega_{nm}t}} \ \ \overline{|\varphi_n(\bm{k}_0)|^2 |\varphi_m(\bm{k}_0)|^2}, 
\label{eq:decoupling}
\end{equation}
The time scale set by this decoupling mechanism is the Heisenberg time $\tau_H$. The correlator $\widetilde{R}_N(\bm{k}_0,\omega_{nm}) = \overline{|\varphi_n(\bm{k}_0)|^2 |\varphi_m(\bm{k}_0)|^2}$, computed above for $n \not= m$, depends only on the eigenenergies difference $\omega_{nm}$ because of the disorder average. Going to Fourier space, we see that: 
\begin{equation}
    \begin{aligned}
        \delta n(\bm{k}_0, t) &\approx \int d\omega \, \widetilde{R}_N(\bm{k}_0,\omega) F_N(\omega) e^{-\mathrm{i}\omega t} \\
        F_N(\omega) &= \overline{\sum_{n \not= m} \delta (\omega-\omega_{nm})}
    \end{aligned}
\end{equation}
From Eq.\eqref{eq:formfactor}, we see that $F_N(\omega)$ is nothing else than the Fourier transform of $N_m (K_N(t)-1)$. From Eq.\eqref{eq:cfscontrast}, and noting that $n_\infty(\bm{k}_0) = 2 n_B(\bm{k}_0)$, we find:
\begin{equation}
    \Lambda(\bm{k_0},t) \approx 1 + \frac{N_m[(K_N-1)\otimes R_N](t)}{n_B(\bm{k}_0)}
    \label{eq:Lambda}
\end{equation}
where $R_N(\bm{k}_0,t)$ is the Fourier transform of $\widetilde{R}_N(\bm{k}_0,\omega)$.

Finally, in the large-time limit, or equivalently in the small-$\omega$ limit, the term $\widetilde{R}_N(\bm{k}_0,0)$ can be factored out of the integrals and we get $\Lambda(\bm{k}_0,t) = 1 + \gamma [K_N(t)-1]$ with $\gamma= N_m\widetilde{R}_N(\bm{k}_0,0)/n_B(\bm{k}_0)$. At this point, it is crucial to realize that $\widetilde{R}_N(\bm{k}_0,0)$ in Eq.\eqref{eq:Lambda} is computed for $n \not= m$ in the limit $\omega \to 0$, see Eq.\eqref{eq:decoupling}. We thus get:
\begin{equation}
   \begin{aligned}
         \widetilde{R}_N(\bm{k}_0,0) &= \lim_{\omega \to 0} \overline{|\varphi_n(\bm{k}_0)|^2 |\varphi_m(\bm{k}_0)|^2}\Big|_{n\neq m} \\
         & = (\overline{|\varphi_n(\bm{k}_0|^2})^2 = n_B(\bm{k}_0)/N_m
   \end{aligned}
\end{equation}
since different eigenstates are statistically independent (note that the same limit $\omega \to 0$ for $n=m$ would have given $n_\infty/N_m$ instead). In turn, $\gamma =1$ and we finally arrive at:
\begin{equation}
    \Lambda(\bm{k}_0,t) \approx K_{\textrm{reg}}(t) \hspace{1cm} t \gtrsim \tau_H.
\end{equation}

\begin{figure*} [!htbp]

        \sidesubfloat[]{%
        \includegraphics[width=.45\linewidth]{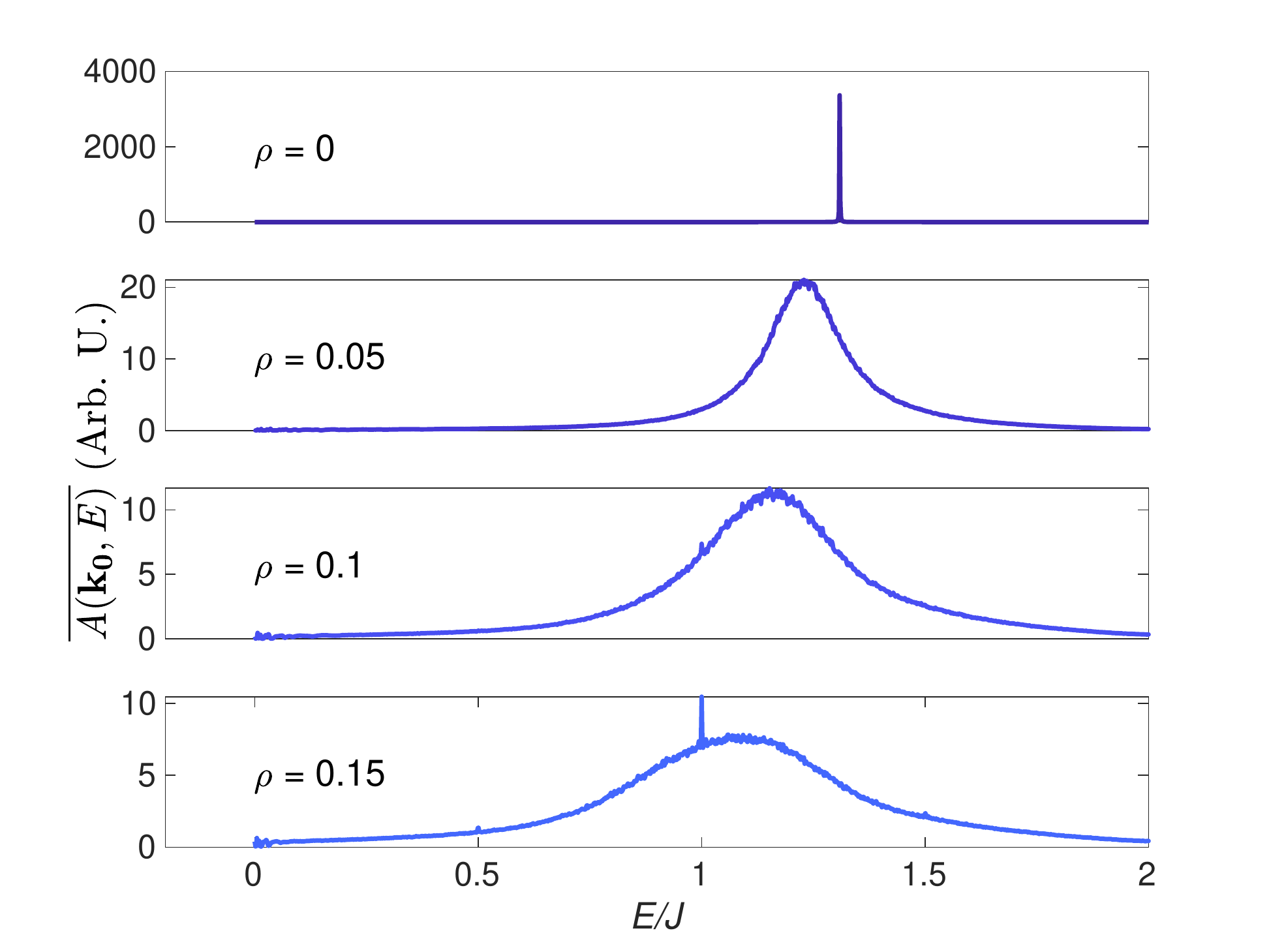}\label{fig:spectralfunctionlowdefect}}\quad
        \sidesubfloat[]{%
        \includegraphics[width=.45\linewidth]{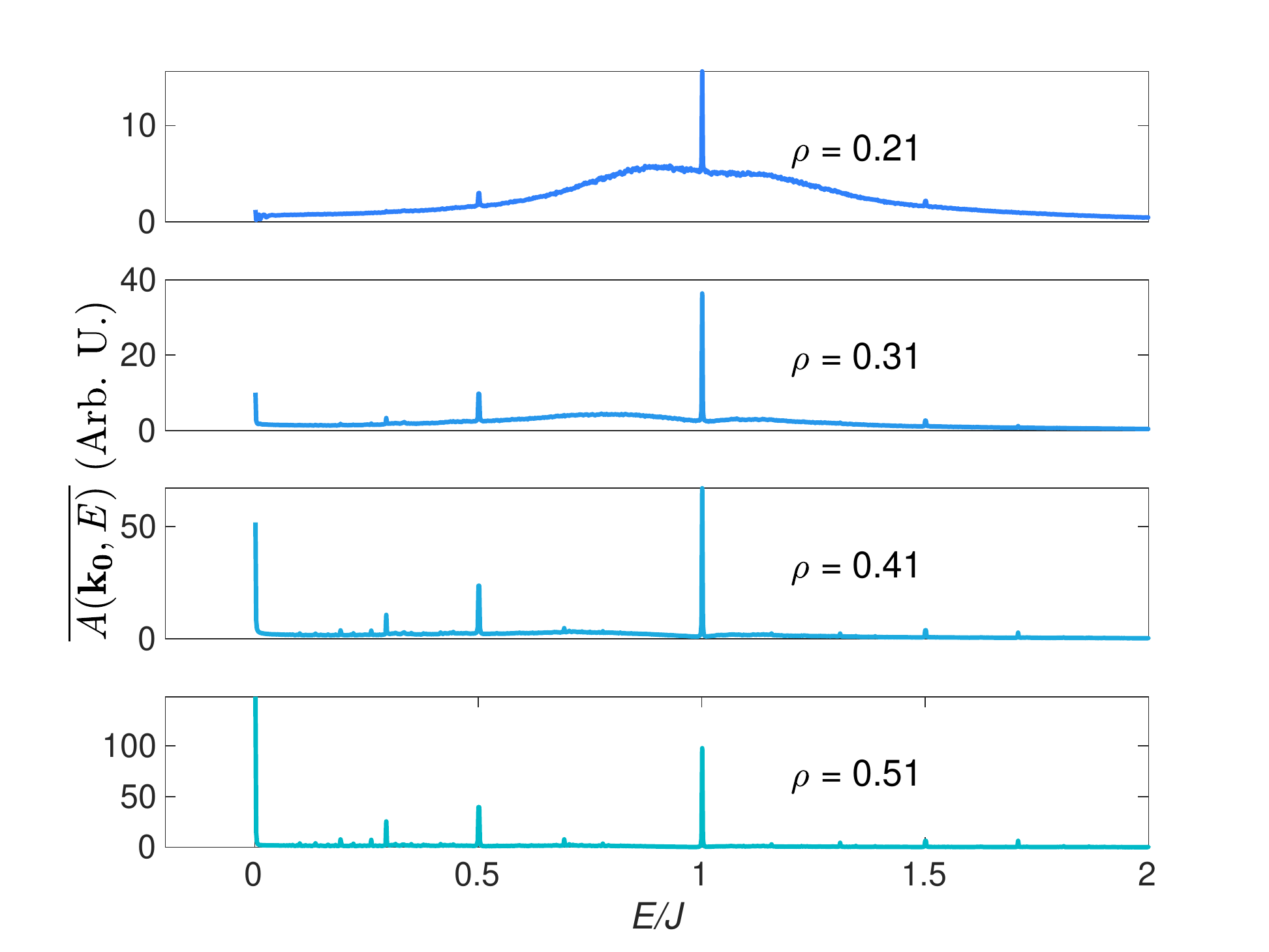}\label{fig:spectralfunctionhighdefect}}
        
        \sidesubfloat[]{%
        \includegraphics[width=.45\linewidth]{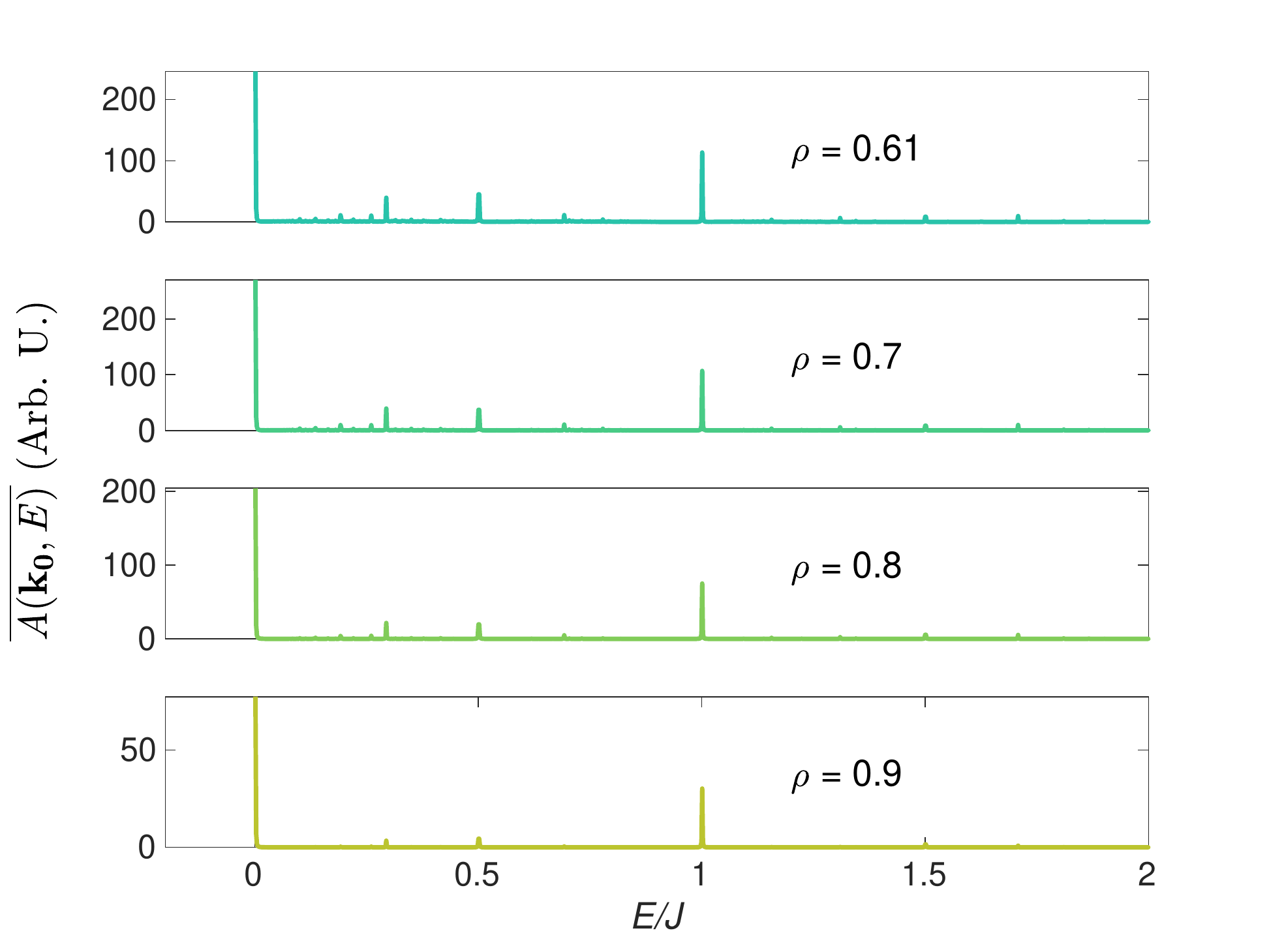}\label{fig:spectralfunctionveryhighdefect}}\quad
        \sidesubfloat[]{%
        \includegraphics[width=.45\linewidth]{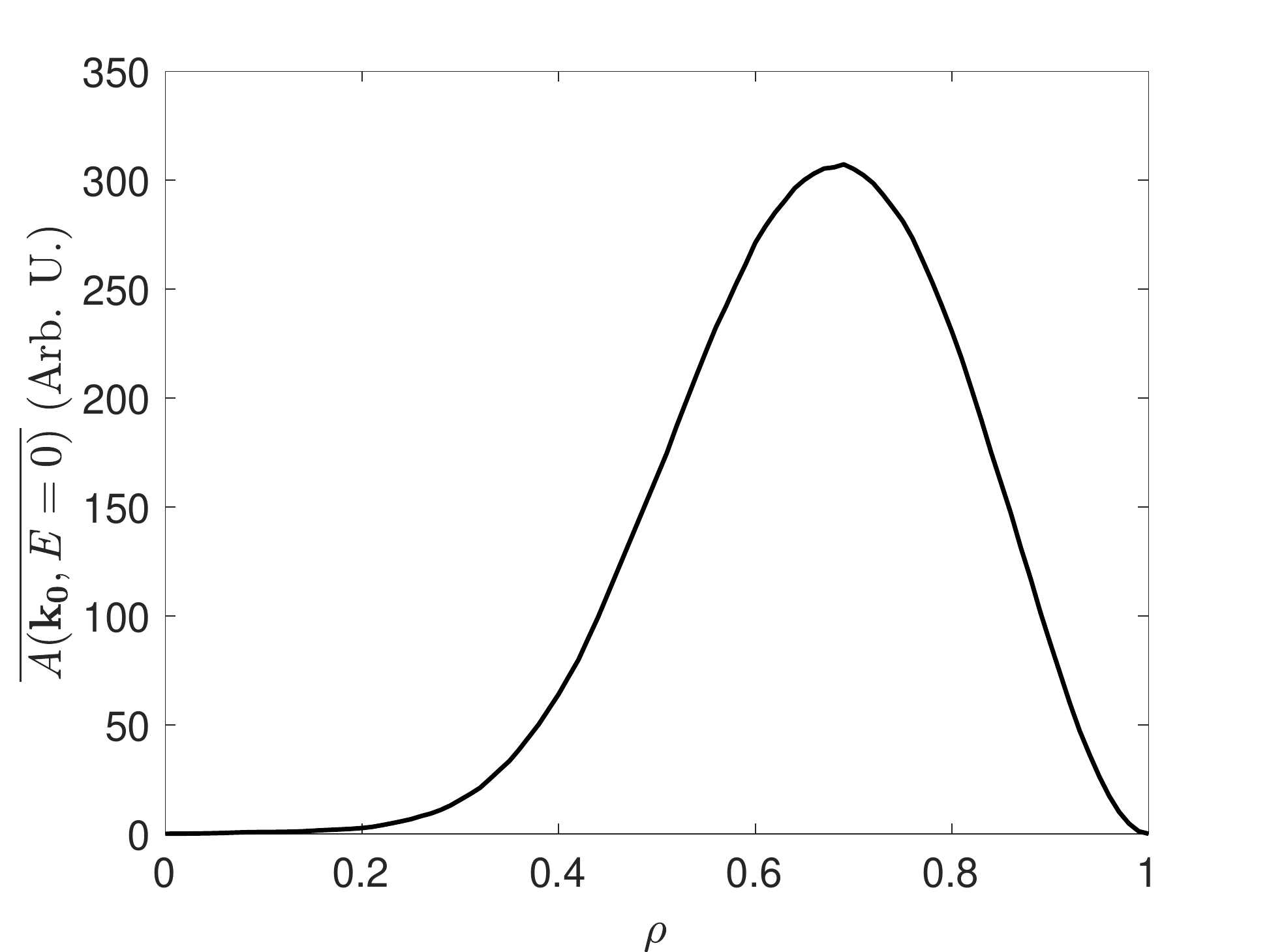}\label{fig:spectralfunctionE0}}\quad        
		\caption{\label{fig:spectralfunctions} Spectral function at CFS point $\overline{A(E, {\bm k}_0)}$ as a function of $E$ for different defect densities $\rho$ and $\bm{k}_0 = k_0 \hat{\bm{e}}_x$ with $k_0a = 0.6\pi $ ($a$. (a) Low-defect regime ($\rho = 0$ to $0.15$) (b) Intermediate-defect regime ($\rho = 0.2$ to $0.5$) (c) High-defect regime ($\rho = 0.6$ to $0.9$) (d) $\overline{A(E=0, {\bm k}_0)}$ as a function of $\rho$.
		}
\end{figure*}

\begin{figure*}[!htbp]
    \centering
\sidesubfloat[]{\includegraphics[width = 0.4\linewidth]{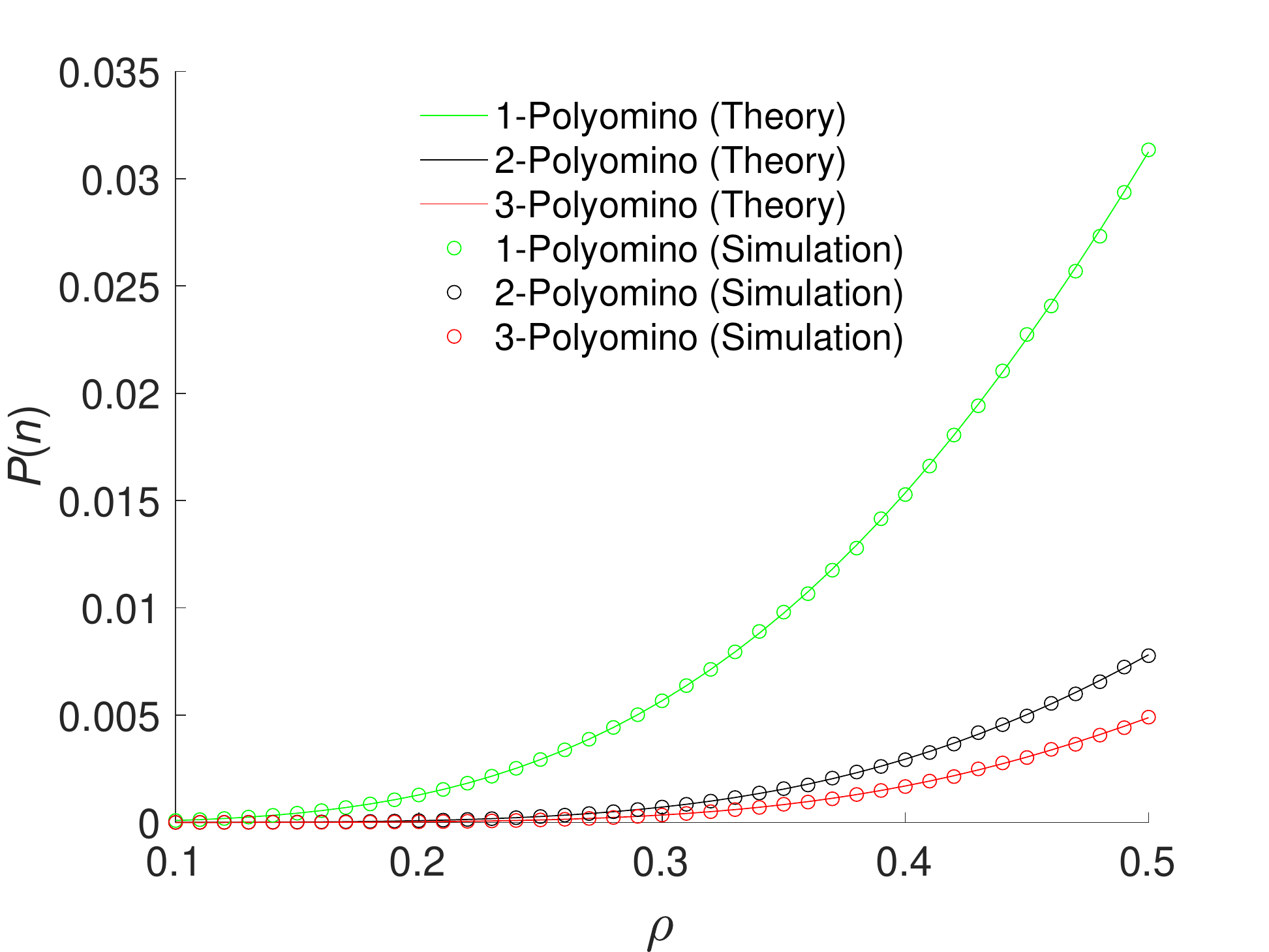}\label{fig:clusterall}}
\sidesubfloat[]{\includegraphics[width = 0.4\linewidth]{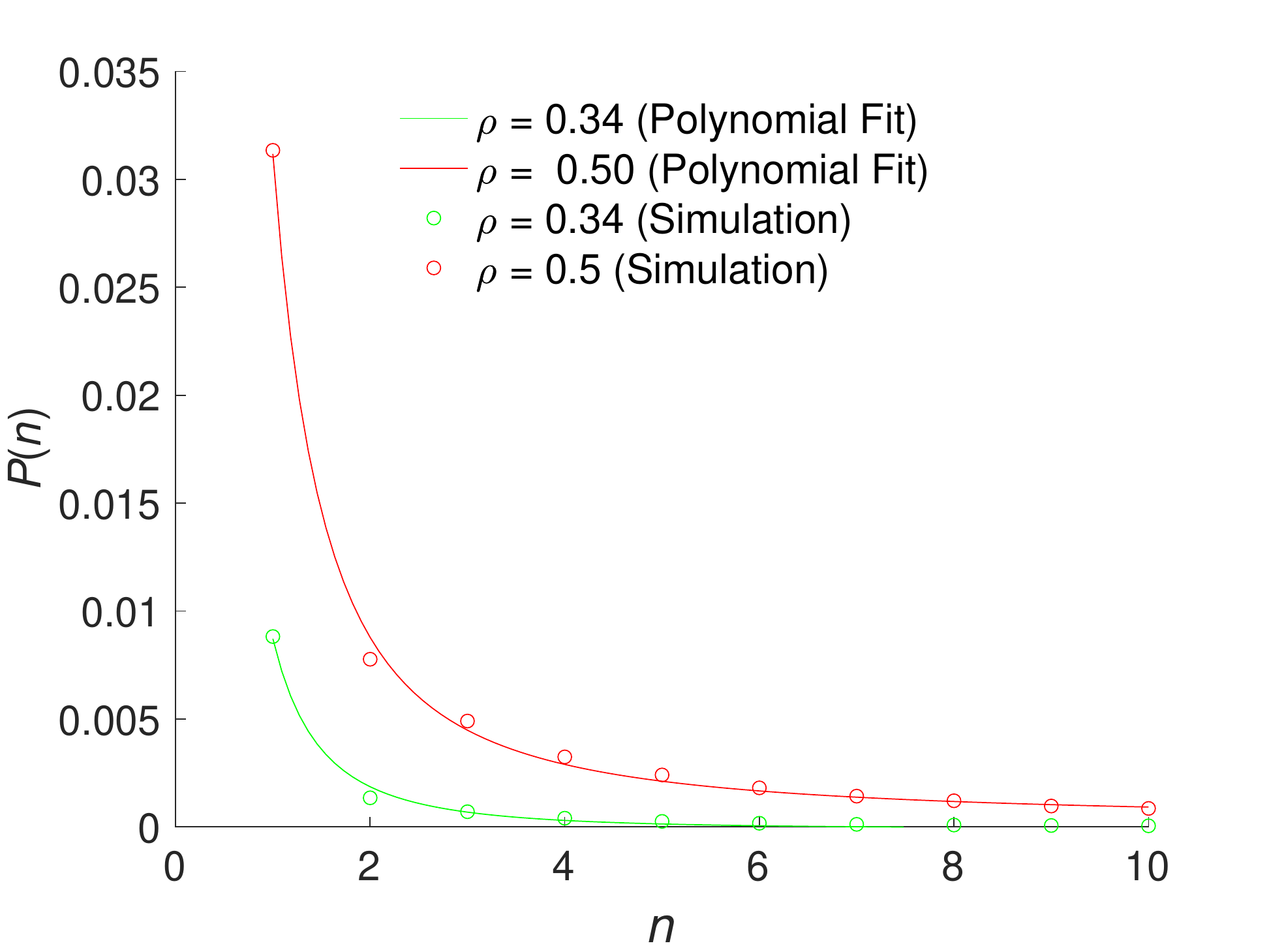}\label{fig:clustersizesmall}}

\sidesubfloat[]{\includegraphics[width = 0.4\linewidth]{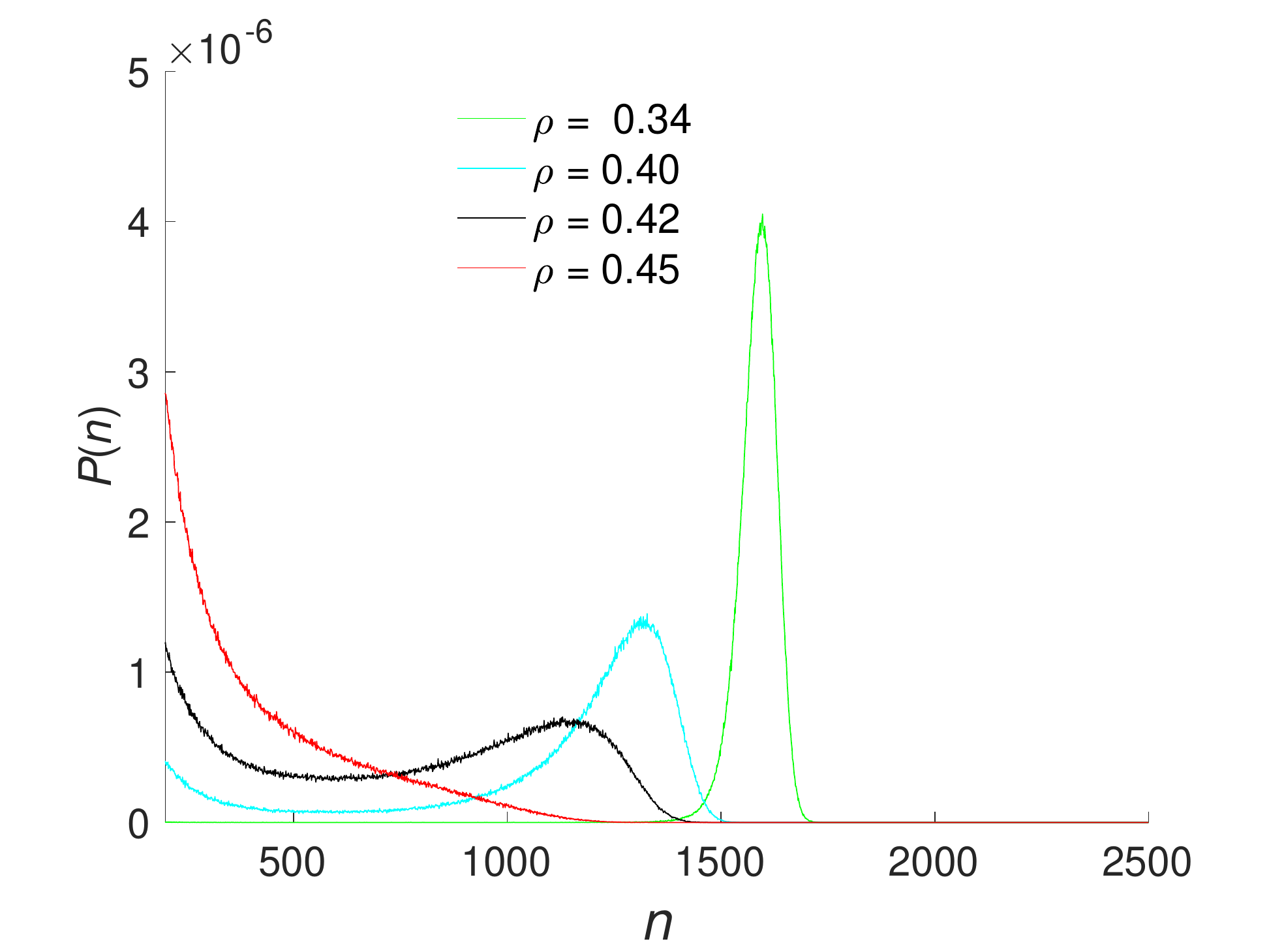}\label{fig:clustersizelarge}}
\sidesubfloat[]{\includegraphics[width = 0.4\linewidth]{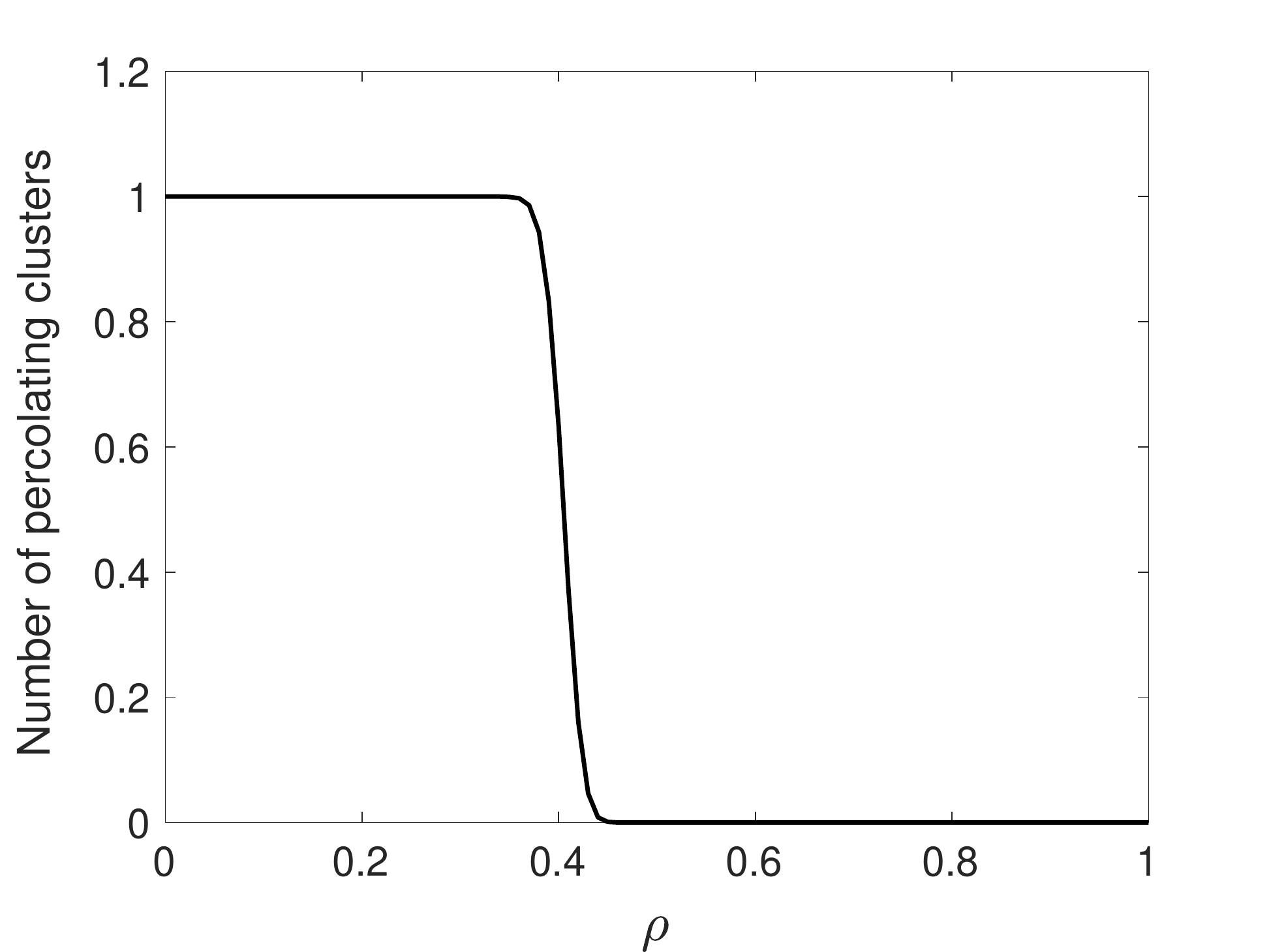}\label{fig:percolationthreshold}}
\caption{(a): Occurrence probability $P(n)$ for defect densities $0.1 \leq \rho \leq 0.5$ ($n=1,2,3$). (b): $P(n)$ behavior for small polyomino sizes $n \leq 20$. The solid line is a power-law (see text). (c): $P(n)$ behavior for large polyomino sizes $n \geq 200$.
(d): Number of $n$-polyominoes in our system with $N=L^2=2500$ sites as a function of $\rho$. (e): Percolation transition. The number of percolating clusters falls from 1 to 0 at the threshold $\rho^*$. Due to finite-size effects, our numerical value $0.44$ overestimates the actual accepted theoretical value $\rho^*\approx 0.41$ for this system.
}
    \label{fig:polyomino}
\end{figure*}

\section{Spectral Function}
\label{SpecFun}

The disorder-averaged spectral function is defined by
\begin{equation}
2 \pi\overline{\bra*{\bm{k}'}{\delta(E-H)\ket*{\bm{k}}}}=(2 \pi)^{d} \delta\left(\bm{k}-\bm{k}^{\prime}\right) \overline{A(\bm{k}, E)}, 
\end{equation}
where we define the plane wave states projected onto the magnetic lattice $\mathcal{M}$
\begin{equation}
    \ket*{\bm{k}} = \frac{1}{\sqrt{N(1-\rho)}}\sum_{i\in \mathcal{M}} e^{\mathrm{i}\bm{k}\cdot\bm{r}_i}\ket*{r_{i}}.
    \label{eq:modifiedplanewaves}
\end{equation}
with $N=L^2$ the number of clean lattice sites. These plane-wave states are normalized to $\bra*{\bm{k}}\ket*{\bm{k}}=1$. Using the $n$-polynomino expansion $\mathcal{H} = \sum_C \mathcal{H}_C$, we see that 
\begin{equation} \label{eq:SpecFun}
     \overline{A(\bm{k}, E)} = \overline{\sum_C \, A_C(\bm{k}, E)}, 
\end{equation}
where
\begin{equation}
\begin{aligned}
   &A_C(\bm{k}, E) = \sum_a \delta[E-E_a(C)] \, |\varphi_a(C, {\bm k})|^2 \\
   &\mathcal{H}_C \ket{\varphi_a(C)} = E_a(C) \, \ket{\varphi_a(C)}
\end{aligned}
\end{equation}
In performing the disorder average, we face the question of the statistical properties of eigenvalues and eigenfunctions of $\mathcal{H}_C$ when the polyomino $C$ changes. Here again, we argue that, under the disorder average, the total spectral function Eq.\eqref{eq:SpecFun} naturally breaks into a regular discrete component $A_d(\bm{k}, E)$ (originating from the regular Hamiltonians possessing eigenenergies immune to disorder) and a smooth component $A_s(\bm{k}, E)$ (originating from the random Hamiltonians):
\begin{equation}
     A_d(\bm{k}, E) = \sum_s A_s({\bm k}) \, \delta(E-E_s). 
\end{equation}
Fig.\ref{fig:spectralfunctions} shows how the spectral function changes when the defect density $\rho$ is varied.

We clearly see that the regular discrete component $A_d(\bm{k},E)$, completely negligible and invisible at $\rho \ll 1$, emerges gradually when $\rho$ is further increased while the smooth component is gradually depleted. When the percolation transition takes place, only small-size $n$-polyominoes survive and the smooth component goes extinct.

Using the identity
\begin{equation}
    U(t) = e^{-i \mathcal{H} t} = \int dE \ \delta(E-\mathcal{H}) \, e^{-iEt},
\end{equation}
it is easy to see that $n({\bm k}_0,t) = \int d\omega \, e^{-i\omega t} \, P({\bm k}_0,t)$ with 
\begin{equation}
    P(\bm{k}_{0},\omega) = \int \frac{dE}{(2\pi)^{2}}  \, \overline{A(E,\bm{k}_{0})A(E - \hbar\omega,\bm{k}_{0})}.
\end{equation}
We can use now the same argument developed above, to infer that the CFS power spectrum $P(\bm{k}_{0},\omega)$ also breaks into a regular discrete component $P_d(\bm{k}_{0},\omega)$, Eq.\eqref{eq:RegCFSPower}, originating from regular Hamiltonians,  and a smooth one $P_s(\bm{k}_{0},\omega)$ originating from random Hamiltonians. 

\section{Distribution of n-polyominoes}
\label{Perco}

The occurrence probability of a $n$-polyomino at defect density $\rho$ writes 
\begin{equation}
    P(n)(\rho)  =  \sum_{t} \, g_{n,t}\, (1-\rho)^{n}\rho^{t}, 
    \label{eq:polyomino}
\end{equation}
where $g_{n,t}$ denotes the number of distinct polyominoes with boundary $t$ and size $n$. Unfortunately, if one can compute $g_{n,t}$ for small-size polyominoes, there is no known analytic formula for this degeneracy factor. It is known that $g_{n,t}$ increases exponentially fast. Table \ref{tab:polyomino} gives the total number of possible $n$-polyomino arrangements as the size $n$ increases. \\
\begin{table}[H]
    \centering
    \begin{tabular}{c|cc}
    \hline
        n	& name & number of  arrangements \\
        \hline\hline
1 &	monomino &	1	\\
2 &	domino	& 	2\\
3 &	tromino	& 6\\
4 &	tetromino &	19\\
5 &	pentomino &	63\\
6 &	hexomino &	216\\
7 &	heptomino &	760\\
8 &	octomino &	2,725\\
9 &	nonomino &	9,910\\
10 &	decomino &	36,446\\
11 &	undecomino & 135,268\\
12 &	dodecomino	& 	505,861 \\
\hline
    \end{tabular}
    \caption{Number of spatial arrangements of n-polyominoes \cite{jensen2000statistics}}.
    \label{tab:polyomino}
\end{table}
One can nevertheless efficiently estimate $P(n)$ numerically by generating a large number of random configurations ($10^{6}$ configurations are used in our numerics) and computing the total fraction of $n$-polyominoes found, see Fig.\ref{fig:clusterall} \cite{redelmeier1981counting}.

Starting from a lattice grid with $N=2500$ sites, the percolation transition is easily seen in Fig.\ref{fig:clustersizelarge}, where $P(n)$, peaked at high cluster sizes for low $\rho$, disappears completely around $\rho \approx 0.45$. Due to finite-size effects, we find a percolation threshold at about $0.44$ instead of the predicted value $\rho^* \approx 0.41$ \cite{newman2000efficient}. We also observe that the polyomino distribution in Fig.\ref{fig:clustersizesmall} can be fit by a power law $P(n) = n_{0}n^{-\tau}+C n^{-\Omega}$ with critical exponents $\tau=-187/91$ and $\Omega = -0.702$ as found in the literature \cite{tiggemann2001simulation}. Hence, despite working with a relatively small system size, finite-size effects do not  significantly alter the polyomino distributions in our system.

\end{document}